\documentclass[twocolumn]{aastex631}

\usepackage{mleftright}
\graphicspath{{./}{figures/}}
\usepackage{graphicx}
\usepackage[caption=false]{subfig}
\usepackage{amsmath}
\usepackage{float}
\usepackage{url}
\usepackage{natbib}
\usepackage{mathtools,nccmath}
\usepackage{placeins}
\usepackage{tablefootnote}
\usepackage{tabularx}

\usepackage{etoolbox}
\makeatletter
\patchcmd\linenumberpar{\@LN@parpgbrk}{\penalty\@LN@parpgpen\relax}{}{}
\makeatother

\begin{document}

\title{Prototype Faraday rotation measure catalogs from the Polarisation Sky Survey of the Universe’s Magnetism (POSSUM) pilot observations}

\correspondingauthor{S. Vanderwoude}
\email{vanderwoude@astro.utoronto.ca}

\author[0009-0004-7773-1618]{S. Vanderwoude}
\affiliation{David A. Dunlap Department of Astronomy and Astrophysics, University of Toronto, Toronto, ON M5S 3H4, Canada}
\affiliation{Dunlap Institute for Astronomy and Astrophysics, University of Toronto, Toronto, ON M5S 3H4, Canada}

\author[0000-0001-7722-8458]{J. L. West}
\affiliation{Dunlap Institute for Astronomy and Astrophysics, University of Toronto, Toronto, ON M5S 3H4, Canada}
\affiliation{Dominion Radio Astrophysical Observatory, Herzberg Astronomy and Astrophysics, National Research Council Canada, P.O. Box 248, Penticton, BC V2A 6J9, Canada}

\author[0000-0002-3382-9558]{B. M. Gaensler}
\affiliation{Dunlap Institute for Astronomy and Astrophysics, University of Toronto, Toronto, ON M5S 3H4, Canada}
\affiliation{David A. Dunlap Department of Astronomy and Astrophysics, University of Toronto, Toronto, ON M5S 3H4, Canada}
\affiliation{Present address: Division of Physical and Biological Sciences, University of California Santa Cruz, Santa Cruz, CA 95064, USA}

\author[0000-0001-5636-7213]{L. Rudnick}
\affiliation{Minnesota Institute for Astrophysics, University of Minnesota, Minneapolis, MN, USA}

\author[0000-0002-7641-9946]{C. L. Van Eck}
\affiliation{Research School of Astronomy \& Astrophysics, The Australian National University, Canberra, ACT 2611, Australia}
\affiliation{Dunlap Institute for Astronomy and Astrophysics, University of Toronto, Toronto, ON M5S 3H4, Canada}

\author[0000-0001-9472-041X]{A. J. M. Thomson}
\affiliation{ATNF, CSIRO Space \& Astronomy, PO Box 1130, Bentley WA 6102, Australia}

\author[0000-0003-4873-1681]{H. Andernach}
\affiliation{Th\"uringer Landessternwarte, Sternwarte 5, D-07778 Tautenburg, Germany}
\affiliation{Permanent Address: Depto.\ de Astronom\'{i}a, Univ.\ de Guanajuato,
Callej\'on de Jalisco s/n, Guanajuato, C.P.\ 36023, GTO, Mexico}

\author[0000-0002-6243-7879]{C. S. Anderson}
\affiliation{Research School of Astronomy \& Astrophysics, The Australian National University, Canberra, ACT 2611, Australia}

\author[0000-0002-3973-8403]{E. Carretti}
\affiliation{INAF, Istituto di Radioastronomia, Via Gobetti 101, 40129 Bologna, Italy}

\author[0000-0002-2155-6054]{G. H. Heald}
\affiliation{ATNF, CSIRO Space \& Astronomy, PO Box 1130, Bentley WA 6102, Australia}

\author[0000-0003-2514-9592]{J. P. Leahy}
\affiliation{Jodrell Bank Centre for Astrophysics, Department of Physics \& Astronomy, The University of Manchester, Oxford Road, Manchester M13 9PL, UK}

\author[0000-0003-2730-957X]{N. M. McClure-Griffiths}
\affiliation{Research School of Astronomy \& Astrophysics, The Australian National University, Canberra, ACT 2611, Australia}

\author[0000-0002-3968-3051]{S. P. O'Sullivan}
\affiliation{Departamento de Física de la Tierra y Astrofísica \& IPARCOS-UCM, Universidad Complutense de Madrid, 28040 Madrid, Spain}

\author[0000-0001-8749-1436]{M. Tahani}
\affiliation{Banting and KIPAC Fellowships: Kavli Institute for Particle Astrophysics \& Cosmology (KIPAC), Stanford University, Stanford, CA 94305, USA}

\author[0000-0002-2173-6151]{A. G. Willis}
\affiliation{Dominion Radio Astrophysical Observatory, Herzberg Astronomy and Astrophysics, National Research Council Canada, P.O. Box 248, Penticton, BC V2A 6J9, Canada}

\begin{abstract}

The Polarisation Sky Survey of the Universe's Magnetism (POSSUM) will conduct a sensitive $\sim$1 GHz radio polarization survey covering 20 000 square degrees of the Southern sky with the Australian Square Kilometre Array Pathfinder (ASKAP). In anticipation of the full survey, we analyze pilot observations of low-band (800--1087 MHz), mid-band (1316--1439 MHz), and combined-band observations for an extragalactic field and a Galactic-plane field (low-band only). Using the POSSUM processing pipeline, we produce prototype RM catalogs that are filtered to construct prototype RM grids. We assess typical RM grid densities and RM uncertainties and their dependence on frequency, bandwidth, and Galactic latitude. We present a median filter method for separating foreground diffuse emission from background components, and find that after application of the filter, 99.5\% of measured RMs of simulated sources are within 3$\sigma$ of their true RM, with a typical loss of polarized intensity of 5\% $\pm$ 5\%. We find RM grid densities of 35.1, 30.6, 37.2, and 13.5 RMs per square degree and median uncertainties on RM measurements of 1.55, 12.82, 1.06, and 1.89 rad m$^{-2}$ for the median-filtered low-band, mid-band, combined-band, and Galactic observations, respectively. We estimate that the full POSSUM survey will produce an RM catalog of $\sim$775 000 RMs with median-filtered low-band observations and $\sim$877 000 RMs with median-filtered combined-band observations. We construct a structure function from the Galactic RM catalog, which shows a break at $0.7^{\circ}$, corresponding to a physical scale of 12-24 pc for the nearest spiral arm.

\end{abstract}


\vspace{5mm}

 \section{Introduction}\label{sec:intro}

While magnetic fields are present everywhere in the Universe, their distribution, strength, and morphology are generally poorly known. This is largely due to the difficulty in detecting these fields. In most cases magnetic fields cannot be detected directly. We rely instead on indirect measurements, one of which is the Faraday rotation of linearly polarized radio sources by intervening magnetized plasma (see \citealt{Beck2015} and \citealt{Han2017} for reviews). A collection of Faraday rotation measures (RMs), quantifiers of the magnitude and direction of the Faraday rotation of polarized extragalactic radio sources, plotted together on a region of the sky, is referred to as an RM grid \citep{Gaensler+04}. RM grids, and the catalogs of measured RMs that are used to construct them, have been an invaluable method of studying magnetic fields in many environments, including the large-scale Galactic magnetic field \citep{Mao+10,VanEck+11,Hutschenreuter+22}, molecular clouds \citep{Tahani+18}, the jets and lobes of radio galaxies \citep{Feain+09,O'Sullivan+18}, galaxy clusters \citep{Bonafede+10,Anderson+21}, and the cosmic web \citep{Vernstrom+19,Amaral+21,Carretti+22}.

The modern era of RM grids opened with the \citet{Taylor2009} RM catalog of the NRAO VLA Sky Survey (NVSS; \citealt{Condon+98}), which contains 37 543 polarized radio sources north of declination $-40^{\circ}$ (J2000) observed at two frequencies, 1364.9 MHz and 1435.1 MHz. With an average sky density of $\sim$1 RM per square degree, this RM grid mapped the broad features of RM structure across the northern sky, providing valuable information about the geometry and direction of the ordered component of the Galactic magnetic field on the largest scales \citep{Sun&Reich2010,Pshirkov+11,Stil+11}. \citet{Oppermann2012} assembled a more extensive set of  41 330 RMs from the \citet{Taylor2009} catalog and other smaller RM catalogs that included RMs in the southern sky, although this latter region remained poorly sampled. \citet{Oppermann2012,Oppermann2015} and \citet{Hutschenreuter+20} used this extended catalog in combination with Bayesian inference to construct a smoother, more detailed map of the RM sky.

Polarization surveys in the southern hemisphere have since helped increase the number of measured RMs in the southern sky. S-PASS/ATCA \citep{Schnitzeler+19} measured the RMs of polarized sources in the southern radio sky at 1300 -- 3100 MHz, with an average density of $\sim$1 RM per 5 square degrees. The POlarized GLEAM Survey (POGS; \citealt{Riseley+18,Riseley+20}) observed the southern sky at lower frequencies, 200 -- 231 MHz, with a density of $\sim$1 RM per 80 square degrees. Recently, the first data release from Spectra and
Polarisation In Cutouts of Extra-galactic sources from RACS
(SPICE-RACS; \citealt{Thomson+23}) mapped 5818 RMs across $\sim$1300 square degrees of the southern sky at 744 -- 1032 MHz. Another low frequency survey in the northern hemisphere, the LOFAR Two-metre Sky Survey (LoTSS, \citealt{Shimwell+17,O'Sullivan+23}), observed from 120 to 168 MHz with an RM sky density of 0.43 RMs per square degree. Most recently, \citet{VanEck+23} consolidated the RM catalogs from 42 publications to produce the largest RM catalog to date, with 55 819 RM measurements across the full sky.  \citet{Hutschenreuter+22} used this catalog to construct the most complete RM sky map to date.

While the collection of available RMs has continued to grow, the average density over the full sky is still just $\sim$1.35 RM per square degree. Low-density RM grids limit our ability to measure weak magnetic fields \citep{Akahori+14a} and smaller-scale structure in our Galaxy and others \citep{Stepanov+08,Tahani+19,Tahani+22}. Furthermore, the RMs from the \citet{Taylor2009} catalog, still by far the largest contributor to the Faraday depth sky map, were derived from a linear fit to the polarization angle as a function of the wavelength squared at just two, relatively close frequencies. This is a problem for two reasons: 1) limited frequencies means the RM measurement may suffer from n$\pi$-ambiguity \citep{Brentjens2005}, meaning that the true RM may have a different magnitude or sign what is calculated \citep{Rand&Lyne1994}, and 2) this method can return an incorrect RM outside of simplest of physical case where there is no depolarization, turbulence, or mixing of synchrotron-emitting and Faraday-rotating plasma. To overcome these problems, broadband spectropolarimetric observations are ideal for use in combination with Faraday rotation measure synthesis (\citealt{Burn1966,Brentjens2005}; see Section \ref{subsec: phi and RM synth} for a more detailed description).

In this paper we compile RM catalogs using pilot observations from the Australian Square Kilometre Array Pathfinder (ASKAP; \citealt{Hotan+21}) and use them to construct prototype RM grids. We use these grids to showcase capabilities of the Polarisation Sky Survey of the Universe's Magnetism (POSSUM; \citealt{Gaensler+10}), and characterise technical aspects of ASKAP data of the POSSUM data processing pipeline. Further, we are able to better understand the limitations of the data, and inform future science projects that will make use of POSSUM RMs. We discuss ``components" in this work instead of ``sources" because a background synchrotron source may be composed of multiple components with individual RMs (see Section \ref{subsec: sourcefinder} for a more detailed discussion). The technical aspects we characterize include:

\begin{itemize}

    \item expected number of measured RMs and RM grid sky densities of the full POSSUM survey
    
    \item typical uncertainties on RM measurements
    
    \item dependence of data quality, RM uncertainties, and component densities on frequency, bandwidth, and Galactic latitude
    
    \item fraction of components that can be used to construct an RM grid

    \item effects of foreground polarized diffuse emission on polarization measurements of background components
    
\end{itemize}

\noindent We present RM catalogs and prototype RM grids of our four observations. We identify and address questions and challenges that the POSSUM survey will face when constructing an RM catalog of this magnitude and showcase the exceptional RM grid density that POSSUM and ASKAP will achieve.

In Section \ref{sec: obs} of this paper we describe the data. In Sections \ref{sec: DE contam} and \ref{sec: data reduc} we describe our data reduction process, including the separation of foreground diffuse emission from background components and extracting polarization properties. In Section \ref{sec: results} we present four prototype RM grids and their corresponding component catalogs, assess the data quality, and compare the properties of the catalogs. We discuss our results and forecast future science with POSSUM in Section \ref{sec: disc}, and we provide a summary of our conclusions in Section \ref{sec: conclusion}.

\subsection{Faraday depth and rotation measure synthesis}
\label{subsec: phi and RM synth}

Synchrotron radiation is emitted by relativistic electrons as they gyrate around magnetic field lines. This emission dominates the radio sky below $\sim$30 GHz and is locally highly linearly polarized, offering an invaluable way to probe Galactic and extragalactic magnetic fields. With Stokes parameters I (total intensity), and $Q$ and U (orthogonal linear polarizations), we can define the total linear polarized intensity, $P$, the polarization position angle (increasing east from north), $\psi$, and the fractional linear polarization, $p$, of the emission as:

\begin{equation}\label{eq: P, psi, p}
    P = \sqrt{Q^2 + U^2}, \quad \psi = \frac{1}{2} \arctan \bigg( \frac{U}{Q} \bigg), \quad p \equiv \frac{P}{I} .
\end{equation}

\noindent This information can then be encoded in the complex linear polarization vector $\widetilde{P} = Q + iU = pI e^{2i\psi}$ \citep{Burn1966}.

When polarized emission passes through a region of magnetized thermal plasma, $\psi$ experiences a wavelength-dependent rotation known as Faraday rotation. In the simplest case, where there is one emitting component and no co-spatial emission and rotation (i.e. the relativistic and thermal electrons are not co-spatial), the amount of rotation experienced by the emission is called the rotation measure, or RM, and is the slope of the linear relationship between polarization angle and the square of the wavelength, $\lambda$,:

\begin{equation}\label{eq: delta PA}
    \Delta \psi = RM \lambda^2 \, .
\end{equation}

\noindent In the more complicated case, synchrotron emission and Faraday rotation can occur within the same region due to the same magnetic field. In this case, we instead define the more general term Faraday depth, $\phi$, to quantify the amount of Faraday rotation from a specific region at a specific distance from the observer, $l$. Integrated along the line of sight, $\phi$ is expressed as:

\begin{equation}\label{eq: phi}
    \phi(l) = 0.812 \int^{l}_{\mathrm{0}} n_e \, \textbf{B} \cdot \mathrm{d}\mathbf{l^{\prime}} \; \; \mathrm{rad} \; \mathrm{ m^{-2}} \, ,
\end{equation}

\noindent where $n_e$ is the thermal electron density in units of cm$^{-3}$, \textbf{B} is the magnetic field vector in units of $\mu$G, and $l$ is the path length in parsecs \citep{Burn1966}. In the simple case, $\mathrm{RM} = \phi$. The convention that we follow in this work is $\phi > 0$ when the line of sight magnetic field is pointing toward the observer and $\phi < 0$ when the magnetic field is pointing away from the observer \citep{Ferriere+21}. As Equation \ref{eq: delta PA} indicates, polarized signals will experience more Faraday rotation at longer wavelengths than at shorter wavelengths.

In this work we use the RM synthesis technique \citep{Burn1966,Brentjens2005} to determine the polarization properties of our components. RM synthesis, combined with denser sampling in wavelength squared space, is a powerful diagnostic tool for studying Faraday rotation and polarization, and overcomes the n$\pi$ ambiguity problem that angle fitting faces. RM synthesis takes the complex polarized fraction $\widetilde{p} \equiv \frac{\widetilde{P}}{I}$ and returns a complex Faraday spectrum  (also known as the Faraday dispersion function):

\begin{equation}
    F(\phi) = \sum_{j=1}^N w_j \widetilde{p_j} \mathrm{e}^{-2i \phi (\lambda^2_j - \lambda^2_0)} \bigg/ \sum_{j=1}^N w_j ,
\end{equation}

\noindent where

\begin{equation}\label{eq: ref lamsq}
    \lambda_0^2 = \sum_{j=1}^N w_j \lambda^2_j \bigg/ \sum_{j=1}^N w_j .
\end{equation}

\noindent $N$ is the number of frequency channels, $\widetilde{p_j}$ and $\lambda_j$ are the complex fractional polarization and wavelength in channel $j$, respectively, and $w_j$ is the weight in channel $j$, which we set as the inverse square of the channel noise. In the real-world case of data with finite wavelength coverage and discrete wavelength channels, the Faraday spectrum is convolved with the rotation measure spread function (RMSF), the normalized response in Faraday depth space to incomplete sampling. Narrower $\lambda^2$ coverage will return a broader RMSF. The RMSF determines the resolution of the Faraday spectrum in Faraday depth space:

\begin{equation}\label{eq: rmsf fwhm}
    \delta \phi  = \frac{3.79}{\Delta \lambda^2} \, \, , \,\,\, \mathrm{where} \,\,\, \Delta \lambda^2 = \lambda_{\mathrm{max}}^2 - \lambda_{\mathrm{min}}^2 .
\end{equation}

\noindent This is equal to the full width at half maximum (FWHM) of the RMSF. The largest detectable value of $\phi$ is given by:

\begin{equation}\label{eq: phi max}
    |\phi_{\mathrm{max}}| = \frac{1.9}{\delta \lambda^2} \, ,
\end{equation}

\noindent where $\delta \lambda^2$ is the wavelength squared channel width  \citep{Dickey+19}. The width of a Gaussian distribution in $\phi$ space at which the sensitivity drops by a factor of two is 

\begin{equation}\label{eq: W max}
    W_{\mathrm{max}} = 0.67 \times (\lambda_{\mathrm{min}}^{-2} + \lambda_{\mathrm{max}}^{-2})
\end{equation}

\noindent where $\lambda_{\mathrm{min}}^2$ and $\lambda_{\mathrm{max}}^2$ are the shortest and longest wavelengths squared of the observation, respectively \citep{Rudnick&Cotton2023}.

We refer to each independent feature in the Faraday spectrum as a ``peak", where the value of $\phi$ associated with the peak is determined by the position of the maximum amplitude of the feature. In the simplest case, there is a single peak in the Faraday spectrum for which $\phi$ = RM. In the more complicated cases, there may be any combination of broadened and multiple peaks (see Section \ref{subsec: complexity descrip} for further discussion of these cases). We refer to performing RM synthesis on a single line of sight as 1D RM synthesis, and this is how we calculate the RM for individual components. 3D RM synthesis involves performing 1D RM synthesis along each line of sight, or at the position of each pixel, in a 3D image cube where the third axis is frequency.

\section{Pilot observations}
\label{sec: obs}

\subsection{ASKAP}\label{subsec: askap}

The Australian Square Kilometre Array Pathfinder (ASKAP; \citealt{Hotan+21}) is located at the Inyarrimanha Ilgari Bundara, the Commonwealth Scientific and Industrial Research Organisation (CSIRO) Murchison Radio-astronomy Observatory in Wajarri Yamaji Country in Western Australia. The telescope consists of 36 12-meter dishes, with a longest baseline of 6440 meters. Each antenna is equipped with a phased array feed (PAF) that consists of 188 individual receivers. The receivers are combined to create 36 formed beams (different from the telescope's synthesized beam), which, when mosaicked together, give ASKAP a $\sim$30 deg$^2$ instantaneous field of view at 800 MHz. The arrangement of these beams in a mosaic is referred to as the observation footprint.

ASKAP observes at 700 -- 1800 MHz with a 288 MHz instantaneous bandwidth. The observations that we analyze in this paper are from three ASKAP Pilot Surveys: the Evolutionary Map of the Universe (EMU; \citealt{Norris+11}) Pilot I Survey \citep{Norris+21}, and the POSSUM Pilot I and Pilot II Surveys (West 2023, in prep). The EMU Pilot I and POSSUM Pilot I Surveys were designed to be commensal, observing the same region of sky at different frequencies to evaluate the polarization capabilities of ASKAP.

\subsubsection{Pilot Surveys}\label{subsubsec: pilot survey descrip}

The EMU Pilot I Survey was conducted from mid-to-late 2019 and is comprised of 10 contiguous fields centered on right ascension (RA J2000) 319.500$^{\circ}$ and declination (Dec J2000) $-55.725^{\circ}$ (Galactic longitude \textit{l} = 340.750$^{\circ}$, Galactic latitude \textit{b} = $-42.526^{\circ}$). The EMU Pilot I Survey was observed at a lower frequency band than the POSSUM Pilot I Survey and as such has a larger field of view. The fields of each survey are contiguous, making their centers somewhat offset. Each field in the EMU Pilot I Survey had an integration time of ten hours and the beams were formed in the \textsc{closepack36} footprint, a trapezoidal configuration with 6 rows of six beams each, with a beam pitch (the separation between the centers of the formed beams) of 0.9$^{\circ}$. This footprint places the formed beams in closer overlap than the \textsc{square6x6} configuration used in other ASKAP observations \citep{Anderson+21,Thomson+23}. The survey was observed at full resolution at the ASKAP low-band frequency range (800 -- 1087 MHz) and averaged to a 288 MHz bandwidth with 1-MHz channel width and a central frequency of 943 MHz.

The POSSUM Pilot I Survey was conducted from mid-to-late 2019 and is comprised of 10 contiguous fields centered on RA (J2000) 321.815$^{\circ}$ and Dec (J2000) $-54.670^{\circ}$ (\textit{l} = 341.683$^{\circ}$, \textit{b} = $-44.074^{\circ}$). Each field had an integration time of ten hours and the beams were formed in the \textsc{closepack36} footprint with a beam pitch of 0.75$^{\circ}$. The survey was observed at full resolution at the ASKAP mid-band frequency range (1152 -- 1439 MHz) and averaged to a 288 MHz bandwidth with 1-MHz channel width and a central frequency of 1377 MHz. In both the EMU and POSSUM pilot I surveys, there is some overlap ($\lesssim$ 5\%) of the edges of the fields to ensure that they are fully contiguous.  The POSSUM pilot I survey is entirely contained within the extent of the EMU pilot I survey. Due to the larger field of view of the EMU pilot I survey (a result of the lower observing band), it covers an additional $\sim$100 deg$^2$ of the sky than the POSSUM pilot I survey.

The POSSUM Pilot II Survey was conducted in 2022 and is comprised of 10 non-contiguous fields pointing at a variety of science targets, including one field near the Galactic plane, which is the field that is analyzed in this work. The Galactic plane observation had an integration time of ten hours and the beams were formed in the \textsc{closepack36} footprint with a beam pitch of 0.9$^{\circ}$. The field was observed at full resolution at the ASKAP low-band frequency range and averaged to a 288 MHz bandwidth with 1-MHz channel width and a central frequency of 943 MHz.

\subsection{Data}

\subsubsection{Low-band Pilot I observation}
\label{subsubsec: SB10635}

The low-band Pilot I observation covers $\sim$30 deg$^2$ centered on RA (J2000) 331.490$^{\circ}$ and Dec (J2000) $-51.192^{\circ}$ (\textit{l} = 343.778$^{\circ}$, \textit{b} = $-50.706^{\circ}$). It was observed on November 24, 2019 as part of the EMU Pilot I Survey (observation SB10635), described in Section \ref{subsubsec: pilot survey descrip}. Two frequency channels are completely flagged, leaving 286 channels for analysis. The Stokes cubes were convolved to a common angular resolution of 21 arcsec across all frequency channels. We will refer to this observation as Extragalactic-Low (EL) throughout the paper.

The formed beams have an approximately Gaussian response and peak sensitivity near the center, and the response becomes distorted toward the edges \citep{Duchesne+23,Thomson+23}. Mosaicking the beams helps achieve approximately uniform sensitivity across the majority of the image by overlapping the lower-sensitivity regions of two beams to increase the overall sensitivity to that of the central beam. We focus our analysis on the central part of the mosaic where the sensitivity is approximately uniform, avoiding the edges of the mosaic that have a higher level of noise. We define a region of uniform sensitivity in the central part of the observation within which we perform all of our analysis. This region was chosen to maximize the area of the observation within which each source has been observed either near the center of a single formed beam or by multiple formed beams. See Figure \ref{fig: footprints} for the location of the region of uniform sensitivity (green) with respect to the beam footprint (yellow). The region has an area of 11.52 deg$^2$ and has a band-averaged sensitivity of $\sim$24 $\mu$Jy/beam in Stokes I.

\subsubsection{Mid-band Pilot I observation}
\label{subsubsec: 10043}

The mid-band Pilot I observation covers $\sim$20 deg$^2$, centered on RA 332.052$^{\circ}$ and Dec $-50.870^{\circ}$ (\textit{l} = 344.044$^{\circ}$, \textit{b} = $-51.154^{\circ}$). It was observed on September 29, 2019 as part of the ten-field POSSUM Pilot I Survey (observation SB10043), described in Section \ref{subsubsec: pilot survey descrip}. 
While the original observation spans 288 MHz, the first 164 channels were contaminated by radio frequency interference (RFI) and were discarded. An additional eight channels are flagged, leaving 116 channels for analysis. We will refer to this observation as Extragalactic-Mid (EM) throughout the paper.

The Stokes cubes were convolved to a common angular resolution of 13 arcsec across all frequency channels. We define a 11.52 deg$^2$ region of uniform sensitivity in this observation. This region is centered on the same location as the region of uniform sensitivity in the EL observation, described in Section \ref{subsubsec: SB10635}, because these two observations are coincident on the sky (see Section \ref{subsubsec: pilot survey descrip}). See Figure \ref{fig: footprints} for the location of the region of uniform sensitivity (green) with respect to the beam footprint (yellow). The band-averaged sensitivity of this region in the EM observation is $\sim$30 $\mu$Jy/beam in Stokes I.

\subsubsection{Combined-band Pilot I observation}

In addition to the individual low- and mid-band observations, we jointly analyzed the combined low- and mid-band regions of uniform sensitivity. To combine the data sets, the mid-band Stokes cubes were convolved to the low-band angular resolution of 21 arcsec and the extracted low- and mid-band spectra of each component were joined together with the lowest mid-band frequency following after the highest low-band frequency (see Section \ref{subsec: sourcefinder}).
The combined data set has 402 1-MHz frequency channels over a bandwidth of 640 MHz, with a central frequency of 1119 MHz and a gap from 1087 to 1316 MHz.  We will refer to this observation as Extragalactic-Combined (EC) throughout the paper.

\subsubsection{Galactic Pilot II observation}
\label{subsec: obs galactic field}

A field near the Galactic plane was observed as part of the POSSUM Pilot II Survey (observation SB43773), described in Section \ref{subsubsec: pilot survey descrip}, and is centered on RA 238.498$^{\circ}$ and Dec $-55.730^{\circ}$ (\textit{l} = 326.680$^{\circ}$, \textit{b} = $-1.514^{\circ}$). The field was observed on September 21, 2022. 
This observation has the same frequency band as the EL Pilot I observation, making these observations ideal for comparison of parameters such as component density as a function of declination. Ten frequency channels are flagged, leaving 278 channels for analysis. The Stokes cubes were convolved to a common angular resolution of 16.5 arcsec across all frequency channels. We will refer to this observation as Galactic-Low (GL) throughout the paper.

\begin{figure*}
\centering 
\subfloat{%
  \includegraphics[width=0.47\textwidth]{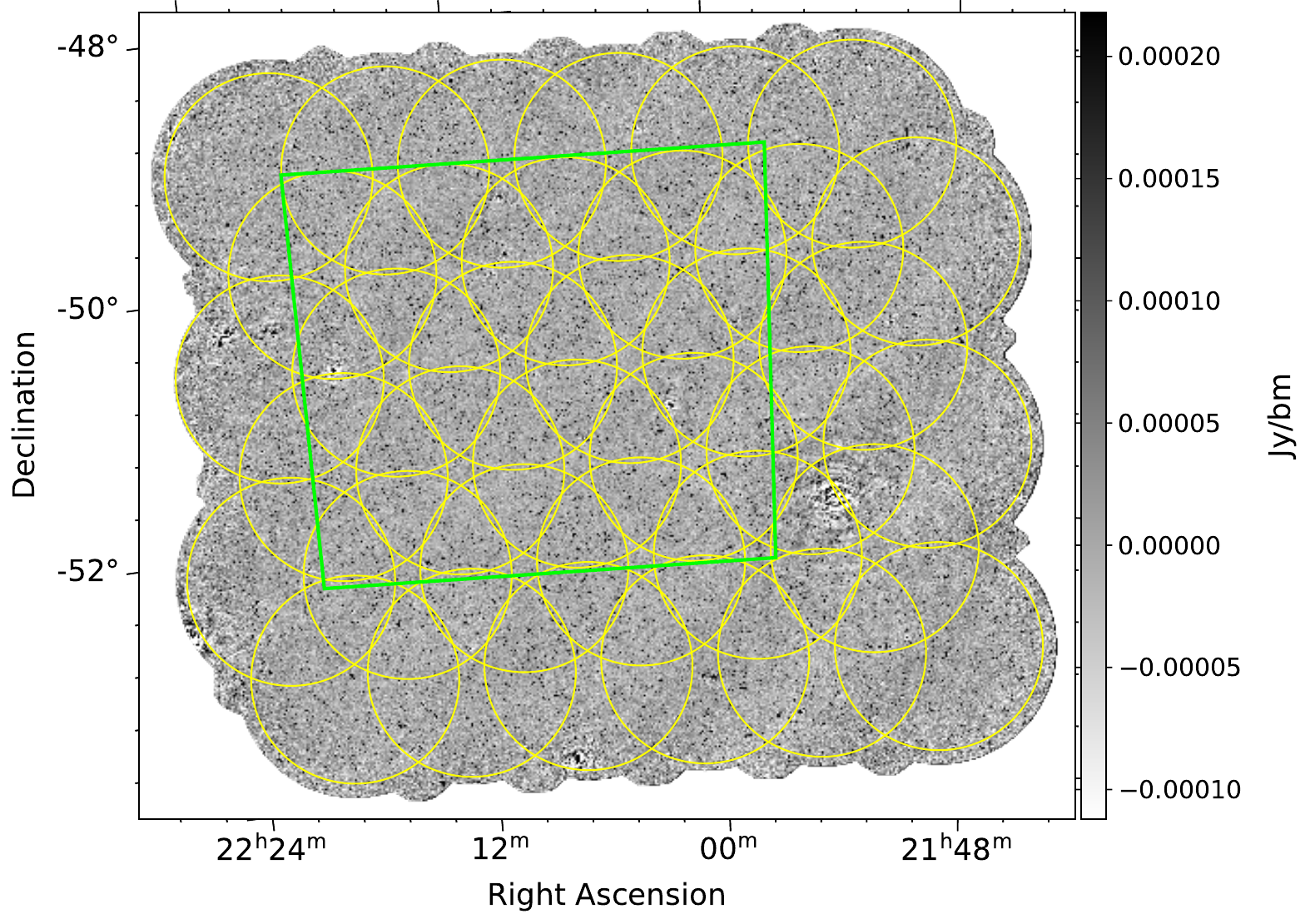}%
}\qquad
\subfloat{%
  \includegraphics[width=0.483\textwidth]{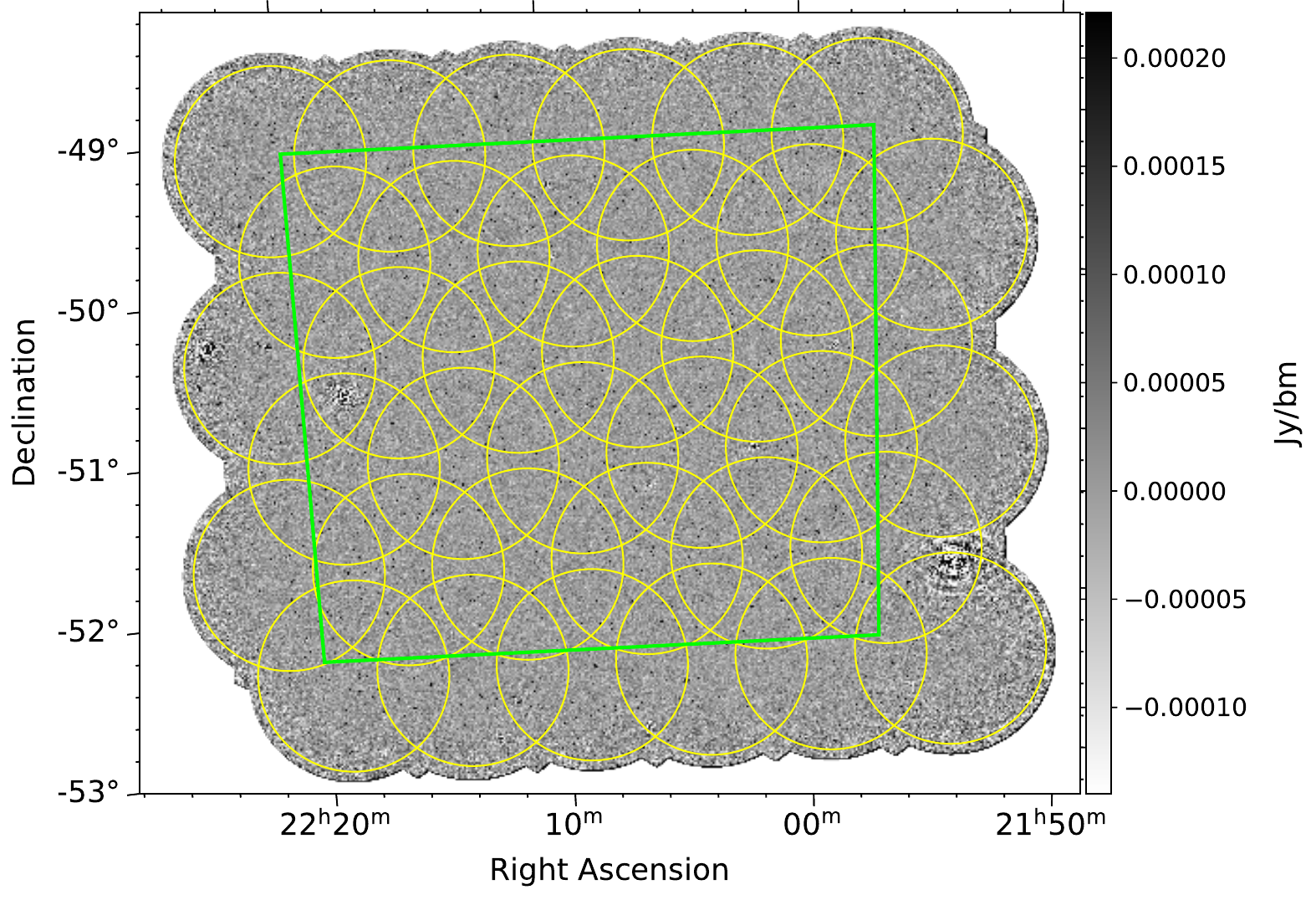}%
}\newline
\subfloat{%
  \includegraphics[width=0.47\textwidth]{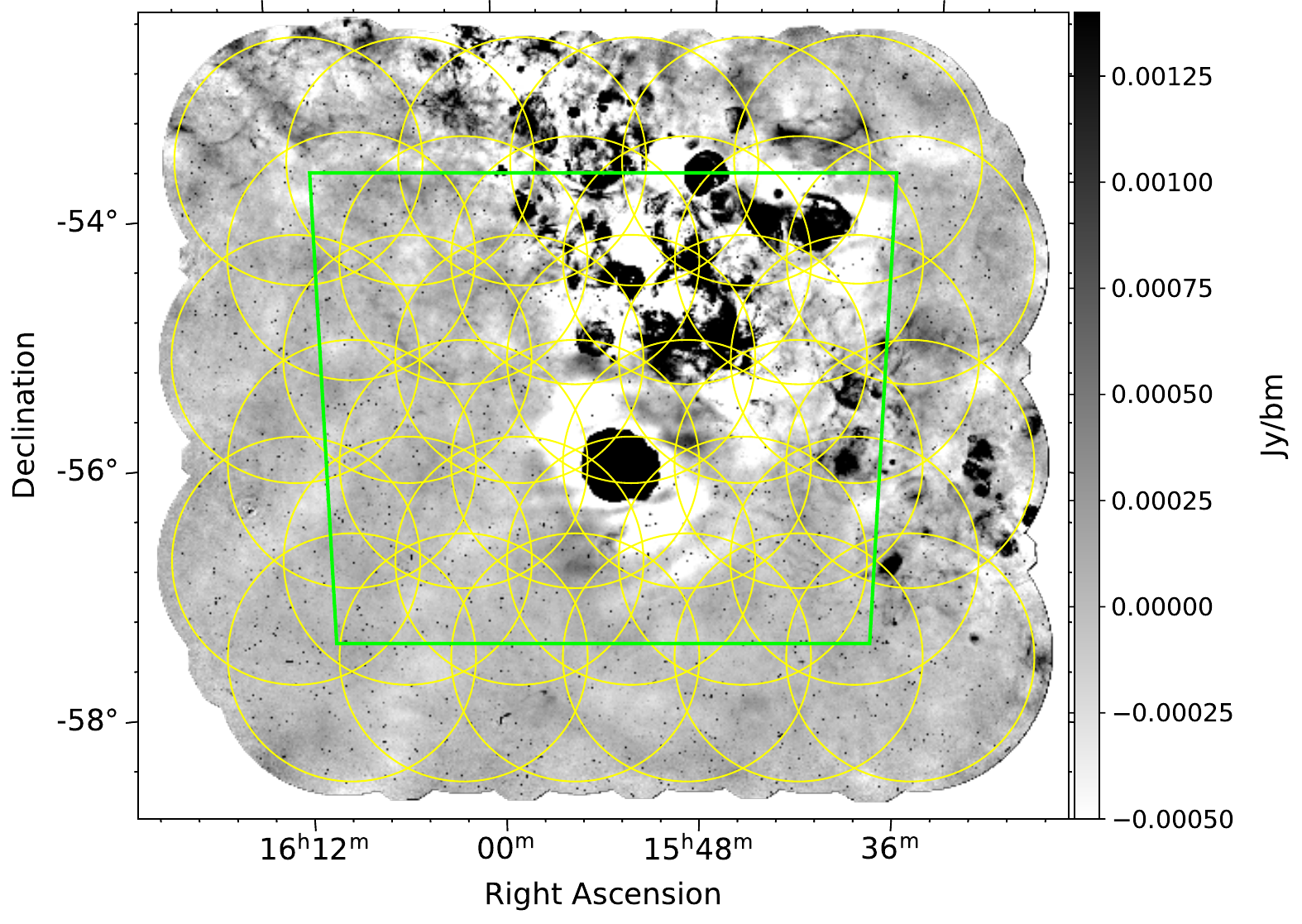}%
}
\caption{Total intensity images of the three observations analyzed in this paper. The yellow circles indicate the formed beam footprint and the green rectangles outline our defined region of uniform sensitivity within which all analysis is performed. \textit{Top row:} Pilot I EL observation (\textit{left}) and Pilot I EM observation (\textit{right}). \textit{Bottom row:} GL Pilot II observation.}
\label{fig: footprints}
\end{figure*}

We define a 17.14 deg$^2$ region of uniform sensitivity in the center of the observation. See Figure \ref{fig: footprints} for the location of the region of uniform sensitivity (green) with respect to the beam footprint (yellow). The band-averaged sensitivity in this region is $\sim$27 $\mu$Jy/beam in Stokes I. The decrease in sensitivity in this observation compared to the EL Pilot I observation is due to foreground emission from the Galactic plane, which we discuss in detail in Section \ref{sec: DE contam}.

\vspace{5mm}

We provide the theoretical RM synthesis properties for our four observations in Table \ref{tab: RMsynth props}. The columns give the expected values of $\delta\phi$, $\lvert \phi_{\mathrm{max}} \rvert$, and $W_{\mathrm{max}}$, corresponding to Equations \ref{eq: rmsf fwhm}, \ref{eq: phi max}, and \ref{eq: W max} respectively. The observations are not uniform in $\lambda^2$ space, so we use the median value of $\delta \lambda^2$ to calculate $\lvert \phi_{\mathrm{max}} \rvert$. We note that Equation \ref{eq: rmsf fwhm} assumes no missing or flagged frequency channels and uniform channel uncertainties. We discuss how the measured value of $\delta\phi$ is calculated in Section \ref{subsec: spec extract}.

\begin{table}
\centering
\caption{Theoretical RM synthesis properties}
\label{tab: RMsynth props}
\begin{tabular}{p{2.0cm}p{1.7cm}p{1.7cm}p{1.7cm}} \hline
Observation & $\delta \phi$ \newline (rad/m$^2$) & $\lvert \phi_{\mathrm{max}} \rvert$ \newline (rad/m$^2$) & $W_{\mathrm{max}}$ \newline (rad/m$^2$) \\\hline
EL  & 58.9  & 8878.5  & 149.8 \\
EM  & 446.3 & 27627.9 & 604.4 \\
EC  & 39.1  & 10713.3 & 389.6 \\
GL  & 58.9  & 8878.5  & 149.8 \\\hline
\end{tabular}
\end{table}

\subsection{Calibration and imaging}\label{subsec: cal and imag}

For all of the observations analyzed in this work, the observed visibilities were reduced by the software package ASKAPsoft, developed by Commonwealth Scientific and Industrial Research Organisation (CSIRO), as part of the ASKAPpipeline. The unpolarized calibration source PKS1934-638 \citep{OAfluxcalref} was used to derive the bandpass correction for each formed beam. The bandpass correction was applied to both the observation and the calibrator visibilities, and the visibilities were then averaged to 1 MHz channel widths. ASKAP uses the technique of self calibration to derive the per beam gains, which were then applied by the ASKAPpipeline to the 1 Mhz-averaged visibilities. The unpolarized bandpass-corrected calibration source was used to derive the on-axis leakage correction, which was derived for each antenna for each formed beam. The on-axis leakage correction was then applied to the bandpass- and gain-corrected visibilities. The residual on-axis instrumental leakage level is expected to be less than 0.1\%.

Imaging of the Stokes $IQU$ parameters was done by ASKAPsoft using these final calibrated visibilities. The point spread functions (PSF) of each formed beam are not expected to be identical. Before combining the formed beams into a single image, each frequency channel was convolved to a common resolution by the ASKAPpipeline. The chosen resolution is the smallest resolution that is common to all of the beams at that frequency.  The convolved beams are then linearly mosaicked using ASKAPsoft to produce the final image cubes that are analyzed in this work.

\subsubsection{Off-axis leakage correction}\label{subsubsec: leakcorr}

Off-axis polarization leakage is typically more difficult to correct for, and increases in magnitude with distance from the pointing axis of the primary beam. We characterize the off-axis leakage in our observations in two different ways: holography and field sources.

For the EL and GL observations, primary beam correction in the Stokes $I$ cube and off-axis leakage in the Stokes $QU$ cubes was characterized using beam models derived from holography observations.
The holography observation derived for the EL observation did not use the same beam weights as the target observation, which is not ideal. Because of this, we expect that the off-axis leakage correction for the EL field will be worse than what is expected for the full POSSUM survey, which will use the same weights for the holography and target observations.

For the EM observation, field sources were used to characterize the off-axis leakage in Stokes $QU$ (Stokes $I$ used holography as described above). Similar methods were used by \citet{Farnsworth+11} and \citet{Lenc+18}, and the same method that is used in this work was used by \citet{Thomson+23}, and is described in detail therein. In the individual formed beams of the 10 POSSUM pilot I mid-band observations (see Section \ref{subsubsec: pilot survey descrip}), the Stokes $QU$ spectra of high signal-to-noise components ($\geq 100 \sigma$ in total intensity) were extracted using the method described in Section \ref{subsec: spec extract}. It is assumed that the majority of these components are intrinsically unpolarized or, when this is not the case, that the mean value of Stokes $QU$ over a large number of sources probing any given part of the beam tends towards zero. The fractional Stokes $q$ ($\frac{Q}{I}$) and $u$ ($\frac{U}{I}$) spectra were fit with polynomial models to avoid effects from spurious noise or intensity spikes from source such as RFI. In each formed beam in each frequency channel, the Stokes $qu$ values of the components in the 10 observations were stacked and the Stokes $qu$ surfaces were fit with Zernike polynomials \citep{Zernike1934} to map intensity as a function of distance from the center of the beam. These maps were multiplied by the Stokes $I$ image to get back $QU$ leakage maps, which were then subtracted from the original $QU$ maps of the formed beam, leaving leakage-corrected images.

Residual off-axis leakage levels in the data are discussed in Section \ref{subsec: resid leakage}. None of the data were corrected for ionospheric Faraday rotation, however we expect that this has a negligible effect on our results.

\vspace{5mm}

We provide a summary of the data specifications for the four observations in Table \ref{tab: data specs}. We note that for the EC observation, the two contributing data sets used different off-axis leakage correction methods (holography for the EL data and field sources for the EM data). We refer to this as ``mixed'' in the \textit{Leakage correction} column of Table \ref{tab: data specs}.

\begin{table*}
\centering
\caption{Summary of observations}
\label{tab: data specs}
\begin{tabular}{p{1.6cm} p{1.3cm} p{1.5cm} p{1.9cm} p{1.3cm} p{1.5cm} p{1.4cm} p{1.8cm} p{1.7cm}}
\hline
Observation & RA \newline J2000 & Dec \newline J2000 & Frequency \newline range & Number \newline of 1 MHz \newline channels & Integration \newline time & Angular \newline resolution & Band-averaged \newline sensitivity & Leakage \newline correction \\
 & (deg) & (deg) & (MHz) & & (hr) & (arcsec) & ($\mu$Jy/beam) &  \\
\hline
EL Pilot I & 331.490 & $-51.192$  & 800 -- 1087 & 286 & 10.0 & 21 & 24 & holography \vspace{2mm} \\ 
EM Pilot I & 332.052 & $-50.870$ & 1316 -- 1439 & 116 & 10.0 & 13 & 30 & field source \vspace{2mm} \\ 
EC Pilot I & 332.146 & $-50.737$ & 800 -- 1439 & 402 & 10.0 & 21 & --- $^a$ & mixed\vspace{2mm} \\
GL Pilot II & 238.498 & $-55.730$ & 800 -- 1087 & 278 & 10.0 & 16.5 & 27 & holography \vspace{1mm} \\  \hline
\multicolumn{9}{c}{$^a$ A band-averaged image of the EC data set from which to estimate the sensitivity is not produced in this work} \\
\multicolumn{2}{c}{(see Section \ref{subsec: sourcefinder}).} \\
\end{tabular}
\end{table*}

\subsection{Source finding}\label{subsec: sourcefinder}

Source finding is performed in each observation by the ASKAP Observatory using the \textit{Selavy} software package \citep{Whiting&Humphreys2012,Whiting+17}. The source finder is run on the individual total intensity multi-frequency synthesis (MFS) images of each observation. In this process, pixels in the MFS image with intensities greater than three times the root-mean-square (rms) in the image are grouped into islands of emission and a Gaussian is fit to peaks in each island to identify individual components. Once the components of an island are identified, only those with intensities above five times the rms are retained. A simple point source, such as an unresolved radio galaxy, would be comprised of a single island with a single component. More extended sources can be comprised of multiple islands and multiple components. All components identified by the source finder are compiled in component catalogs.

We discuss components throughout this work instead of sources because we do not perform any crossmatching with optical or infrared catalogs to determine host galaxies for our components. As such, we are unable to determine whether two neighboring components are part of the same source or are merely projected close together on the sky.

We use the EL total intensity component catalog for for spectra extraction in the EC observation.  A deeper total intensity catalog could potentially be produced by creating a new Stokes $I$ MFS image from the combined data sets, which would have better sensitivity due to the broader Stokes $I$ band and would presumably show components that were below the detection limit in the EL MFS image. However, this is would not be consistent with the plans for the full POSSUM survey.

\section{Diffuse emission contamination of background components}
\label{sec: DE contam}

In this section we determine the effects of foreground diffuse emission from the Galaxy on the polarization properties of the background components that pass through it, and we propose a method for separating the large-scale diffuse emission from these more compact background components. We perform this separation as our first step of data reduction before proceeding with extracting the polarization parameters of the components, which we describe in Section \ref{sec: data reduc}.

The presence of foreground polarized diffuse emission can affect polarization measurements of background components, whose properties are the desired products for RM catalogs and RM grids.  Because polarization is a vector quantity, there will be interference between the diffuse and compact polarizations in the Faraday spectrum \citep{Farnsworth+11}, possibly producing multiple peaks, or shifting the amplitude or Faraday depth of the main peak.

\begin{figure}
\centering
    \includegraphics[width=0.47\textwidth]{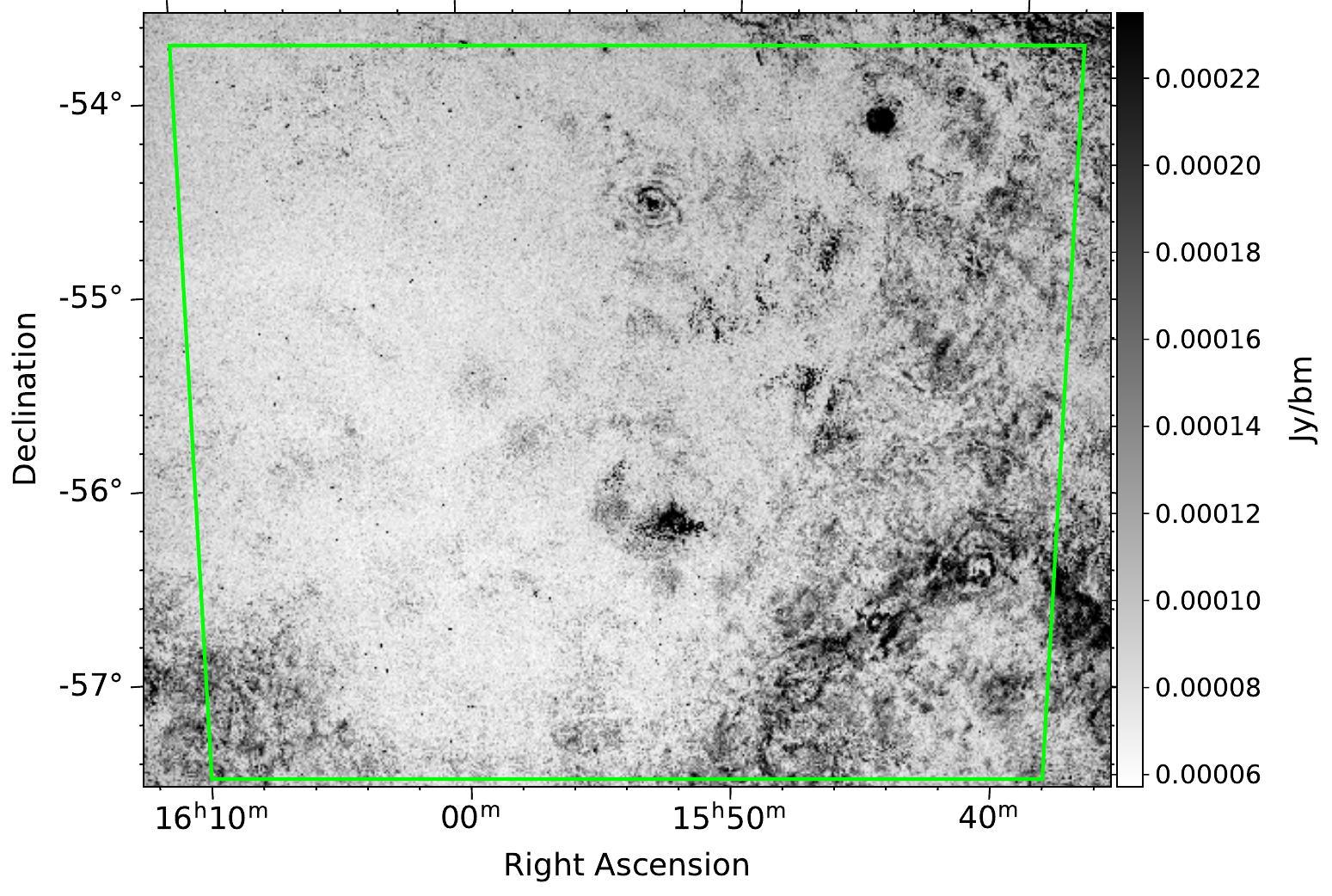}
    \caption{Peak polarized intensity map of the GL observation highlighting extensive diffuse emission present along the line of sight. The green box outlines the region of uniform sensitivity, as per Figure \ref{fig: footprints}.}
\label{fig: gal PI map}
\end{figure}

To determine whether any of our observations contain foreground polarized diffuse emission, we perform 3D RM synthesis to produce maps of peak polarized intensity, using the \texttt{RM-Tools} software package \citep{RM-tools2020}. We find that the GL observation contains a significant level of polarized diffuse emission (see Figure \ref{fig: gal PI map}) while the other observations do not. We therefore introduce a method for removing the diffuse emission and show below the results of applying it to the GL observation.

\subsection{Median filter method for diffuse emission removal}
\label{subsec: medfilt method}

\subsubsection{Median filter method description}
\label{subsubsec: med filt description}

We apply a median filter to each frequency channel of the Stokes $IQU$ cubes using the \texttt{median\_filter} function from \texttt{SciPy} \citep{SciPy2020}. The filter places a box of user-defined size around each pixel in the image and replaces the value of the central pixel with the median value in the box. The resulting median image is an estimate of the larger-scale structure in the observation, which we refer to as the \textit{diffuse map}. The smallest scales that the filter is sensitive to is determined by the box size. Subtracting this diffuse cube, channel by channel, from the original image cube removes the large-scale structure, leaving a map of the smaller-scale structure, or the background components, which we refer to as the \textit{component map}.

For the median filter to correctly remove foreground diffuse emission in Stokes \textit{Q} and \textit{U}, the RM of the emission must not vary significantly within the median filter box. A moderate gradient in RM results in a more substantial gradient in polarization angle (e.g. $\Delta$RM = 5 rad m$^{-2}$ results in $\Delta$PA = 32$^{\circ}$), and the median value of the pixels in the filter box will not be representative of the foreground emission in that region. The median filter method described here is best applied to fields where the RM of the foreground diffuse emission varies smoothly on scales larger than the chosen box size.

Initial tests of 7 different box sizes from 60$\times$60 arcsec to 180$\times$180 arcsec indicated that a 120$\times$120 arcsec box size recovered most of the large scale structure of the diffuse emission and did not seriously compromise the quality of the RMs. We perform all of the testing of the median filter method with this box size.

\subsubsection{Testing with simulated components}
\label{subsubsec: testing filt}

To better quantify the accuracy of RM and polarized intensity measurements after the median filter is applied, we inject simulated compact components, modeled as 2D Gaussians with a FWHM of 16.5 arcsec and known polarized intensity and RM, into two regions of the GL observation, one with substantial diffuse emission and one with little to no diffuse emission. Component positions, RMs, and polarized intensities were randomly generated from uniform distributions and 100 components were injected into each region 50 times, yielding 5000 simulated components in each region. After the components are injected into the cubes, the 120 arcsec median filter was applied, the component spectra were extracted, and 1D RM synthesis was performed.

\begin{figure*}
\centering
    \includegraphics[width=0.9\textwidth]{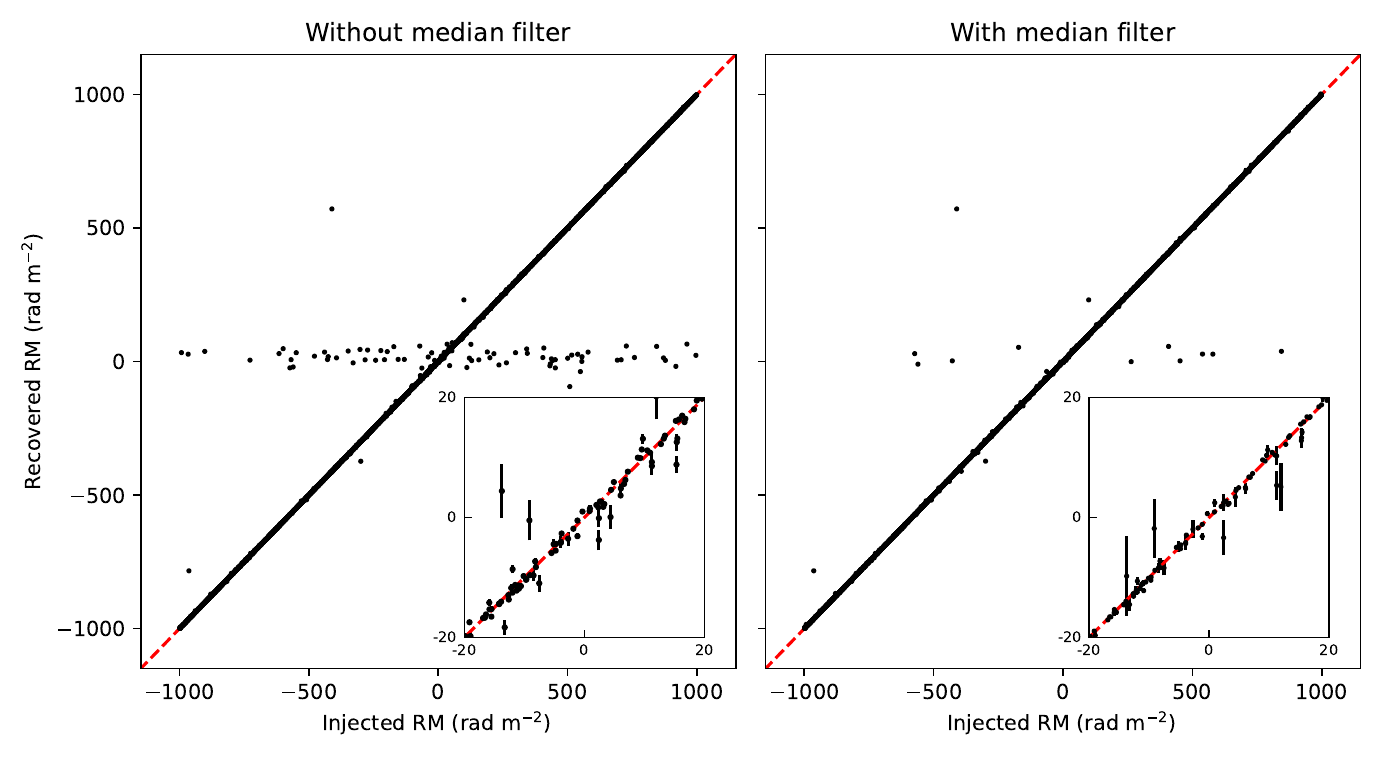}
    \caption{\textit{Left panel}: recovered RM versus injected RM for 5000 simulated compact components before applying the median filter. The red dashed line is the one-to-one line. The extreme outliers are components that are fainter than the local foreground diffuse emission. The inset is a close up of $\phi$ $\pm$ 20 rad m$^{-2}$. \textit{Right panel}: same as the left panel but after applying the median filter. There is an 80\% reduction in the most extreme outliers and an overall increase in the accuracy of the recovered RMs of all components.}
\label{fig: sim RM vs RM}
\end{figure*}

Figure \ref{fig: sim RM vs RM} plots the RMs of the recovered components versus the corresponding injected RMs in the region where diffuse emission is present before the median filter is applied (left panel) and after (right panel). These plots show the effects of the median filter in two situations: where the diffuse emission is brighter or fainter in peak polarized intensity than that of the compact component.

The left panel of Figure \ref{fig: sim RM vs RM} shows a large number of outliers along a horizontal line centered around RM $= 0$ rad m$^{-2}$. From 3D RM synthesis of the GL observation, we find that the peak RM of the foreground diffuse emission is typically $\pm$50 rad m$^{-2}$. These outliers around RM = 0 rad m$^{-2}$ are components where, prior to filtering, the local diffuse emission is brighter in polarized intensity than the background component (see Figure \ref{fig: gal nofilt DE faint} in Appendix \ref{app: example spectra} for an example of this case). In these cases, the Faraday spectrum has peaks at multiple values of $\phi$ and RM synthesis identifies the brightest peak associated with the low-RM diffuse emission as the component RM. In the right panel of Figure \ref{fig: sim RM vs RM}, we can see an 80\% reduction in the number of these outlier points after the median filter is applied, from 68 components to 13. The components where the recovered RM lies close to the one-to-one line in the left panel are those where the component is brighter in peak polarized intensity than the foreground diffuse emission. 

The insets of each panel of Figure \ref{fig: sim RM vs RM} show a close up of the one-to-one line from $\pm$20 rad m$^{-2}$. We can see in the left panel inset that the RMs of these components are also often not the same as the injected value within their uncertainties. This may be due to interference effects between the diffuse emission and component peaks in the Faraday spectrum or the Stokes spectra. The right panel inset shows that the median filter increases the accuracy of the measured RMs of these components. Without filtering, 97.9\% of the recovered RMs are within 3$\sigma$ of the injected value, and after filtering 99.5\% are within 3$\sigma$.

We also calculate the percent difference in recovered peak polarized intensity from the injected value of our simulated components after applying the median filter. We find a median loss in polarized intensity of 5\% with a standard deviation of 5\%. The difference in polarized intensity comes from a small contribution of the source to the median of the box, which biases the median high or low depending on the sign of the polarized signal. The same tests described above are also performed on simulated components injected into a region of the GL observation with no diffuse emission and they show similar results for the peak polarized intensity loss.

We perform testing on a smaller scale to determine if the box size should vary with the PSF width of the observation. The EM observation has a PSF that is 0.79$\times$ the size of the GL observation. We inject 200 simulated compact components in the EM Stokes cubes, and perform median filtering with two different box sizes: 120 arcsec and 95 arcsec. We then calculate the fractional difference in injected and recovered polarized intensity and RM. We find no significant change in the distributions of the fractional differences. In principle, box size should vary with PSF such that the PSF does not take up a substantial portion of the box size. However, we find that the PSF width between our four observations does not vary enough to have significant impact on our results. We note that our testing has been done on compact components, and that further testing on the impact of box size on polarized intensity loss and RM should be done for more extended components.

\subsection{Application of the median filter to the data}

\subsubsection{GL observation}
\label{subsec: medfilt application Gal}

\begin{figure*}
\centering 
\subfloat{%
  \includegraphics[width=0.85\textwidth]{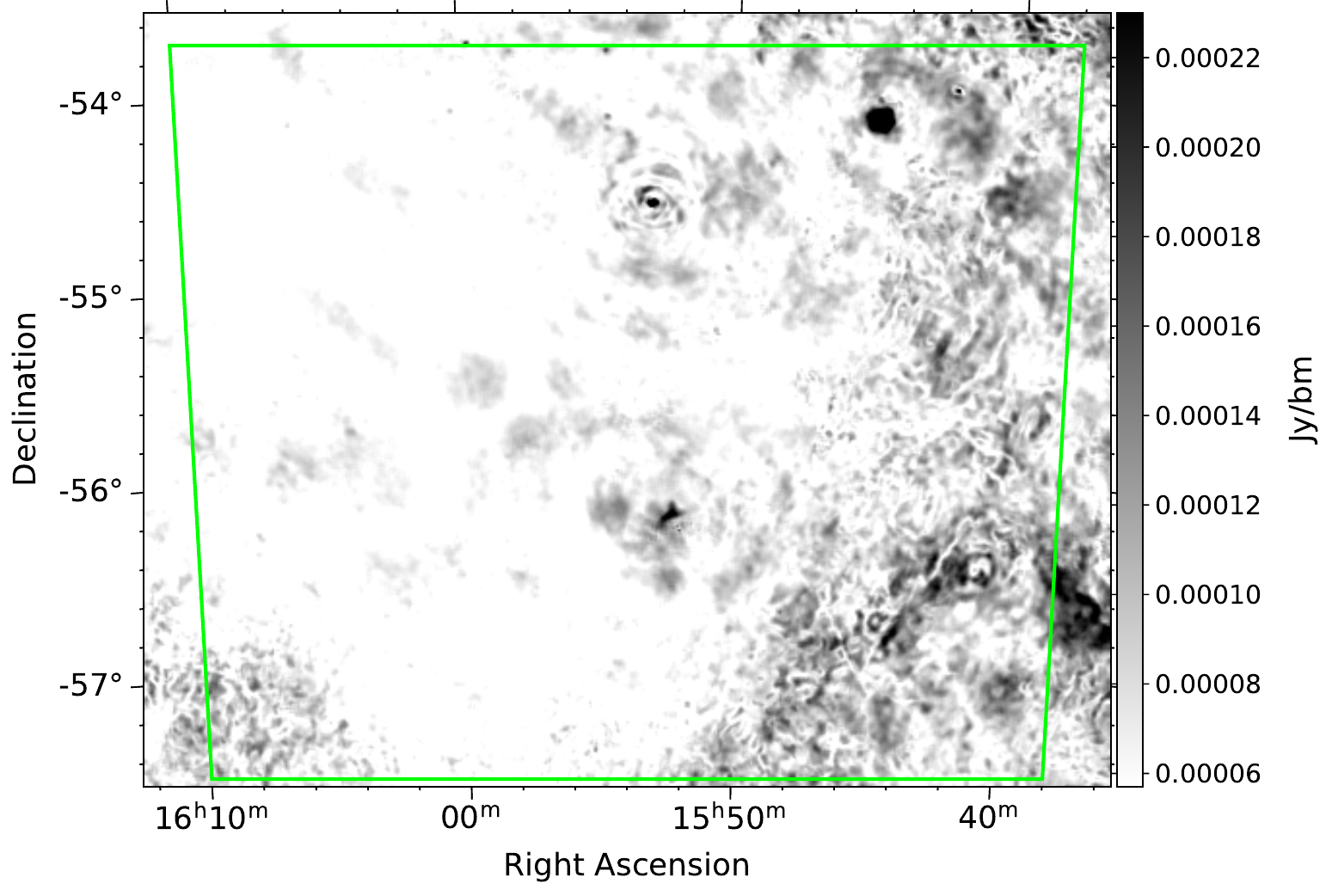}%
  \label{fig: gal medfilt DE}%
}\\
\subfloat{%
  \includegraphics[width=0.85\textwidth]{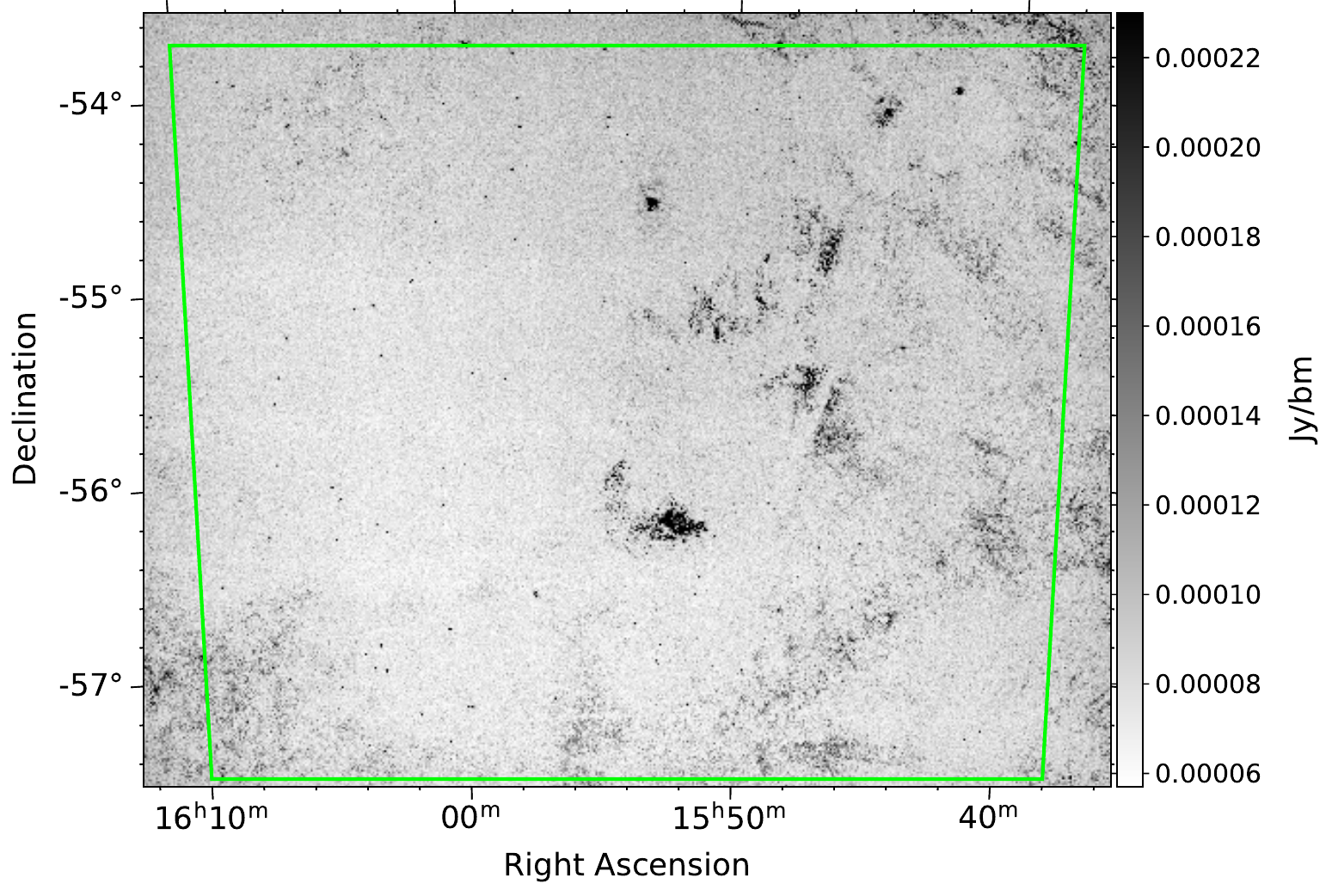}%
  \label{fig: gal medfilt srcs}%
}
\caption{Peak polarized intensity maps from the diffuse cubes (\textit{top}) and component cubes (\textit{bottom}) of the GL observation after applying a 120 arcsec median filter. Faint residual diffuse emission can be seen in the component map, particularly on the right half of the image, where the diffuse emission or the polarization angle varies on smaller scales. The unfiltered peak polarized intensity map of the GL observation is shown in Figure \ref{fig: gal PI map}.}
\label{fig: gal medfilt plots}
\end{figure*}

We now apply the median filter to the GL observation Stokes $IQU$ cubes using the 120 arcsec box. This returns what we will refer to as a \textit{diffuse cube} (a diffuse map of each spectral channel in the Stokes cube), and subtracting this from the original image produces what we will refer to as a \textit{component cube} (a foreground-subtracted component map of each spectral channel). In Figure \ref{fig: gal medfilt plots} we plot the peak polarized intensity images from 3D RM synthesis of the diffuse (top panel) and component (bottom panel) cubes. The diffuse peak polarized intensity map contains the large-scale diffuse emission that has been separated from the background components. Comparing this map to Figure \ref{fig: gal PI map}, we see that the overall structure of the diffuse emission is retained. The component map in Figure \ref{fig: gal medfilt plots} shows that most of the diffuse emission is removed by the median filter, however some faint residual diffuse emission can be seen in the component map in places where the diffuse emission or the polarization angle has finer structure in the unfiltered data.

In Figures \ref{fig: gal nofilt DE faint} and \ref{fig: gal medfilt DE faint} in Appendix \ref{app: example spectra} we show how the median filter corrects for the presence of foreground diffuse emission in front of a faint background component. We confirm that this is in fact a situation of diffuse emission dominating the Faraday spectrum by visual inspection of the diffuse and component maps from the median filter process. Before filtering, 1D RM synthesis detects the RM of the diffuse emission because it is brighter in polarized intensity than the faint background component. After filtering, the diffuse emission is reduced to a level that is below that of the background component, and the component becomes the brightest detectable peak in the Faraday spectrum, as desired.

\subsubsection{Artefacts in the EL observation}\label{subsec: medfilt application 10635}

In addition to removing foreground diffuse emission, we test the ability of the median filter to remove a particular type of imaging artefacts in the EL observation. Imaging artefacts can cause artificial enhancements or losses in total and polarized intensity measurements.
Removing these artificial signals from an observation makes data reduction and analysis simpler and more reliable.

\begin{figure*}
\centering 
\subfloat{%
  \includegraphics[width=0.49\textwidth]{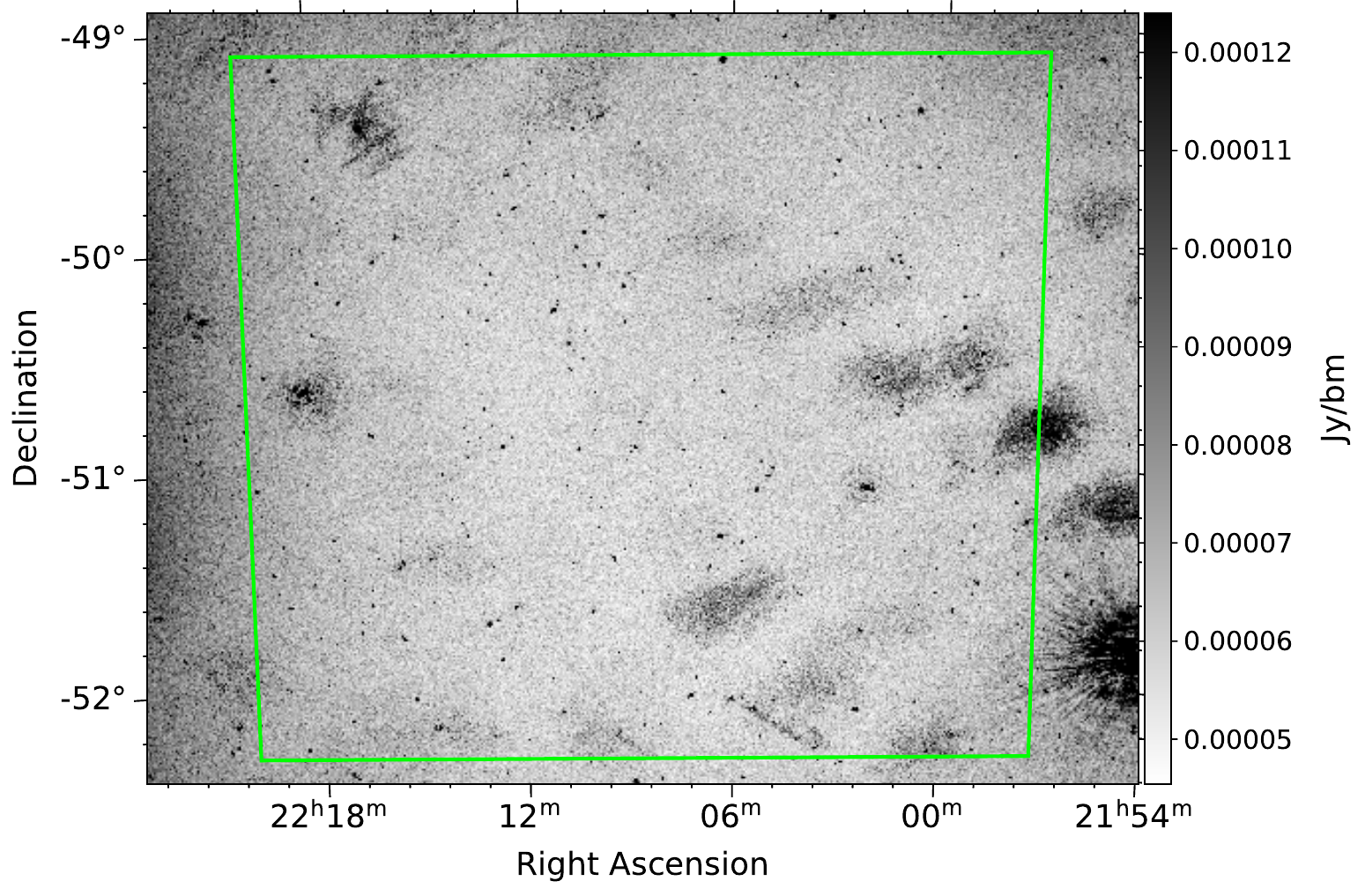}%
}\quad
\subfloat{%
  \includegraphics[width=0.49\textwidth]{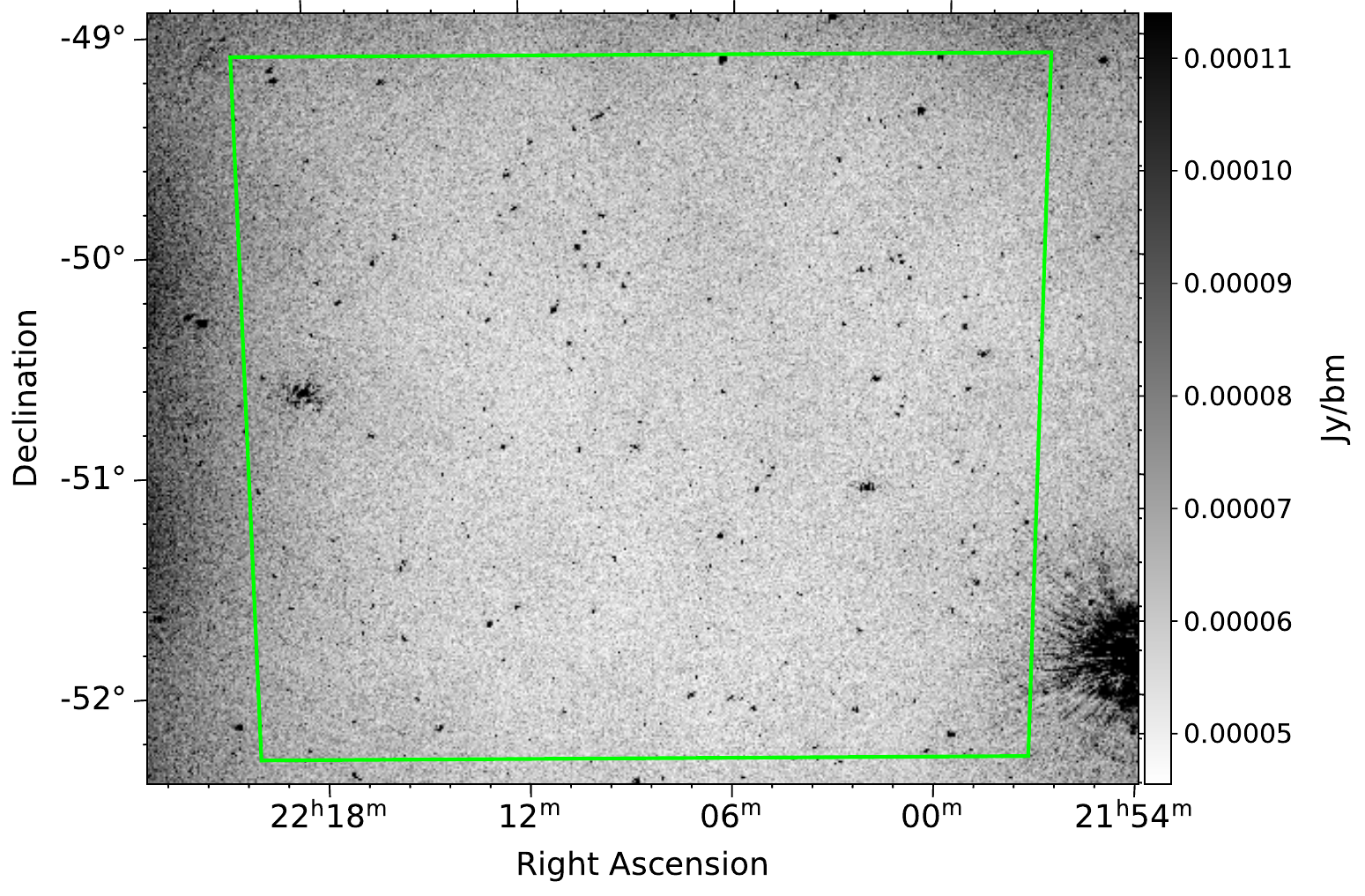}%
}
\caption{\textit{Left}: The peak polarized intensity map of the EL observation before applying the median filter. The ripple-like bright patches are presumed to be artefacts of solar interference during observing. \textit{Right}: The peak polarized intensity map of the EL observation after the application of the median filter with a 120 arcsec box. The median filter is able to completely remove the ripple-like artefacts from the data. }
   \label{fig: 10635 medfilt plots}
\end{figure*}

The left panel in Figure \ref{fig: 10635 medfilt plots} shows the peak polarized intensity map of the unfiltered EL observation. We can see large, bright patches of emission with a ripple-like pattern, which are presumed to be artefacts from solar interference during the observation and not real polarized emission. Since these patches of artificial emission are much larger than our background components, we test whether the median filter is able to remove them in the same way that it removes large-scale diffuse emission features. The brightest patch of emission in the lower right of the image, outside of the region of uniform sensitivity, is a sidelobe artefact surrounding a particularly bright background component. We do not expect the median filter to remove this artefact due to the more small-scale variations in its structure.

We apply the median filter to the EL Stokes cubes with a 120 arcsec box size. The right panel of Figure \ref{fig: 10635 medfilt plots} shows the peak polarized intensity image of the component map after the median filter is applied. We can see that the median filter is indeed able to remove the ripple artefact from the image, although we note that flux errors due to calibration and imaging artefacts are not corrected by applying the median filter.

\subsection{Other filtering methods}

In the course of our analysis, we also tested removing the largest angular scales in the Stokes cube channels with a spatial low-pass filter. We applied the spatial low-pass filter with both a 2D Gaussian window and a Tukey window \citep{Tukey1967}. The median percent difference between the injected and the recovered peak polarized intensity was nearly 50\% for all components using the Gaussian window filter, while the Tukey window filter had a median difference of $\sim$10--15\%. In both cases this difference is more than the difference we see when using the median filter, which is typically $\sim$5\%.

Different variations of the median filter method were also investigated, including applying the median filter with sigma-clipping and iterative median filtering with a smaller box size of 60 arcsec. We saw no significant difference in the results between median filtering with and without sigma-clipping. We found that the output of iterative median filtering was very similar to a single pass of the median filter with the 120 arcsec box. 

As a result of the limitations of the spatial low-pass method and the variations of the median filter method, we decided that the median filter method described in Section \ref{subsubsec: med filt description} was the best approach to separate diffuse emission from background components.

\subsection{Advantages and disadvantages of applying the median filter to all observations}

Here we suggest the advantages and disadvantages of applying the median filter to all observations in the full POSSUM survey as opposed to only those observations with extensive diffuse emission contamination. An advantage of a uniform application of the median filter to all observations is that the RMs in the final catalog will all be extracted from the same data reduction process. We have also shown that the filter is able to remove some types of imaging artefacts in the data that can affect polarization measurements, while not affecting the RMs of components that are not contaminated by diffuse emission.

The primary disadvantage of applying the median filter to observations that do not contain diffuse emission is that the systematic polarized intensity loss described in Section \ref{subsubsec: testing filt} will likely reduce the number of polarized components in the final POSSUM catalog and the reported polarized intensity values will be inaccurate. To further assess the impact of the median filter on the data, we apply the filter to all four observations and perform the rest of our analysis on both the filtered and unfiltered versions of the observations.

\section{Data reduction and polarized component selection}
\label{sec: data reduc}

A primary goal of this work is to characterize the effects of frequency, bandwidth, Galactic latitude, and the application of the median filter on expected component densities and RM uncertainty. To do this, we extract the polarization parameters of components from each of our observations, both with and without the median filter applied, and perform the remainder of our data reduction on these components.

\subsection{Extracting polarization parameters}
\label{subsec: spec extract}

The Stokes $IQU$ spectra of each component are extracted from the image cubes following the method employed in the POSSUM pipeline, which will be described in detail in a future paper (Van Eck, in prep). The spectra are extracted from the position of the peak total intensity as defined in the source finder component catalogs in the EL, EM and GL observations, and we use the EL component catalog positions to extract source spectra in the EC observation (see Section \ref{subsec: sourcefinder}). The per-channel intensity is calculated by averaging a 5$\times$5 pixel box (corresponding to 15$\times$15 arcsec for the EL observation and 10$\times$10 arcsec for the EM and GL observations) around the peak pixel and normalizing the corresponding sum by the sum over the same region for the PSF model. This extraction method accounts for any small offset in pixel position between the peak total intensity and the peak polarized intensity of a component.

The associated uncertainty of the channel intensity measurement is estimated from the local noise in an annulus centered on the position of the peak total intensity. The inner and outer radii  of the annulus are fixed at 10 and 31 pixels respectively. These were the radii used by the POSSUM pipeline at the time of analysis, however these values have since changed and will be described in Van Eck (2023, in prep). These radii correspond to 2.9--8.9 times the PSF radius for the EL observation, 4.5--14.1 times the PSF radius for the EM observation, and 2.5--7.5 times the PSF radius for the GL observation. The median absolute deviation from the median (MADFM) of the pixels within the annulus is calculated and is converted to a standard deviation by multiplying by 1.4826. The MADFM is used in place of a direct standard deviation measurement because the MADFM is more robust to outliers such as neighboring components (e.g. close double sources) within the annulus or frequency channel anomalies (see Appendix B in \citealt{Thomson+23}).

We use the \texttt{RM-Tools} package to perform 1D RM synthesis on our components and 3D RM synthesis on our regions of uniform sensitivity. The dimensionless Faraday spectrum is multiplied by the value of Stokes $I$ at $\lambda_0^2$ to return the intensity units of the Stokes spectra. \texttt{RM-Tools} uses a 3-point parabolic fit to the brightest peak in the Faraday spectrum and reports the value of $\phi$ at which the maximum of occurs as the RM and the amplitude at the maximum as the peak polarized intensity of the component. The polarized intensity is corrected for polarization bias using the method of \citet{George+12}. Equation \ref{eq: rmsf fwhm} for $\delta\phi$ assumes uniform weighting and spacing of frequency channels, however we use variance weighting when performing RM synthesis and we need to take into account any flagged channels. To do this, \texttt{RM-Tools} reports $\delta\phi$ as the FWHM of a Gaussian fit to the central peak of the RMSF. The uncertainty in RM, $\delta$RM, is calculated as:

\begin{equation}\label{eq: deltaRM}
    \delta\mathrm{RM} = \frac{\delta\phi}{2 \, \mathrm{S/N_{\mathrm{pol}}}} \, ,
\end{equation}

\noindent where $\delta\phi$ is the FWHM of the RMSF (see Section \ref{subsec: phi and RM synth}) and S/N$\mathrm{pol}$ is the signal-to-noise of the peak polarized intensity measurement \citep{Brentjens2005}. This is calculated as:

\begin{equation}\label{eq: s/n_pol}
    \mathrm{S/N}_{pol} = \frac{\mathrm{PI}}{\sigma_{\mathrm{FS,th}}},
\end{equation}

\noindent where PI is the peak polarized intensity and $\sigma_{\mathrm{FS,th}}$ is the theoretical noise in the Faraday spectrum, calculated from the $QU$ channel uncertainty using inverse variance weighting. A complete list of the outputs of RM synthesis with \texttt{RM-Tools} can be found on the package website\footnote{\url{https://github.com/CIRADA-Tools/RM-Tools/wiki}}.

At the time of this analysis, the POSSUM pipeline restricts the search range in Faraday depth to $\pm$2000 rad m$^{-2}$, so we impose that limit as well. Limiting the search range in Faraday depth reduces both compute time and the chance of false detections. Due to the exploratory nature of this work, the small number of components with $\lvert \phi \rvert >$ 2000 rad m$^{-2}$ that may be excluded from our analysis by this limit are not crucial to our results. We provide some example plots of Stokes $IQU$, polarization angle $\psi$ versus wavelength squared, and Faraday spectra for a variety of components in Figure \ref{fig: example spectra} in Appendix \ref{app: example spectra}.

A sinusoidal ripple can be seen in some of the Stokes $I$ spectra in Figure \ref{fig: example spectra}, most clearly in Figures \ref{fig: low medfilt bright simple}, \ref{fig: mid medfilt bright simple}, and \ref{fig: comb medfilt bright simple}. The ripple is also present in the Stokes $QU$ spectra, however it is less obvious due to the lower S/N and more complicated intrinsic structure of the $QU$ spectra. The ripple is believed to be the result of a standing wave between the surface of the telescope dish and the PAF, which manifests in the instrumental gains \citep{Sault15}. The ripple has a periodicity of approximately 25 MHz and can present in the Faraday spectrum as peaks at $\sim$300--800 rad m$^{-2}$ in the ASKAP low-band and $\sim$1700--2000 rad m$^{-2}$ in the ASKAP mid-band. The amplitude of the ripple and the corresponding peaks in the Faraday spectrum increase with increasing signal-to-noise (S/N). 
Section 4.3 of \citet{Thomson+23} provides a more detailed discussion of the amplitude of the ripple in the Faraday spectrum. 
While we leave a full examination of the effects of the ripple on ASKAP data for future work, we interpret any RM peaks found at these values of $\phi$ with caution, particularly for high-S/N components, and suggest future users of the data presented here do the same.

1D RM synthesis is performed on the fractional Stokes $qu$ spectra to mitigate the effect of the Stokes $I$ behavior. \texttt{RM-Tools} fits the Stokes $I$ spectrum with a log-log space polynomial:

\begin{equation}\label{eq: stokes I model}
    I(\nu) = C_0 \bigg(\frac{\nu}{\nu_{ref}}\bigg)^{C_1 + C_2 \log(\frac{\nu}{\nu_{ref}}) + C_3 \log(\frac{\nu}{\nu_{ref}})^2} \, ,
\end{equation}

\noindent where $C_i$ are the coefficients of the polynomial and $\nu_{ref}$ is the reference frequency calculated as the mean of the frequency channels. $C_1$ can be interpreted as the spectral index of the Stokes $I$ spectrum. The Stokes $QU$ spectra are divided by this model to get the fractional Stokes $qu$ spectra, which we then perform 1D RM synthesis on.

After performing 1D RM synthesis on all components, we want to select components for our polarized catalogs where the measurements of polarization parameters such as RM are reliable. To do this we impose a threshold on S/N$_{\mathrm{pol}}$. Work by \citet{Brentjens2005} and \citet{Macquart+12} have shown that measurements of polarization properties below a S/N threshold of $\sim$7 are unreliable. We set a more conservative threshold of S/N$_{\mathrm{pol}} \geq 8$ to avoid these unreliable detections.

In the GL observation we manually removed four components with anomalously large values of $\delta\phi$ ($\gtrsim$30\% larger than the expected value from Table \ref{tab: RMsynth props}). These components appeared to be bright spots associated with supernova remnants, and not background extragalactic radio components. The large angular extent of the supernova remnants causes a frequency dependence in the per-channel noise values of the spectra taken at the position of these bright spots, which caused high sidelobe peaks in the RMSF leading to a bad Gaussian fit to the central RMSF peak. This incorrect measurement of $\delta\phi$ affects other measured properties such as polarization angle and $\delta$RM, and so we remove these components from our catalog.

\subsubsection{Signal-to-noise threshold and Faraday depth search range}\label{subsubsec: Faraday spectrum range test}

\begin{figure}
\centering
    \includegraphics[width=0.47\textwidth]{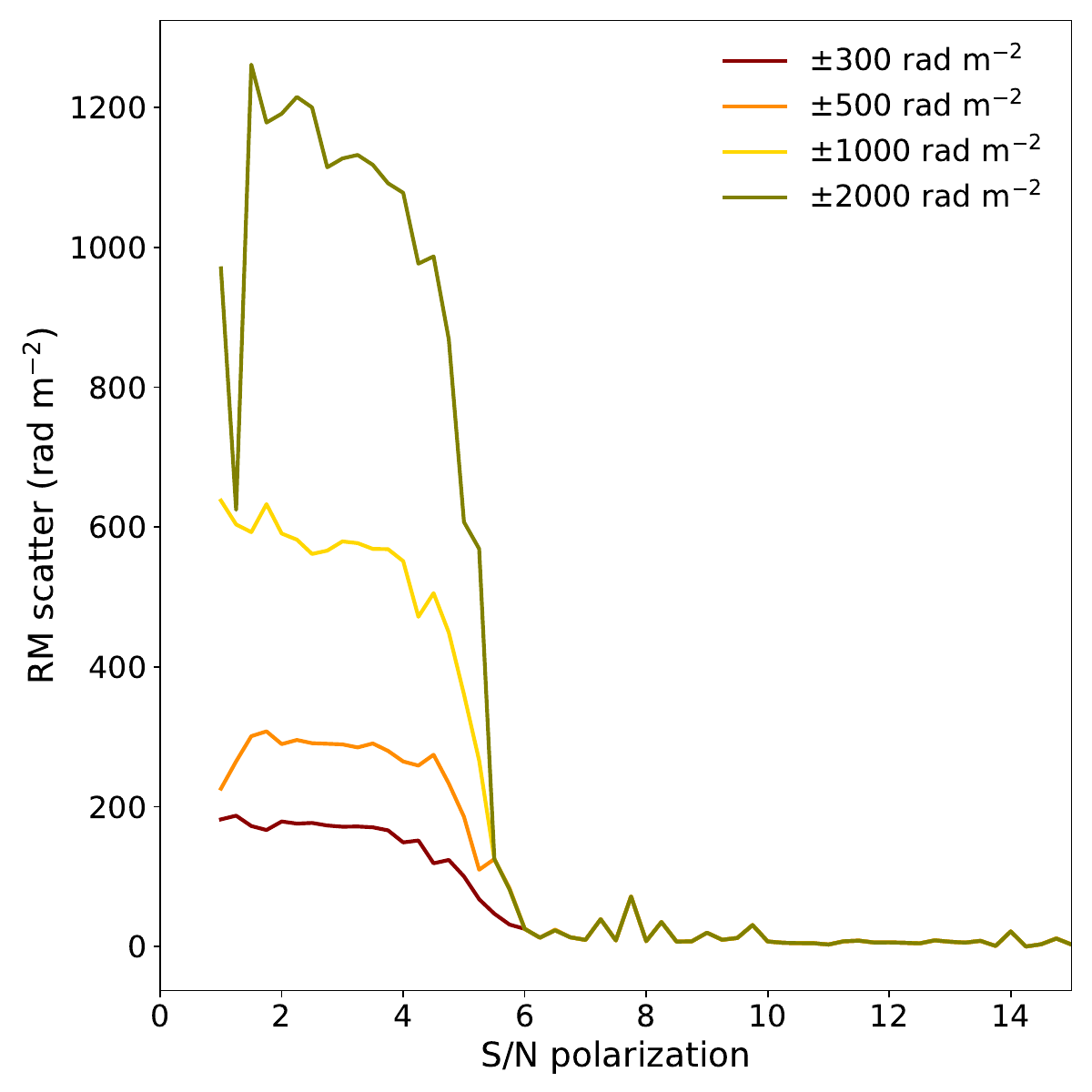}
    \caption{Scatter in RM as a function of S/N$_{\mathrm{pol}}$ for four Faraday depth search limits.}
\label{fig: rm scatter vs snr}
\end{figure}

Here we test our S/N$_{\mathrm{pol}}$ threshold of 8 which we use to select polarized components in our observations. In Figure \ref{fig: rm scatter vs snr} we plot the standard deviation in RM (RM scatter) as a function of S/N$_{\mathrm{pol}}$ for four ranges of $\phi$ over which we limit the search for peaks in the Faraday spectrum with 1D RM synthesis. Above S/N$_{\mathrm{pol}} \approx 6$, the scatter is identical for all search ranges, and below this threshold the scatter in RM sharply increases for all but the smallest search range. For the smallest search range, the scatter in RM increases more gradually, with the same value of scatter ($\sim$125 rad m$^{-2}$) at S/N$_{\mathrm{pol}} = 4.7$ as the other search ranges have at S/N$_{\mathrm{pol}} = 5.5$. This suggests that if the search range in $\phi$ is greatly restricted ($\lvert \phi \rvert\ \lesssim 300$ rad m$^{-2}$), the magnitude of the scatter in RM is reduced at lower S/N. For science goals that require large numbers of RMs or low S/N sources, restricting the search range in $\phi$ when performing 1D RM synthesis will allow for a lower S/N$_{\mathrm{pol}}$ threshold when selecting polarized components.

Figure \ref{fig: rm scatter vs snr} also suggests that our polarization threshold of S/N$_{\mathrm{pol}} = 8$ maybe be somewhat conservative, however there is a spike in RM scatter of $\sim$65 rad m$^{-2}$ just below this threshold. We maintain our S/N$_{\mathrm{pol}} = 8$ threshold over $\phi$ $\pm$ 2000 rad m$^{-2}$ for the rest of the analysis, however we keep in mind that this will result in conservative estimates for RM sky densities and total RMs in Section \ref{sec: results}.

\subsection{Residual off-axis leakage estimation}\label{subsec: resid leakage}

We calculate the residual Stokes $qu$ off-axis leakage in our EL, EM, and GL observations using the unpolarized components (S/N$_{\mathrm{pol}} < 8$) with S/N in total intensity greater than 8$\sigma$ in the respective catalogues. For each unpolarized component, we plot the absolute real (imaginary) value of the complex Faraday spectrum at peak $\phi$, corresponding to Stokes $Q$ ($U$), divided by the median Stokes $I$ value, as a function of distance from center of the region of uniform sensitivity. We then bin the data to calculate the median value of the leakage as a function of distance from the region center.

In all three observations we find that the leakage in both $q$ and $u$ is approximately constant with distance out to the edge of our regions of uniform sensitivity. The typical estimated residual Stokes $q$ ($u$) off-axis leakage in the median-filtered observations is 0.5\% (0.6\%) in the EL observation, 1.1\% (1.2\%) in the EM observation, and 0.6\% (0.5\%) in the GL observation. In the unfiltered observations, the typical residual off-axis leakage levels were the same for both Stokes $q$ and $u$ in all observations, and were 0.6\% in the EL observation, 1.2\% in the EM observation, and 0.7\% in the GL observation. The slight increase in residual leakage in the unfiltered EL Stokes $q$ data is may due to the presence of the ripple artefact, and the increase in residual leakage in the unfiltered GL observation is most likely due to contamination from the foreground diffuse emission. Components with a polarized fraction at or below the residual leakage level in each observation should be treated with caution when interpreting the polarization properties. We discuss this further in Section \ref{subsec: pol'd src cats}.

We also use the median leakage to estimate the individual $qu$ leakage levels local to each polarized component by calculating the distance of each component from the field center and interpolating the median curve to find the leakage value at that distance. The larger of the residual $qu$ leakage values for each component is included in the RM catalogs of each observation (see Section \ref{subsec: pol'd src cats}). Since the EC observation uses two methods of off-axis leakage correction (see Section \ref{subsubsec: leakcorr}), for components in this observation we assign the local residual leakage level of the EL portion of the data. We report the EL leakage level for the EC data because the EL data is the dominant contributor to the EC spectra in terms of number of frequency channels (see Section \ref{subsec: data quality}).

\subsection{Faraday complexity}\label{subsec: complexity descrip}

In the construction of our RM grids and RM catalogs, we want to differentiate between two broad types of polarized components: ``Faraday simple" and ``Faraday complex". A Faraday simple component has just one associated value $\phi$, or a single peak in the Faraday spectrum, which we can equate to its RM. Examples of Faraday simple Faraday spectra are provided in Appendix 
\ref{app: example spectra} in Figures \ref{fig: low medfilt bright simple}, \ref{fig: mid medfilt bright simple}, and \ref{fig: comb medfilt bright simple}. A Faraday complex component is any component whose Faraday spectrum is not Faraday simple. 

There are three general scenarios that give rise to Faraday complexity. The first scenario is multiple components with different intrinsic RMs that are unresolved within the synthesized beam. This scenario will result in either multiple peaks in the Faraday spectrum, or non-Gaussian broadening of a single peak if the individual peaks cannot be resolved due to insufficient Faraday resolution. The second scenario is multiple synchrotron-emitting regions along the line of sight with different intrinsic RMs, such as the combination of a background component and a foreground region of co-spatial emission and rotation (e.g. a supernova remnant). Similar to the first scenario, the second scenario will also manifest as either multiple or broadened peaks in the Faraday spectrum, depending on the Faraday resolution. Depolarization can also occur due to differential Faraday rotation across an emitting and rotating region.  The third scenario is tangled magnetic fields due to turbulence in a foreground Faraday-rotating screen (e.g. the ISM). This occurs when the angular size of the background component is larger than the typical angular scale of the foreground turbulent magnetic field structure, and is therefore most relevant for more distant Faraday screens. This scenario also causes depolarization of the signal from the background component and will appear as a broadening of the main peak in the Faraday spectrum. Since Faraday complexity will manifest as any combination of broadened and multiple peaks in the Faraday spectrum, it is difficult to ascribe a single value of $\phi$ to a Faraday complex component, making these components less ideal for use in an RM grid. See \citet{Alger+21} and \citet{Thomson+23} for a more detailed description of Faraday complexity. Examples of Faraday complex Faraday spectra are provided in Appendix \ref{app: example spectra} in Figures \ref{fig: gal nofilt DE faint}, \ref{fig: gal medfilt DE faint}, \ref{fig: low no filt complex}, and \ref{fig: low medfilt complex}.

An automated method for identification and classification of Faraday complexity has become increasingly necessary for large radio surveys like POSSUM that will yield hundreds of thousands of polarized radio components. We use two of these automated methods to quantify the Faraday complexity of our components: the normalized second moment of the clean peaks, $m_2$, and the $\sigma_{\mathrm{add}}$ complexity metric. We describe how the two metrics are calculated below and compare their results for components in the median-filtered GL observation. We use the GL observation as an example because we expect a greater number of complex components to be present due to the line of sight through the Galactic plane, allowing us to evaluate how well the two metrics agree in their identification of Faraday complex components.

\subsubsection{RM-CLEAN and the second moment of the clean peaks metric}
\label{subsubsec: M2}

The first Faraday complexity metric that we calculate is the second moment of the clean peaks, M$_2$, proposed by \citet{Brown11}. Examples of the application of this metric can be found in \citet{Anderson+15} and \citet{Livingston+22}. To calculate M$_2$, we perform RM-\textsc{clean} \citep{Heald+09} on our polarized components. RM-\textsc{clean} deconvolves the Faraday spectrum with the RMSF, returning a list of the cleaned peak amplitudes, $|F_c(\phi_i)|$, and $\phi$ values of peaks therein, known as clean components. M$_2$ is calculated as:

\begin{equation}
    \mathrm{M_2} = \bigg[ \frac{\sum^N_i \big(\phi_i -\bar{\phi}\big)^2 |F_c(\phi_i)|}{\sum_i^N |F_c(\phi_i)|}\bigg]^{1/2} \, ,
\end{equation}

\noindent where

\begin{equation}
    \bar{\phi} = \frac{\sum_i^N \phi_i |F_c(\phi_i)|}{\sum_i^N |F_c(\phi_i)|} \, 
\end{equation}

\noindent for $N$ clean components. M$_2$ is the square root of the centered second moment of the clean component distribution in a Faraday spectrum. A single clean component corresponds to M$_2$ = 0. The magnitude of M$_2$ will depend on the number of clean components, their amplitudes $|F_c(\phi_i)|$, and their separations in $\phi$ space.

We use \texttt{RM-Tools} to perform the cleaning. A minimum threshold in polarized intensity is required for the deconvolution. To choose this threshold, we make use of the Gaussian-equivalent significance (GES) formalism from \citet{Hales+12}, which quantifies the significance of a peak in the Faraday spectrum in terms of Gaussian statistics. For example, the likelihood of an 8$\sigma$ detection in our Faraday spectrum being noise is equivalent to an 8$\sigma$ detection in Gaussian noise. We choose the same threshold here as we do for our S/N$_{\mathrm{pol}}$ threshold in Section \ref{subsec: spec extract} of 8$\sigma$, or 8 times the GES. Choosing the clean threshold in this way helps avoid cleaning too deep, which can generate spurious clean peak detections.

Previous studies have shown that measuring complex structure in Faraday depth that is smaller than the FWHM of the RMSF is difficult \citep{Farnsworth+11,Kumazaki+14,Sun+15}, so the typical value of $\delta\phi$ of an observation will also be a factor in the ability to measure complexity. As the final measurement of Faraday complexity, we normalize M$_2$ by the component's measured $\delta\phi$:

\begin{equation}
    m_2 = \frac{\mathrm{M}_2}{\delta\phi} \, .
\end{equation}

\noindent Normalizing the M$_2$ value takes the resolution of the Faraday spectrum into account when quantifying complexity and also makes the values of $m_2$ comparable between components with different values of $\delta\phi$.

\subsubsection{Sigma add complexity metric}\label{subsec: sigadd metric}

The second metric that we use to quantify the Faraday complexity of our polarized components is the $\sigma_{\mathrm{add}}$ metric, which is described by \citet{Purcell+17} and which will be presented formally in a future \texttt{RM-Tools} paper by Van Eck (2024, in prep). This metric quantifies how different the fractional Stokes $qu$ spectra are from those of a Faraday simple model. After 1D RM synthesis is performed on a component, a Faraday simple model is created using the component RM, polarized intensity, and polarized fraction, and the fractional Stokes $qu$ spectra of the model is subtracted from the component. If the component is simple, the residuals should be purely noise with a Gaussian distribution and mean of zero. If the component is complex, due to any of the Faraday complex scenarios described in Section \ref{subsec: complexity descrip}, the residuals will retain some structure that is not expected to be normally distributed around a mean value of zero, and which indicates that the component is Faraday complex. We describe the method of calculating the $\sigma_{\mathrm{add}}$ metric for the Stokes $q$ spectrum below.

Structure in the residuals in the Stokes $q$ spectrum is modeled as an additional noise term, $\sigma_{\mathrm{add},q}$, added to the total noise of the input data:

\begin{equation}\label{eqn: sigadd var total}
    \sigma_{\mathrm{total},q,i}^2 = \sigma_{\mathrm{noise},q,i}^2 + \sigma_{\mathrm{add},q}^2 \, ,
\end{equation}

\noindent where $\sigma_{\mathrm{noise},q,i}^2$ is the uncertainty of the data in the $i^{\mathrm{th}}$ spectral channel. The Stokes $q$ spectrum is normalized by the uncertainty spectrum and $\sigma_{\mathrm{noise},q}^2$ is set to a unit vector, which allows a dimensionless $\sigma_{\mathrm{add},q}$ to be compared between different components and observations. We use Bayesian inference to estimate the value of $\sigma_{\mathrm{add},q}$ for a component, where the posterior probability of a given value of $\sigma_{\mathrm{add},q}$ being true is the product of the likelihood, $\mathcal{L}$, and the prior probability, $ \pi$. 
The likelihood of $\sigma_{\mathrm{add},q}$, assuming that the residuals are Gaussian-distributed, is given by

\begin{equation}
    \mathcal{L}(\sigma_{\mathrm{add},q}) = \frac{1}{\sqrt{(2\pi)^M \Pi^M_i (\sigma_{\mathrm{noise},q,i}^2 + \sigma_{\mathrm{add},q}^2})} \exp \bigg(- \frac{\chi^2}{2}\bigg) \, ,
\end{equation}

\noindent with:

\begin{equation}
    \chi^2 = \sum^M_i \frac{(x_{q,i} - x_{\mathrm{med},q})^2}{\sigma_{\mathrm{total},q,i}^2} \, .
\end{equation}

\noindent $M$ is the number of frequency channels, $x_{q,i}$ is the normalized Stokes $q$ residual in the $i_{\mathrm{th}}$ channel, and $x_{\mathrm{med},q}$ is the median of the normalized residuals. Assuming no prior knowledge of what $\sigma_{\mathrm{add},q}$ should be, we use the scale-invariant Jeffrey's prior:

\begin{equation}
    \pi(\sigma_{\mathrm{add},q}) = 
        \begin{cases}
            \sigma_{\mathrm{add},q}^{-1} & \text{if $\sigma_{\mathrm{add},q} > \langle \sigma_{\mathrm{total},q} \rangle$/10$^3$} \\
            0 & \text{otherwise} \, ,
        \end{cases}
\end{equation}

\noindent where $\langle \sigma_{\mathrm{total},q} \rangle$ is the average value of $\sigma_{\mathrm{total},q}$ over the $i$ frequency channels. $\sigma_{\mathrm{add},q}$ is reported as the 50$\mathrm{th}$ percentile of the posterior probability distribution, and the uncertainties $\delta \sigma_{\mathrm{add},q,\pm}$ are reported as the 16$\mathrm{th}$ and 84$\mathrm{th}$ percentiles. A value of $\sigma_{\mathrm{total},q} = 0$ indicates that the normalized residuals have a Gaussian distribution with $\mu = 0$ and $\sigma = 1$, and no Faraday complexity is determined to be present. A value of $\sigma_{\mathrm{total},q} > 0$ indicates that Faraday complexity is present in the spectrum and that the residuals are deviating from Gaussian, with the value of $\sigma_{\mathrm{total},q}$ quantifying the magnitude of the deviation.

This calculation is repeated for the Stokes $u$ spectrum, and we calculate the total value of $\sigma_{\mathrm{add}}$ as 

\begin{equation}\label{eq: sigadd total}
    \sigma_{\mathrm{add}} = \sqrt{\sigma_{\mathrm{add},q}^2 + \sigma_{\mathrm{add},u}^2}.
\end{equation}

\noindent $\delta \sigma_{\mathrm{add},\pm}$ are calculated by propagating the errors of $\sigma_{\mathrm{add},qu}$. $\sigma_{\mathrm{add},qu}$ is an output of 1D RM synthesis with the \texttt{RM-Tools} package.

This metric has been used by \citet{Allison+17} in modeling quasar spectral variability, by \citet{Purcell+15} to characterize additional systematic uncertainties on RM measurements, and most recently by \citet{Thomson+23} to measure Faraday complexity in polarized radio sources. \citet{Thomson+23} highlight error modes in $\sigma_{\mathrm{add}}$ that are important to understand if using the metric as the sole quantifier of Faraday complexity. An advantage of using the $\sigma_{\mathrm{add}}$ metric to quantify Faraday complexity is that the calculation requires no decision on the part of the user. Other methods of quantifying complexity require the user to make choices, such as model selection with $QU$ fitting (e.g. \citealt{O'Sullivan+12}), or selecting a clean threshold with the second moment of the clean peaks, which we discussed in Section \ref{subsubsec: M2}.

\subsubsection{Threshold for complexity}
\label{subsubsec: complex thresh}

The complexity metrics described above attempt to quantify the Faraday complexity of the components in our catalogs in an automated way, where the values of $\sigma_{\mathrm{add}}$ and $m_2$ are $\geq$ 0, and larger values indicate increased Faraday complexity. While a  Faraday simple component will have a $\sigma_{\mathrm{add}}$ or $m_2$ value of $\sim$ 0, individual science cases will have unique tolerances to Faraday complexity. Here we discuss the tolerance for Faraday complexity for our work and determine thresholds of $\sigma_{\mathrm{add}}$ and $m_2$, above which we consider a component too complex to be included in the construction of our RM grids.  $m_2$ measures complexity in the Faraday spectrum, while $\sigma_{\mathrm{add}}$ measures complexity in the Stokes $qu$ spectra, which manifest as any behaviour that is not purely sinusoidal. The two methods should return equivalent results since the two are Fourier conjugates (a single delta function peak in the Faraday spectrum corresponds to sinusoidal Stokes $qu$ spectra).

Faraday simple components are ideal for many science cases because there is no ambiguity in the RM measurement. A simple component’s RM can be written as the sum of individual contributions of all intervening Faraday-rotating structures along the line of sight:

\begin{equation}\label{eq: RM sum}
\mathrm{RM} = \mathrm{RM}_{\mathrm{int}} + \mathrm{RM}_{\mathrm{IGM}} + \mathrm{RM}_{\mathrm{MW}} \, , 
\end{equation}

\noindent where RM$_{\mathrm{int}}$ is the intrinsic RM of the source (due to the local environment), RM$_{\mathrm{IGM}}$ is the contribution from the intergalactic medium (IGM), and RM$_{\mathrm{MW}}$ is the contribution from the Milky Way. $\lvert$RM$_{\mathrm{int}}\rvert$ is expected to be $\sim$6 rad m$^{-2}$ \citep{Schnitzeler2010} and $\vert$RM$_{\mathrm{IGM}}\rvert$ is estimated to be $\sim$1–10 rad m$^{-2}$ \citep{Akahori&Ryu2010,Vernstrom+19,Amaral+21}. RM$_{\mathrm{MW}}$ varies with the position on the sky, but it is expected to be the dominant contribution to the total RM for most components \citep{Schnitzeler2010,Hutschenreuter+22,O'Sullivan+23}. Faraday complexity can be introduced by any one of the contributions in Equation \ref{eq: RM sum}.

We aim to quantify Faraday complexity because it can be present to varying degrees, causing anywhere from a small amount to total depolarization. The tolerance for Faraday complexity of a given user of our RM catalog will be determined by their unique science case. For example, isolating the relatively small IGM magnetic field contribution to a component RM will ideally require precise RM measurements (small $\delta$RM) and a single peaked Faraday spectrum showing little-to-no effects of complexity in the Faraday spectrum (a Faraday simple spectrum). In this case, a strict complexity threshold of the larger of $m_2$ $\leq \delta$RM/$\delta\phi$ and $m_2$ $\leq$ 1/$\delta\phi$ may be desired, which constrains the Faraday complexity level of a component to be at most either the level of uncertainty in RM or the lowest current estimate of the contribution of RM$_{\mathrm{IGM}}$ to the total RM (note that the sample spacing of the Faraday spectrum, $\Delta\phi$, must be chosen to be less than or equal to twice the complexity threshold). Alternatively, if an RM catalog is being used to measure the distribution of RMs in H II regions, where RMs have been observed to have magnitudes of $\sim$50–1000 rad m$^{-2}$ \citep{Harvey-Smith+11,Costa&Spangler2018}, the tolerance for complexity in the Faraday spectrum may be higher than in the case of the IGM due to the significantly larger RMs, and a higher threshold for complexity allows for a denser RM grid. In this case, the user may select an $m_2$ threshold of the larger of $\delta$RM and some acceptable spread in the clean peaks of RMs.

\citet{Hutschenreuter+22} use Bayesian inference to construct an all-sky RM map from the \citet{VanEck+23} RM catalogue, correlating many lines of sight to determine the Galactic contribution to the RM in a region of sky. This method of mapping the RM sky increases the tolerance to complexity because variations in RM of nearby components on the sky will be damped by the combining of many lines of sight. This is the default RM catalog use case that we adopt here for the construction of our RM grids, although we do not go on to apply the method of \citet{Hutschenreuter+22} to construct a smoothed RM sky map over our two regions of sky. Two clean components of equal amplitude in the Faraday spectrum separated by $\delta\phi$ will give $m_2 = 0.5$. We choose this value of $m_2$ as our threshold for Faraday complexity because this is the value where multiple, independent peaks in the Faraday spectrum will start to be resolved, and assigning a single RM to a component becomes difficult. As such, we set a complexity threshold of the larger of $m_2$ $\leq \delta$RM/$\delta\phi$ and $m_2$ $\leq$ 0.5. We use the same threshold for all of our data (EL, EM, EC, and GL) since we have normalized by $\delta\phi$, making $m_2$ comparable between observations.

The relationship between the magnitude of $\sigma_{\mathrm{add}}$ and the type and degree of Faraday complexity in a component is not well understood yet and requires deeper investigation, which we suggest for future work. 
When calculating $\sigma_{\mathrm{add}}$, the Stokes $qu$ spectra are normalized by the uncertainty spectra and the $\sigma_{\mathrm{noise}}$ term in Equation \ref{eqn: sigadd var total} is normalized to a unity vector. This means that after subtracting the Stokes $qu$ spectra of a Faraday simple model from a Faraday simple component, the distribution of the normalized residuals should have $\sigma = 1$ when $\sigma_{\mathrm{add}} = 0$. 
Since we are allowing for some tolerance to Faraday complexity, we can relax our threshold on $\sigma_{\mathrm{add}}$ to include more than just the components at this low $\sigma_{\mathrm{add}}$ peak.
After manual inspection of component spectra in the different observations, we set a threshold for complexity of $\sigma_{\mathrm{add}} - \delta \sigma_{\mathrm{add,-}} <$ 1, or where $\sigma_{\mathrm{add}}$ is below 1 within uncertainty. From our inspection, components with values of $\sigma_{\mathrm{add}}$ below this threshold show limited Faraday complexity, and we apply this threshold to the rest of our analysis.

\subsubsection{Comparison of sigma add and the second moment of the clean peaks}
\label{subsubsec: M2 vs sigadd}

\begin{figure}
\centering
    \includegraphics[width=0.47\textwidth]{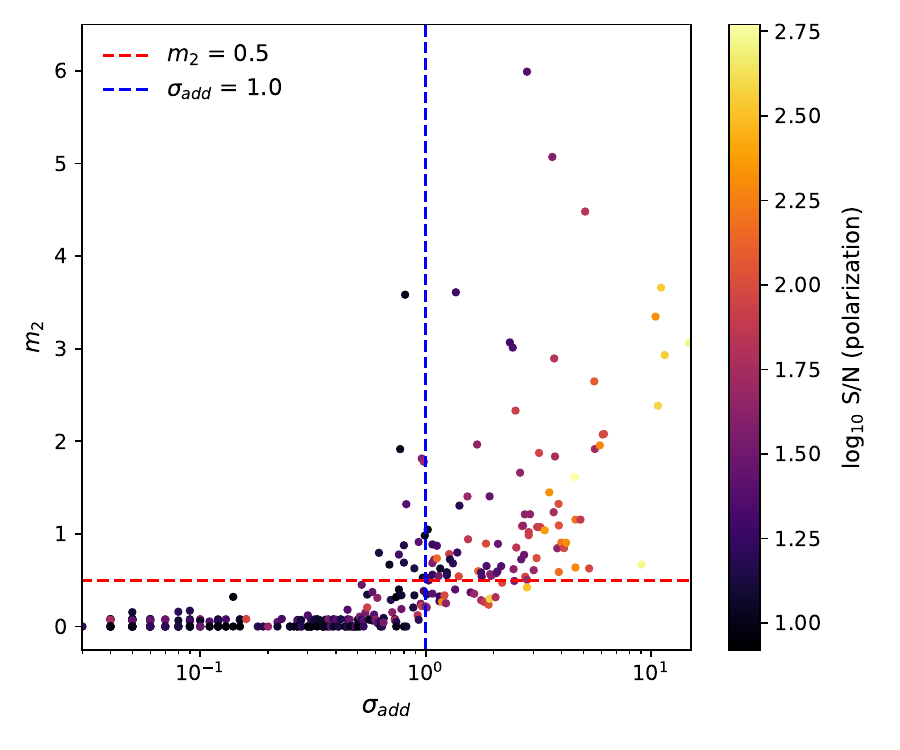}
    \caption{Comparison of the $\sigma_{\mathrm{add}}$ and $m_2$ complexity metrics for polarized components in the median-filtered GL observation. The points are colored by $\log_{10}$(S/N$_{\mathrm{pol}}$), and the thresholds for complexity for both metrics are shown by dashed lines: $\sigma_{\mathrm{add}}$ = 1 (blue vertical) and $m_2$ = 0.5 (red horizontal). Note that the $\sigma_{\mathrm{add}}$ axis is on a log scale, while the $m_2$ axis is on a linear scale.}
\label{fig: sigadd vs m2}
\end{figure}

In Figure \ref{fig: sigadd vs m2} we plot the value of $m_2$ versus $\sigma_{\mathrm{add}}$ for the polarized components in the median-filtered GL observation. We choose the GL observation as our example because it has more Faraday complex components than the other observations. There are 20 components out of the 347 total polarized components in the GL observation for which $m_2$ was unable to detect any peaks in the Faraday spectra at an 8$\sigma$ GES cleaning threshold. All of these components had $8.0 \leq$ S/N$_{pol} \leq$ 8.3 and fall below our $\sigma_{\mathrm{add}}$ complexity threshold. These components are not included in Figure \ref{fig: sigadd vs m2}.  The upper right quadrant of Figure \ref{fig: sigadd vs m2} contains components which both metrics identify as complex, and the lower left quadrant contains components which both metrics identify as simple. The other two quadrants contain components where the two metrics disagree on their complexity classification.

Most components in Figure \ref{fig: sigadd vs m2} fall either above or below both thresholds, indicating that the two metrics are identifying similar levels of complexity in the same components. There are 16 components that are not above our $m_2$ complexity threshold but are above our $\sigma_{\mathrm{add}}$ complexity threshold. These components typically have a higher S/N (all but four have S/N$_{\mathrm{pol}} > 43$), suggesting that $\sigma_{\mathrm{add}}$ may have a stronger S/N dependence than $m_2$. There are 22 components that are not above our $\sigma_{\mathrm{add}}$ threshold but are above our $m_2$ threshold, all but three of which have S/N$_{\mathrm{pol}} \leq 40$. \citet{Thomson+23} plot the same comparison of $m_2$ and $\sigma_{\mathrm{add}}$ and also find generally good agreement between them.

Both metrics also show a more general dependence on S/N$_{\mathrm{pol}}$, where components with a higher S/N$_{\mathrm{pol}}$ tend to have larger complexity values and components with a lower S/N$_{\mathrm{pol}}$ tend to have a lower complexity value. In the median-filtered GL observation, all components below a S/N$_{\mathrm{pol}}$ of 9.9 for $m_2$ and 11.5 for $\sigma_{\mathrm{add}}$ lie below our complexity thresholds, and all except two components with a S/N$_{\mathrm{pol}}$ greater than 96 lie above our complexity thresholds with both metrics. This dependence on S/N may suggest a genuine increase in complexity with increasing brightness or that below some S/N threshold, we cannot detect the complexity that is present, or both (see \citealt{Anderson+15}). \citet{Thomson+23} also see this S/N dependence in these complexity metrics with their data. They suggest that the ripple in the Stokes spectra (see Section \ref{subsec: spec extract}) is the primary cause of this dependence. Moving forward, we interpret complexity in high-S/N components with caution.

\section{Results}
\label{sec: results}

\subsection{Polarized component catalogs}
\label{subsec: pol'd src cats}

We present RM catalogs for the EL, EM, EC, and GL observation, each of these both with and without the median filter applied. We include all columns in the \texttt{RM-\textsc{Table}}\footnote{\url{https://github.com/CIRADA-Tools/RMTable}} \citep{VanEck+23} standard convention for RM catalogs plus some additional columns beyond this standard that include information from the source finder catalog. A description of the table columns is provided in Appendix \ref{app: cols and cats} along with the first two rows of the median-filtered GL observation catalog as an example. The basic properties of the polarized (S/N$_{\mathrm{pol}}$ $\geq$ 8) components in the catalogs are summarized in Table \ref{tab: numbers}. These include: the median value of $\delta\phi$ (Equation \ref{eq: rmsf fwhm}), the mean and median value of $\delta$RM (Equation \ref{eq: deltaRM}), the sky density of all polarized components, the sky density of Faraday simple polarized components (defined in Section \ref{subsubsec: complex thresh}), the total number of polarized components, the typical residual off axis leakage level (the larger of the Stokes $qu$ leakage estimates), and the fraction of the polarized components with a polarized fraction at or below the residual leakage estimate. We focus the majority of our presentation of the results and our discussion on the median-filtered catalogs.

\begin{table*}
\centering
\caption{Summary of polarized component catalogs}
\label{tab: numbers}
\begin{tabular}{p{0.6cm} p{0.9cm} p{1.5cm} p{1.5cm} p{1.5cm} p{1.6cm} p{1.6cm} p{1.2cm} p{1.5cm} p{1.2cm} p{1.2cm}}
\hline
 & Median \newline filter & Median \newline $\delta\phi$ & Median $\delta$RM & Mean \newline $\delta$RM & Polarized \newline component \newline sky density & Simple \newline component \newline sky density & Fraction  \newline complex & Total \newline polarized \newline components & Residual \newline off-axis \newline leakage$^a$ & Fraction \newline below \newline leakage \\
 & & (rad m$^{-2}$) & (rad m$^{-2}$) & (rad m$^{-2}$) & (deg$^{-2}$) & (deg$^{-2}$) &  &  & (\%)  &  \\
\hline
EL & N & 60.8 & 1.56 & 1.70 & 45.0 & 37.8 & 0.160 & 518 & 0.6 & 0.04 \\
EL & Y & 61.0 & 1.55 & 1.65 & 42.0 & 35.1 & 0.165 & 484 & 0.6 & 0.05 \\
EM & N & 452.7 & 12.74 & 13.12 & 31.6 & 30.7 & 0.027 & 364 & 1.2 & 0.05 \\
EM & Y & 452.8 & 12.82 & 13.28 & 31.4 & 30.6 & 0.028 & 362 & 1.2 & 0.06 \\
EC & N & 42.6 & 1.13 & 1.22 & 51.5 & 40.5 & 0.214 & 593 & --- & --- \\
EC & Y & 42.5 & 1.06 & 1.17 & 48.0 & 37.2 & 0.226 & 553 & --- & --- \\
GL & N & 62.1 & 2.71 & 2.42 & 31.4 & 23.6 & 0.249 & 539 & 0.7 & 0.17 \\
GL & Y & 61.8 & 1.89 & 1.92 & 20.2 & 13.5 & 0.334 & 347 & 0.6 & 0.20 \\ \hline
\multicolumn{11}{c}{$^a$ We do not calculate residual off-axis leakage for the EC observation because it is a combination of the EL and EM data.} \\
\end{tabular}
\end{table*}

\vspace{5mm}

The final data products, including the RM catalogs, can be accessed via the CSIRO ASKAP Science Data Archive (CASDA; \citealt{Chapman+17,Huynh+20}). The data are split over two collections:

\begin{itemize}

\item the EM Stokes cubes can be accessed along with the full POSSUM pilot I data collection at \url{https://data.csiro.au/collection/csiro%3A62003v1}

\item the EL, EC, and GL Stokes cubes and the RM catalogs can be accessed at \url{https://data.csiro.au/collection/csiro%3A62005v1}.

\end{itemize}

\noindent Descriptions of the data collections can be found on the respective web pages.

\subsection{Catalog reliability}

\subsubsection{Data quality}\label{subsec: data quality}

\begin{figure}
\centering
    \includegraphics[width=0.47\textwidth]{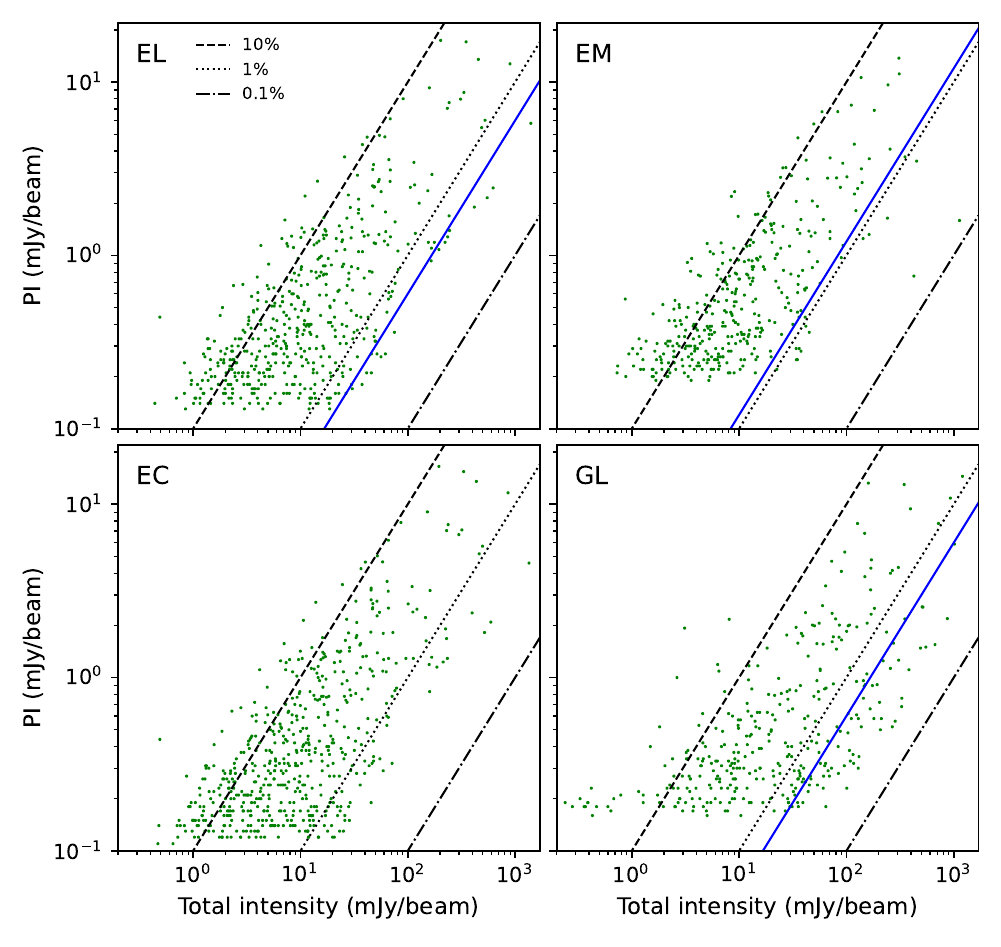}
    \caption{Polarized intensity versus total intensity for components in the four median-filtered polarized component catalogs. Lines of constant polarized fraction are plotted at 0.1\%, 1\%, and 10\%, and the solid blue line indicates the typical estimated residual off-axis leakage level in the observation.}
\label{fig: PI vs I}
\end{figure}

Here we assess the quality of the polarization data in our four median-filtered observations. We plot polarized intensity versus total intensity for components in each observation in Figure \ref{fig: PI vs I}. The distribution of the components in all cases is similar to those seen in other studies, both with ASKAP \citep{Anderson+21,Thomson+23} and other telescopes \citep{Banfield+14,Hales+14,Anderson+15,O'Sullivan+23}. The unfiltered catalogs show a very similar distribution of components and leakage levels as the median-filtered catalogs, with the exception of a significant number of additional components in the GL observation with high polarized fraction ($\gtrsim$50\%) and low intensity. These components are diffuse emission detections that are mostly removed by the median filter (see Section \ref{subsubsec: testing filt} and Figure \ref{fig: sim RM vs RM} therein). 

In the EL and EM observations, we see a small fraction of components that lie below the typical residual off-axis leakage level (0.04 and 0.05 in the median-filtered EL and EM fields, respectively). The median-filtered GL observation has a much higher fraction of components below the leakage level, 0.17. This might suggest that there is residual diffuse emission contaminating these components, or that applying the holography leakage correction to fields with diffuse emission is more difficult. Further work needs to be done to understand the reason for the increased number of components below the typical residual leakage level in the GL field, and we interpret the polarization properties of these components with caution.

\begin{figure}
\centering
    \includegraphics[width=0.47\textwidth]{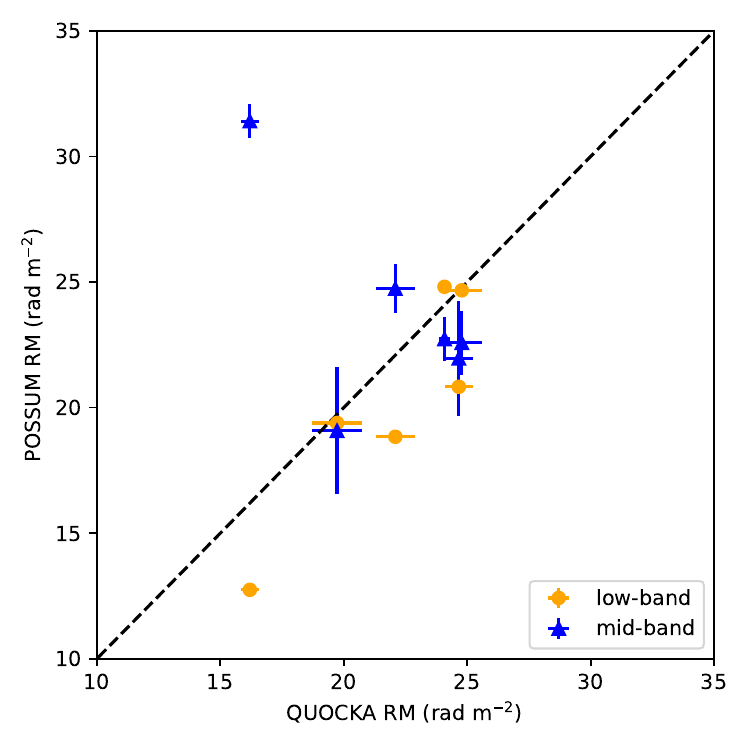}
    \caption{Comparison of EL and EM POSSUM RMs to RMs from the QUOCKA survey. POSSUM EL RMs are plotted as orange circles and POSSUM EM RMs are plotted as blue triangles.}
\label{fig: QUOCKA vs POSSUM}
\end{figure}

In Figure \ref{fig: QUOCKA vs POSSUM}, we compare the RMs of six components in our median-filtered EL and EM observations to RMs of the same components observed with the QU observations at cm wavelength with km baselines using ATCA (QUOCKA\footnote{\url{https://research.csiro.au/quocka/}}; Heald et al, in prep) survey. The QUOCKA data are observed over two frequency ranges:  $\sim$1.3--3 GHz and $\sim$4.6--8.4 GHz. The angular resolution of the original QUOCKA data ranges from 15.1$\times$7.2 arcsec$^2$ to 19$\times$8.1 arcsec$^2$. All of the QUOCKA data are convolved to the angular resolution of the EL data before we perform RM synthesis. The Faraday resolution ($\delta\phi$) of the QUOCKA data ranges from 72--110 rad m$^{-2}$.

Each QUOCKA RM has two associated POSSUM RMs: one in the EL data and one in the EM data.
Four of the six EL POSSUM components that were matched to the QUOCKA data are Faraday complex according to our threshold, which somewhat complicates the direct comparison of RMs. None of the six EM POSSUM components that were matched to the QUOCKA data are Faraday complex.
We find a correlation between Faraday complexity and RM agreement, where components with lower values of $m_2$ typically have better agreement between the POSSUM and QUOCKA RMs. In the EL data, the two RMs that are within 3$\sigma$ of the QUOCKA RM have $m_2 \leq 0.3$, while the other components all have $m_2 \geq 0.4$. In the EM data, the five RMs that are within 3$\sigma$ of the QUOCKA RM have $m_2 \sim 0.01$, while the remaining component has $m_2 = 0.23$. As discussed in previous sections, Faraday complexity can result in incorrect RM measurements, and this is likely the primary source of disagreement between the POSSUM and QUOCKA RMs. A primary goal of QUOCKA is to investigate Faraday complexity in greater detail to better understand components such as these.

\begin{figure*}
\centering
    \includegraphics[width=\textwidth]{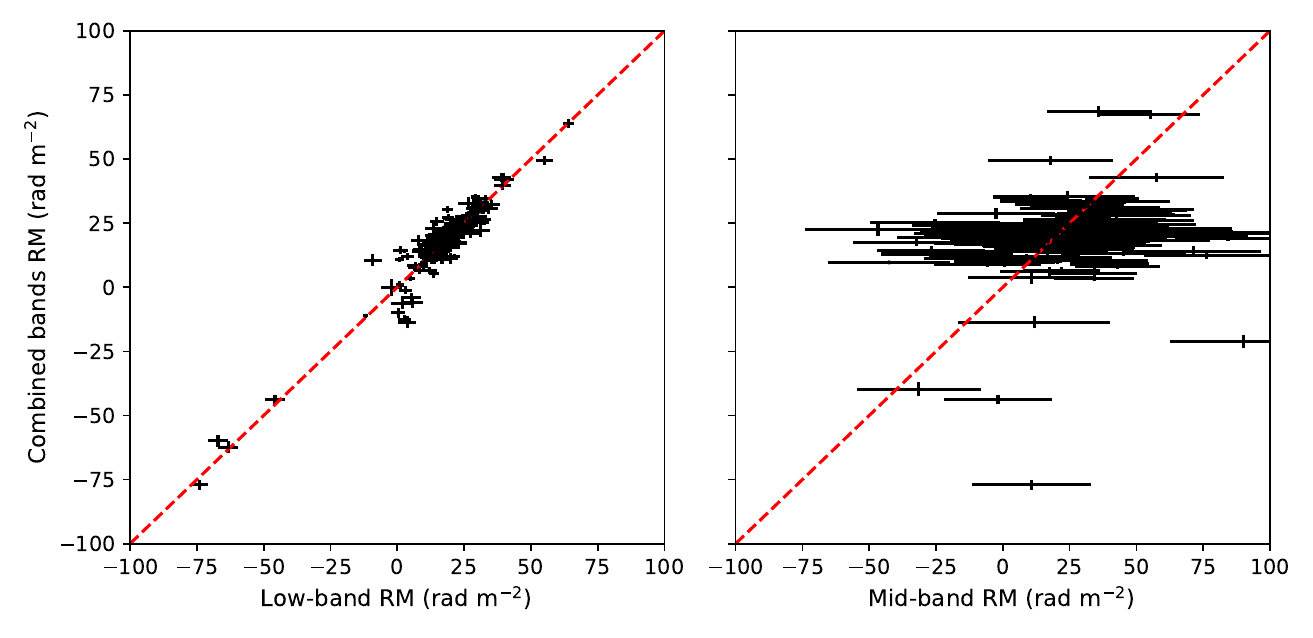}
    \caption{\textit{Left:} EC RMs versus EL RMs. \textit{Right:} EC RMs versus EM RMs. The dashed red line in both panels is the one-to-one line.}
\label{fig: low mid vs comb}
\end{figure*}

Next we compare the median-filtered EL and EM RMs to the median-filtered EC RMs. We compare only components that are below our complexity threshold in the two bands being compared to avoid issues with Faraday complexity. This is a total of 362 Faraday simple components in the EL and EC RM catalogs and 226 Faraday simple components in the EM and EC RM catalogs. All observations were convolved to 21-arcsec angular resolution before spectral extraction for proper comparison. We plot the comparison to the EC data in Figure \ref{fig: low mid vs comb}, with the EL RMs on the left panel and the EM RMs on the right panel.

\begin{figure}
\centering
    \includegraphics[width=0.47\textwidth]{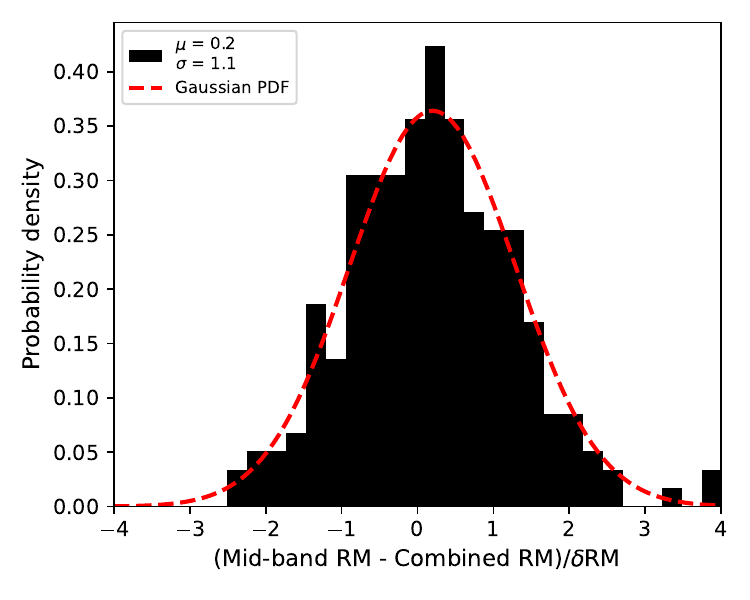}
    \caption{Distribution of the difference between the EM and EC RMs in units of $\delta$RM.  We overplot a Gaussian fit to the data, shown by the dashed red line. There are three points beyond $\pm 3\sigma$ and no points beyond the bounds of the graph. The distribution has $\mu \sim 0$ and $\sigma \sim 1$, suggesting that the large spread in RMs in Figure \ref{fig: low mid vs comb} is due to the large RM uncertainties in the EM and is not an issue with data quality.}
\label{fig: RM diff dist}
\end{figure}

In Figure \ref{fig: RM diff dist} we plot the distribution of the difference in RM between the EM and EC RMs divided by the uncertainty of the two RM measurements added in quadrature, $\delta$RM. The distribution has $\mu = 0.20 \pm 0.07$ and $\sigma = 1.10$, suggesting that the RM measurements in the two bands typically agree, and that the apparent disagreement in the right panel of Figure \ref{fig: low mid vs comb} is not a data quality issue, but that the large uncertainties on the EM RMs make extracting precise RMs difficult. The same calculations for the EL and EC RMs (362 components) give a distribution with $\mu = -0.06 \pm 0.08$ and $\sigma = 1.59$, and we get a distribution with $\mu = -0.16 \pm 0.08$ and $\sigma = 1.16$ for the comparison of the EL and EM RMs (235 components). All three distribution have $\sigma > 1$, which suggests that the RM uncertainties may be consistently underestimated.

\subsubsection{Polarized fraction dependence on signal-to-noise}\label{subsubsec: pf vs snr}

\begin{figure}
\centering
    \includegraphics[width=0.47\textwidth]{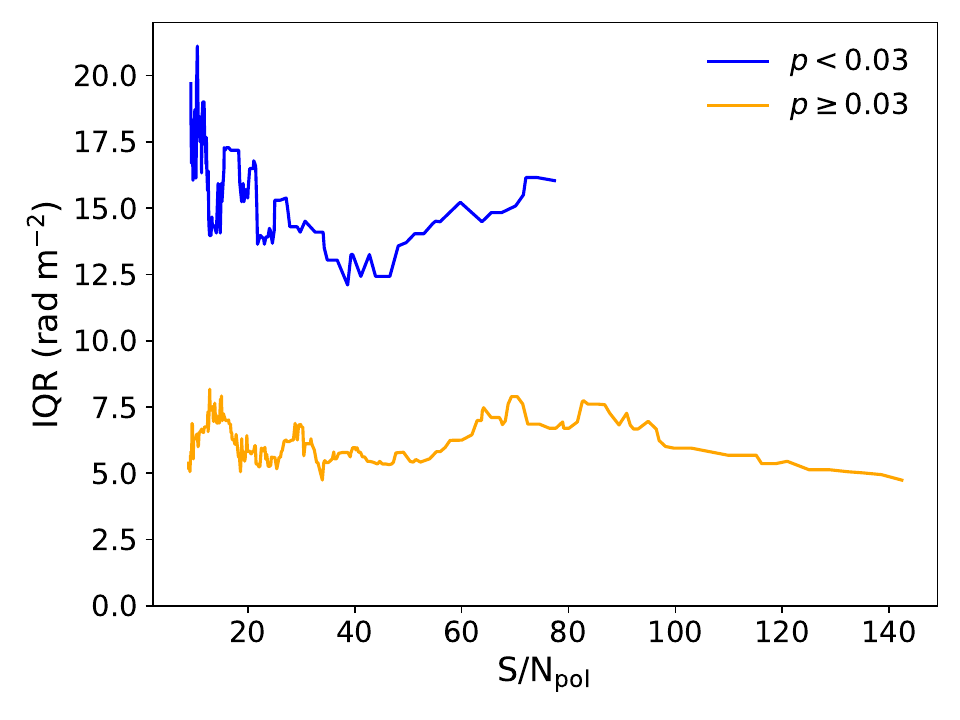}
    \caption{Scatter in RM as measured by the interquartile range (IQR) versus S/N$_{\mathrm{pol}}$ for both low ($p < 0.03$) and high ($p \geq 0.03$) polarized fraction components.}
\label{fig: pf vs snr}
\end{figure}

In Figure \ref{fig: pf vs snr} we plot the scatter in RM as a function of S/N$_{\mathrm{pol}}$, with components separated into low and high fractional polarization populations. We calculate the interquartile range (IQR) of a running sample of 50 components as a measure of the RM scatter. We will show in Section \ref{subsubsec: SFs} that the Galactic magnetic field exhibits no effects of turbulence in the extragalactic field at any scale probed by the EC data. As such, the scatter in RMs of the components in this field will be due to extragalactic factors, primarily intrinsic RM differences. 

Figure \ref{fig: pf vs snr} shows a dependence of RM scatter on polarized fraction. The high polarized fraction sample has a typical scatter in RM of $\sim$6 rad m$^{-2}$ while the low polarized fraction sample  has a typical value of $\sim$15 rad m$^{-2}$. 
This marked difference highlights the need to be more cautious with the use of low polarized fraction components when constructing RM grids. These preliminary results suggest that low polarized fraction components should be weighted relative to the magnitude of their RM scatter \citep{Rudnick2019} before being included in an RM grid. Further analysis of RM scatter as a function of polarized fraction with a larger sample is required to understand this relationship in more detail and to better inform the use of low polarized fraction components in future science cases such as RM grids.

\subsection{Features of the catalogs}

\subsubsection{RM precision}\label{subsubsec: rm precision}

Precise RM measurements will be of particular importance to many science cases that will make use of the POSSUM full survey RM catalog. A broad RMSF decreases the precision with which the RM of a component can be measured, as $\delta$RM $\propto$ $\delta\phi$ (see Equation \ref{eq: deltaRM}). $\lambda^2$ coverage is the primary factor in determining $\delta\phi$ for our data (see Equation \ref{eq: rmsf fwhm}), which is evident in the increasing magnitude of $\delta$RM with $\delta\phi$ in the EL, EM, and EC catalogs in Table \ref{tab: numbers}.

The values of $\delta$RM for the median-filtered GL observation are $\sim$20\% larger than the median-filtered EL values of $\delta$RM despite the similarities in bandwidth, frequency, and median $\delta\phi$ of the two observations. This difference is due to both the lower sensitivity of the observation (see Table \ref{tab: data specs}) and the higher rms in this observation from Galactic foreground contamination.
This difference in RM precision between two low-band observations highlights the approximate range in the precision of RM measurements that ASKAP will achieve in this observing band. For the full POSSUM survey that will observe both Galactic and extragalactic fields, we expect typical $\delta$RM values of $\sim$1.5--2 rad m$^{-2}$ for EL observations and $\sim$1 rad m$^{-2}$ for EC observations.

The EC has the smallest mean and median values of $\delta$RM, making it the optimal data set for extracting precise RMs for the POSSUM catalog. The mean and median EM values of $\delta$RM are an order of magnitude larger than the EL or EC RM uncertainties,
which make this observing band suboptimal for constructing an RM catalog or an RM grid because extracting a precise RM for many components will be difficult. Low-band observations are sufficient for the construction of a POSSUM RM catalog and RM grid, however the addition of the EM data increases both the precision with which we can measure RMs (see Section \ref{subsec: medfilt method}), and the sensitivity of the observations (see Section \ref{subsec: medfilt application Gal}), which increases the total polarized component count. The addition of the EM data also increases the sensitivity to a more strongly depolarized source population, which is important for certain science applications (e.g. clusters). As such, we suggest that combined-band observations should be used whenever possible, and that the low-band is sufficient when combined data are not available.

\subsubsection{Polarized component sky densities}
\label{subsubsec: results pol densities}

POSSUM will achieve an unprecedented sky density of polarized components across 20 000 deg$^2$ (Dec $< 0^{\circ}$). Here we estimate the expected RM sky densities of the full POSSUM survey.

The median-filtered (unfiltered) EL observation has a polarized sky density of 42.0 (45.0) RMs per square degree, while the EM observation has 31.4 (31.6) RMs per square degree, a difference of over 10 RMs per square degree. A major difference between the EL and EM observations is their sensitivities, which in the case of our two data sets is caused by a difference in bandwidth. The EM observation has only 40\% of the number of channels of the EL observation, which decreases the S/N of the polarization detections in the EM data.

Additionally, while synchrotron radiation is typically brighter at lower frequencies due to its negative spectral index \citep{Condon&Ransom2016}, depolarization effects are stronger at these frequencies. To investigate the properties of the components in the two observations, we crossmatch the median-filtered EL and EM polarized component catalogs. Selecting Faraday simple components, we take the ratio of the EL to EM polarized fractions and peak polarized intensities. The median value of the polarized fraction ratio is 0.774 with a MADFM of 0.249, and the median value of the polarized intensity ratio is 1.149 with a MADFM of 0.401. The low-to-mid band polarized intensity and fraction ratios $\frac{P_L}{P_M}$ and $\frac{p_L}{p_M}$ give a total intensity ratio of $\frac{I_L}{I_M} = 1.484$. This gives a spectral index in total intensity of $-1.043$, which is reasonable. Altogether this suggests that while polarized components are typically more depolarized in the EL observation, they are also typically brighter in polarized intensity, and the higher typical polarized fraction in the EM data may account for the discrepancy in percent decrease in total intensity components and polarized sky density between the two bands. In addition, a single polarized component in the EL observation may be resolved into two components in the EM observation due to the higher resolution of the EM observation.

The highest polarized component sky density of our four observations is found in the EC observation, with 48.0 (51.5) RMs per square degree in the median-filtered (unfiltered) observations. The wider bandwidth increases the sensitivity of polarization measurements, allowing detection of fainter polarized signals and increasing the total number of polarized components by 69 as compared with the EL observation. However, combining the EL and EM observations has the disadvantage that we lose resolution in EM, and decreased resolution can lead to stronger beam depolarization effects (depolarization due to multiple interfering polarized sources within the beam). To estimate the total number of individual RMs detected in the EL and EM observations at their original resolutions (21 and 13 arcsec, respectively), we crossmatch the EL and EM polarized component catalogs with a 1 arcsec match radius and take the total number of individual RMs to be the number of matches (199 RMs) plus the number of unmatched RMs in the EL and EM catalogs (285 and 163 RMs, respectively). This gives a total of 647 RMs, which is 94 more than the total number of RMs from the EC polarized component catalog.

As discussed in Section \ref{subsec: data quality}, the EM polarized component catalog has typical RM uncertainties that are an order of magnitude larger than the EL catalog, which significantly decreases RM precision when using the EM data alone. Whether it is ideal to use the combined or individual data sets will depend on the particular science goals of the use. In the context of using RM grids to map the Faraday depth sky, observations at lower frequencies also improve RM grid capabilities (assuming that RM sky density is maintained) beyond the increased RM precision described above. Radio galaxies have been shown to have a substantial range of RM values and fractional polarizations across their extent \citep{Anderson+18,Sebokolodi+20,Baidoo+23}, which leads to depolarization starting in the high RM regions. At decreasing frequencies, observed polarization is coming from lower-RM (and more Faraday simple) regions of the source \citep{O'Sullivan+23}, making the dominant contribution to the RM the Galactic foreground, which is ideal for using RM grids to probe the Galactic magnetic field.

The lowest polarized component sky density is found in the GL observation with 20.2 (31.4) RMs per square degree in the median-filtered (unfiltered) observation. Although the observation has the same bandwidth and frequency range as the EL observation and a similar number of frequency channels, the polarized component sky density in the median-filtered observation is less than half of the EL sky density. One reason for this drastic difference is that depolarization effects are presumably stronger in the GL observation compared to the EL and EM observations (which have a line of sight well off the Galactic plane), since Faraday rotation has been shown to increase at low Galactic latitudes \citep{Schnitzeler2010,Hutschenreuter+20,VanEck+21}. The higher fraction of complex components in the median-filtered GL observation (0.334) compared to the EL observation (0.165) points to an increased amount of Faraday structure in these regions. In addition, $\sim$60\% fewer total intensity components per square degree were identified by the source finder in the GL observation (4988 components in the GL observation compared to 12 124 in the EL observation). This is most likely due to source and sidelobe confusion associated with complex emission in the Galactic plane.

The GL observation sees a significantly larger decrease in polarized component density, from 31.4 to 20.2 RMs per square degree, due to the extensive diffuse emission present in the observation. This indicates that nearly half of the components included in our unfiltered polarized component catalog for the GL observation are not intrinsically polarized above our reliable polarization detection threshold, and that their polarization properties are being altered by the intervening diffuse emission. This again highlights the importance of removing foreground polarized diffuse emission in obtaining reliable polarization measurements of background components for use in RM grids.

A reduction in polarized component sky density after the application of the median filter also occurs in our other observations. The decrease in polarized component sky density is $\sim$7\% for the EL and EC observations, while the EM observation sees only a $\sim$0.6\% decrease. If this decrease was due solely to the systematic loss of polarized intensity discussed in Section \ref{subsubsec: testing filt}, we would expect to see a similar percent decrease in all three observations. The greater decrease in sky density seen in the EL and EC observations suggests that the ripple artefact present in the EL observation was artificially increasing the number of polarized components being identified, and that the median filter is removing this artificial polarized intensity contribution. The small decrease in the polarized component sky density in the EM observation also suggests that the filter has very little effect on polarized component densities in observations with no foreground diffuse emission present.

Due to the limited effect that the median filter has on the polarized sky density in the EM observation (where there is no diffuse emission or artefacts present) recommend applying the median filter to all observations to maintain consistent data reduction and RM extraction. A version of the median filter is expected to be integrated into the POSSUM pipeline for the full survey. This will be described in more detail by Van Eck et al. (2023, in prep).

\subsubsection{Faraday simple component sky densities and Faraday complexity}

The median-filtered EL, EM, EC, and GL observations have Faraday simple component sky densities of 35.1, 30.6, 37.2, and 13.5 components per square degree, respectively. We see the relationship between a narrower RMSF and increased Faraday complexity in Table \ref{tab: numbers}, where the median-filtered EL, EM, and EC observations have $\delta\phi$ values of 61.0, 452.8, and 42.5 rad m$^{-2}$ and fractions of Faraday complex components of 0.165, 0.028, and 0.226, respectively. The increasing rate of complexity in the EL and EC data is due to the broader bandwidth (smaller $\delta\phi$), which allows for a greater chance of detecting complexity in a component’s Stokes spectra. The increase in Faraday complex fraction with increasing bandwidth results in a 2.1 RM per square degree increase in Faraday simple RM sky density in the EC observation, as compared with the EL observation. This implies that combining ASKAP low- and mid-band observations of the same field may not necessarily significantly increase the total density of the POSSUM full survey RM grid.

The EL and EC observations have a similar complexity rate to that reported by \citet{Anderson+15}, who found a Faraday complexity fraction in polarized components of 0.12 (as measured by the $m_2$ metric) in a 30 deg$^2$ field at RA = $52.5^{\circ}$ and Dec = $-36.2^{\circ}$ and over a 1.3--2 GHz band, and seen by \citet{Livingston+22}, who found a Faraday complexity fraction of 0.37 at 1.4--3 GHz in RMs of background components passing through the Small Magellanic Cloud (SMC). We note, however, that both of these studies had a somewhat broader frequency coverage than the observations we analyze in this work, and so a direct comparison of results is not possible.

The GL observation sees the largest decrease in RM sky density after the application of the median filter, from 20.2 to 13.5 RMs per square degree, due to the enhanced Faraday complexity in the Galactic plane as compared to the extragalactic line of sight of the Pilot I observations. The combination of fewer total intensity components identified by the source finder and enhanced Faraday complexity indicate that observations on and near the Galactic plane will likely have a significantly lower RM sky density than those that are well off of the plane, such as our extragalactic Pilot I field. This will have a notable impact on the total number of RMs that the full POSSUM catalog with contain as well as on the RM grid density near the Galactic plane. We provide estimates of total RMs in the POSSUM catalog in Section \ref{subsec: survey densities}.

Each observation shows an increased rate of complexity in the median-filtered catalog when compared to their corresponding unfiltered catalog. The components that are present in the unfiltered catalogs and not in the filtered catalogs are all identified as Faraday simple by the $\sigma_{\mathrm{add}}$ metric, likely due to their low S/N$_{\mathrm{pol}}$ ($\lesssim 15)$. Thus, in all observations, the increase in the fraction of Faraday complex components after applying the median filter is due to a decrease in the number of Faraday simple components (and in the total number of components in the catalog), not an increase in the number of Faraday complex components.

We note that the fractions of Faraday complex components quoted in Table \ref{tab: numbers} are lower limits. As mentioned in Section \ref{subsubsec: complex thresh}, measuring complex structure below the resolution of the Faraday spectrum is difficult. Increasing both the Faraday resolution, through broader-band observations, and the angular resolution of the observations would increase the prevalence of Faraday complexity in our data by resolving more structure in the Faraday spectrum and in the polarization across the source itself. The complexity analysis in this work is interpreted in the context of deriving RM grids from current and future POSSUM data.

\subsubsection{Expected survey component sky densities}
\label{subsec: survey densities}

The current most complete all-sky RM catalog used to map the Galactic Faraday depth sky by \citet{Hutschenreuter+22} is from \citet{VanEck+23} and has average RM sky density of $\sim$1.35 RMs per square degree. This low angular resolution limits our ability to map structure on smaller scales. Recent denser RM grids of smaller regions of the sky have illustrated some of the science that will be possible with data from surveys like SPICE-RACS and POSSUM, including more accurate measurements of the ordered and random magnetic field components of the SMC \citep{Livingston+21}, and probing the magnetoionic structure of the Fornax cluster in remarkable detail \citep{Anderson+21}.

We have determined that the observing strategy that will return the greatest number of polarized components in the POSSUM full survey catalog is the combined low- and mid-band observation strategy, with mitigation of diffuse foreground emission.  The median-filtered EC data set returns $\sim$13\% more RMs compared to the median-filtered EL data set, due to increased sensitivity and the broader bandwidth.
However, while maximizing RM sky density is important for RM grid studies, we note that this strategy has disadvantages. Observing the same field in two bands increases both the necessary observing time and the compute time and resources required to image and process the observations. In addition, there is a gap between the low- and mid-band data due to RFI, which results in brighter and more structured sidelobes in the RMSF, making real structures in the FS more difficult to identify. When determining whether to use the combined-band strategy, the observer should balance the expected gain in polarized components with the additional costs to resources and data quality.

Approximately 85\% of the full POSSUM survey will be located at $\lvert b \rvert \geq 10^{\circ}$ and 15\% will be located at $\lvert b \rvert \leq 10^{\circ}$. If we estimate similar polarized component sky densities as are seen the EC and GL observations for these two regions respectively, we anticipate that the full 20 000 deg$^2$ POSSUM survey will yield an RM catalog of $\sim$877 000 polarized components. If we make the same calculation with the EL and GL observation polarized component densities instead, we anticipate an RM catalog of $\sim$775 000 polarized components. From our Faraday simple sky densities, we estimate that a 20 000 square degree POSSUM RM grid will contain $\sim$675 000 Faraday simple RMs in the case of combined-band observations and $\sim$637 000 Faraday simple RMs in the case of low-band only observations.

\citet{Basu+19} showed that MHD simulations of a turbulent synchrotron-emitting and Faraday-rotating plasma with some spatial correlation have complicated features in the Faraday spectra of sightlines passing through it. These features include both broadened peaks and localized, narrow peaks that are due to either regions of strong synchrotron emissivity, or a build-up of emissivity at similar $\phi$ along the line of sight, instead of Faraday thin structures like Faraday-rotating screens as is generally assumed. They also show that the standard simple models of turbulence and depolarization do not adequately describe the Stokes \textit{QU} spectra at frequencies $\lesssim 1$ GHz. This is potentially relevant to fields in the full POSSUM survey near the Galactic plane. The results of \citet{Basu+19} suggest that extracting an RM from components with lines of sight through turbulent foreground ISM is particularly difficult, through either RM synthesis or modeling the Stokes spectra. As discussed in Section \ref{subsec: complexity descrip}, components such as these will be flagged as Faraday complex and are not ideal for use in an RM grid. The interpretation of their corresponding RMs that will be included in the POSSUM catalogue will need to be done with caution.

\subsection{Recommended component selection thresholds}

Here we recommend some criteria for selecting subsets of data from our catalogs for interested users. To select the polarized components used in this work from the catalogs, users should select \texttt{snrPIfit} $\geq$ 8. This is a conservative threshold (see Section \ref{subsec: spec extract} and Section \ref{subsubsec: Faraday spectrum range test}), so users may wish to set a lower threshold to select more polarized components, but we recommend doing this with caution because this increases the chance of spurious RM detections. We also recommend removing components in the GL catalog with $\delta\phi$ $\geq$ 30\% the median value of polarized catalog (see Section \ref{subsec: spec extract}).

Users interested in components that are defined as Faraday simple in this work from the subset of polarized components, which is recommended for constructing RM grids, should select those components for which \texttt{complex\_flag} is `N'. This selects components that fall below one or both of our thresholds for Faraday complexity discussed in Section \ref{subsubsec: complex thresh}. Conversely, for users interested in studying Faraday complexity in more detail, components where \texttt{complex\_flag} is `Y' should be selected. To select Faraday complex components using a user defined threshold on the $m_2$ metric (not included in the catalogue, but can be calculated from \texttt{rm\_width} and \texttt{rmsf\_fwhm}), we recommend that is the greater of the component $\delta$RM/$\delta\phi$ and the determined tolerance to complexity of the science case. We suggest some thresholds for specific cases in Section \ref{subsubsec: complex thresh}. The observed correlation between complexity and S/N$_{\mathrm{pol}}$ discussed in Section \ref{subsubsec: M2 vs sigadd} should be considered when selecting components based on complexity level.

In Section \ref{subsec: resid leakage} we describe the residual off-axis leakage in the four observations. Users looking to select components that are above the residual leakage levels should select components where \texttt{fracpol} is greater than the residual leakage levels reported in Section Section \ref{subsec: resid leakage} for the observation of interest. Alternatively, users may select components where Stokes Q (\texttt{stokesQ}) and  Stokes U (\texttt{stokesU}) divided by Stokes I (\texttt{stokesI}) is greater than the local residual leakage estimate given by \texttt{leakage}. This second option cuts on local residual leakage levels while the first open cuts on global leakage levels in the observation of interest.

\subsection{Rotation measure grids}\label{subsec: RM grids results}

In Figure \ref{fig: RMgrids} we present four RM grids constructed from the polarized component catalogs in each of our median-filtered observations. 
We also construct RM grids for the four observations without the application of the median filter. These plots are provided in Appendix \ref{app: nofilt RMgrids}. Figure \ref{fig: SB10635 RMgrid}, \ref{fig: SB10043 RMgrid}, and \ref{fig: comb RMgrid} show the RM grids of the median-filtered EL, EM, and EC RM grids, respectively. In Table \ref{tab: RMgrid stats} we give the median and standard deviation of the RM distributions in the RM grids.

\begin{figure*}
\centering 
\subfloat[Median-filtered EL observation.]{%
  \includegraphics[width=0.75\textwidth]{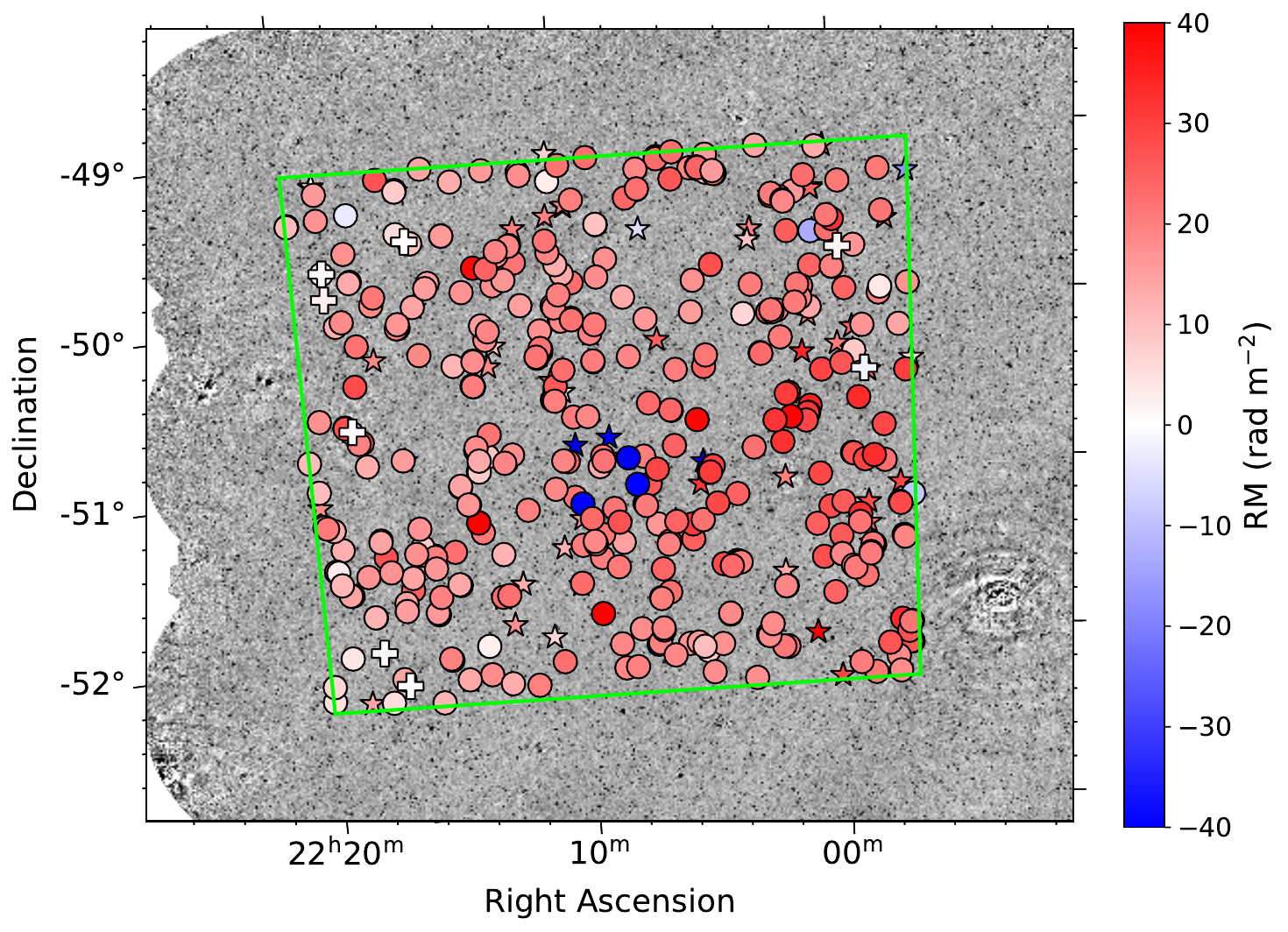}%
  \label{fig: SB10635 RMgrid}%
}\\
\subfloat[Median-filtered EM observation.]{%
  \includegraphics[width=0.75\textwidth]{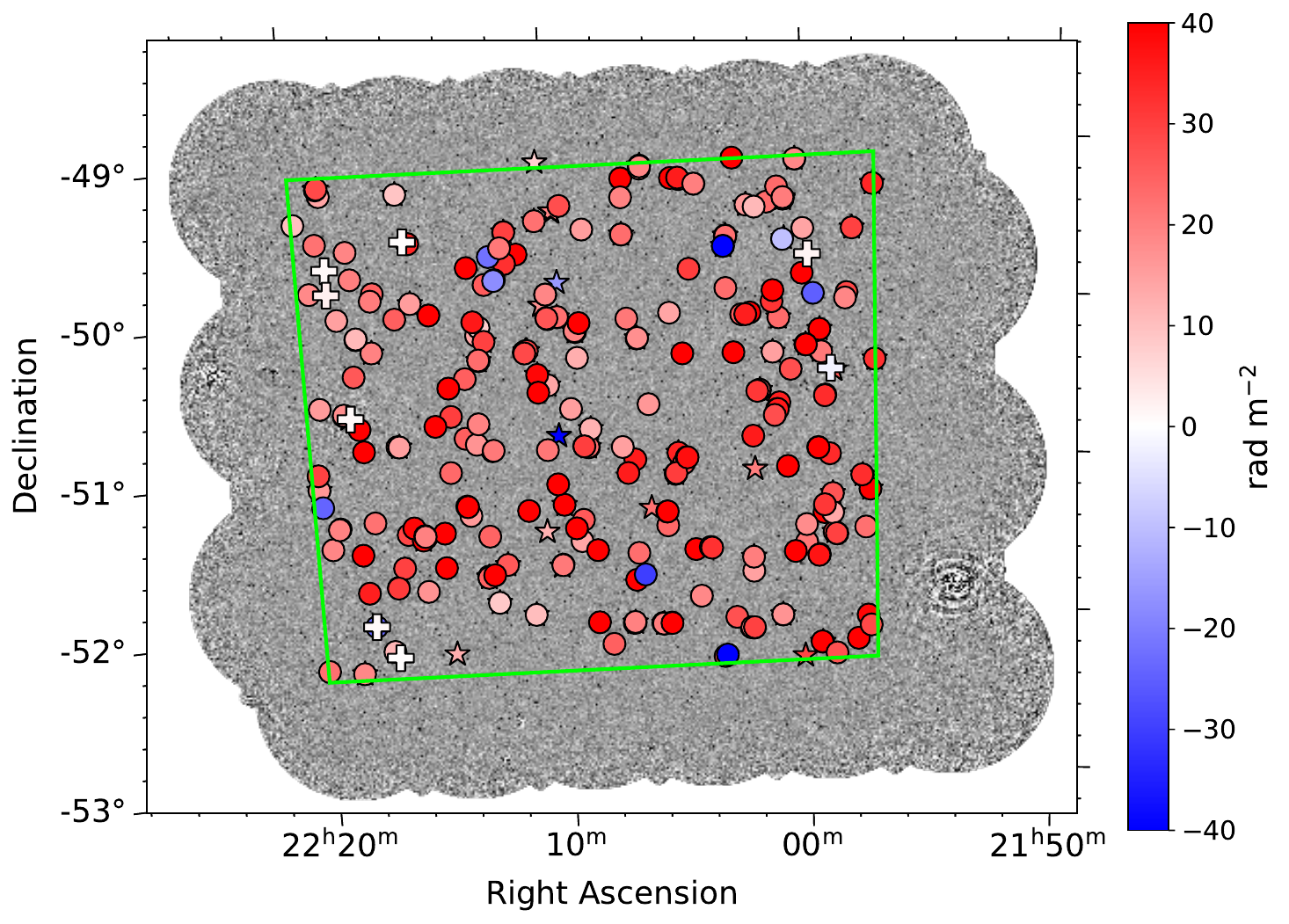}%
  \label{fig: SB10043 RMgrid}%
}
\caption{}
\end{figure*}
\begin{figure*}
\ContinuedFloat
\centering
\subfloat[Median-filtered EC observation.]{%
  \includegraphics[width=0.75\textwidth]{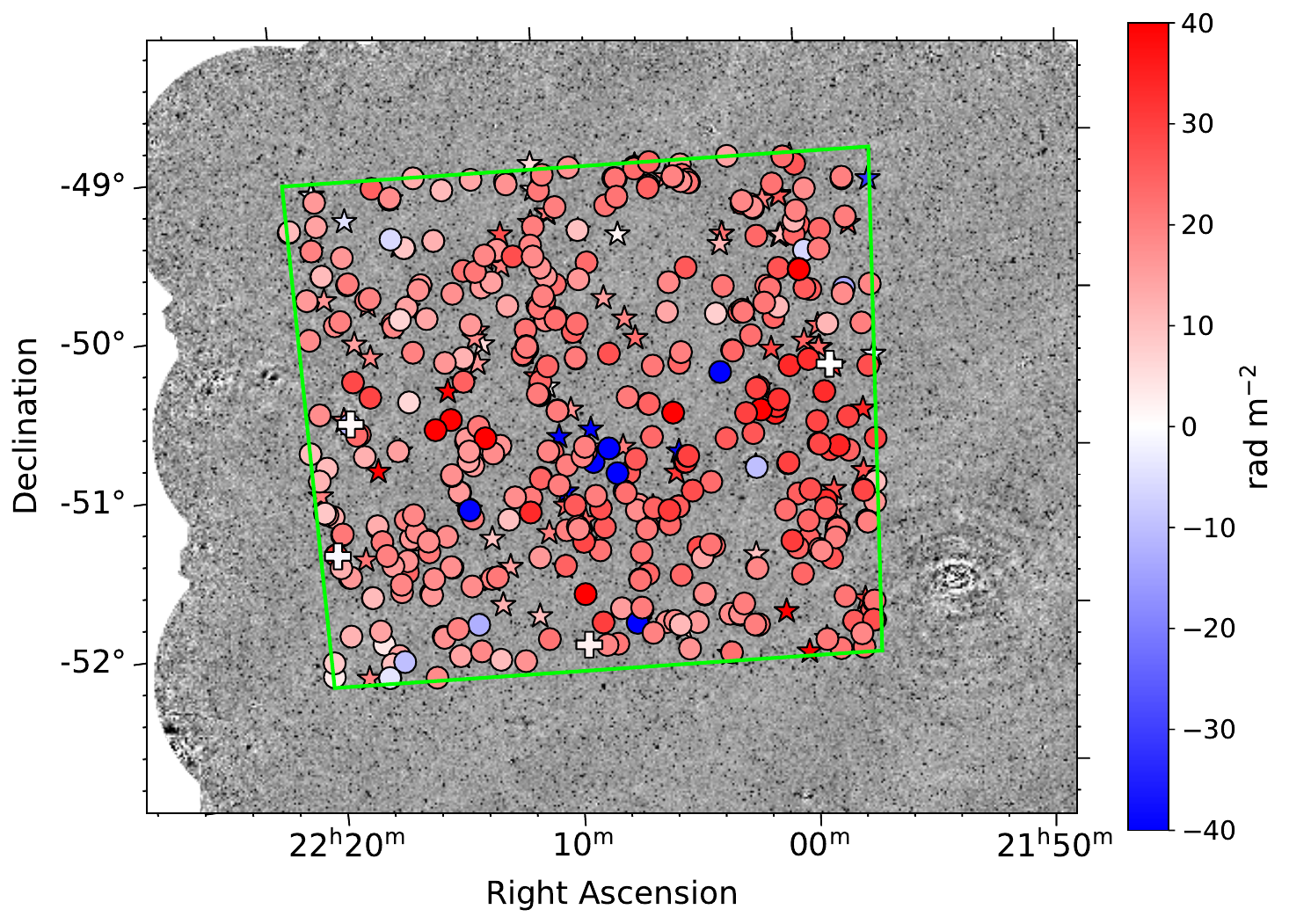}%
  \label{fig: comb RMgrid}%
}\\
\subfloat[Median-filtered GL observation.]{%
  \includegraphics[width=0.75\textwidth]{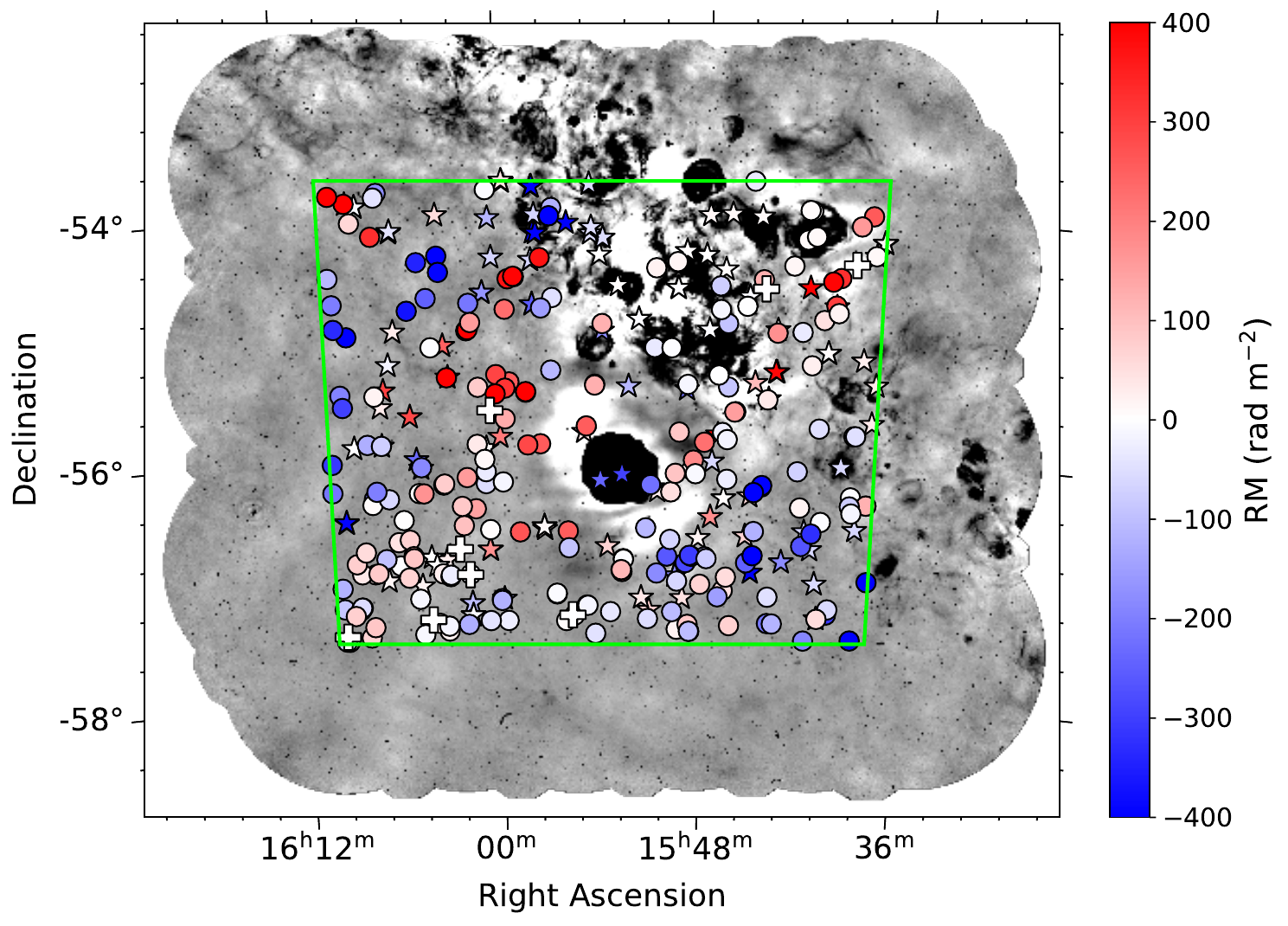}%
  \label{fig: gal RMgrid}%
}
\caption{RM grids polarized components in the median-filtered EL, EM, EC, and GL observations. The markers indicate the presence of a polarized component at that location in the observation. Positive RMs are red and negative RMs are blue, while the depth of the color indicates the magnitude of the RM. Circle and star markers denote Faraday simple and complex components, respectively. Crosses indicate Faraday simple components where the RM is consistent with zero within uncertainty. The colorscale of the background total intensity image is the same as Figure \ref{fig: footprints}.}
\label{fig: RMgrids}
\end{figure*}

The line of sight of the EL, EM, and EC observations is at a high Galactic latitude ($b \approx -51^{\circ}$) and looks through the Galactic halo, where previous RM studies have seen smaller magnitude RMs \citep{Taylor2009,Mao+10,Schnitzeler2010}. The EL and EC RM grids show a similar relatively smooth RM variation, with predominantly positive RMs and a median RM of $\sim$+20 rad m$^{-2}$. One feature of note is the several large negative RMs ($\leq -40$ rad m$^{-2}$) near the center of the observations, which are in contrast to the generally positive and moderate RM background. We discuss the reliability of these RM measurements and a potential origin in Section \ref{subsubsec: compare Fsky}.

\begin{table}
\centering
\caption{RM distribution in RM grids}
\label{tab: RMgrid stats}
\begin{tabular}{p{1.4cm} p{1.34cm} p{1.34cm} p{1.34cm} p{1.34cm}}
\hline
    & Median RM \newline (rad/m$^2$) & stdev \newline RM \newline (rad/m$^2$) & Median $\lvert$RM$\rvert$ \newline (rad/m$^2$) & stdev $\lvert$RM$\rvert$ \newline (rad/m$^2$) \\ \hline
EL      & $+20.0$      & 12.4   & $+20.1$        & 9.2     \\
EM      & $+21.3$      & 19.8   & $+22.4$        & 15.7    \\
EC      & $+20.2$      & 18.5   & $+20.5$        & 14.5    \\
GL      & $-4.6$       & 201.4  & $+80.1$        & 148.9   \\
\hline
\end{tabular}
\end{table}

The EM RM grid shows more variation in RM magnitude throughout the observation when compared with the EL and EC grids, although the observation also has an overall positive RM and median RM value of +21.2 rad m$^{-2}$. As discussed in Section \ref{subsec: data quality}, the relatively large uncertainties on RM due to the smaller bandwidth and lower Faraday resolution of the EM observation is the cause of this larger variation in RM values.

The line of sight of the GL observation, shown in Figure \ref{fig: gal RMgrid}, passes through the Galactic plane and is looking near to the Scutum-Centaurus spiral arm of the Galaxy. We see significantly more amplified RMs in this observation and an overall change in sign in RM from negative to positive and back to negative moving diagonally from the top left to the bottom right of the figure. The Galactic plane is a more complicated region of sky than the EL line of sight, and so we expect to see more structure in the RMs in the former.

\subsection{Types of components in data and limits on independent lines of sight}\label{subsec: src types}

\begin{figure}
\centering
    \includegraphics[width=0.47\textwidth]{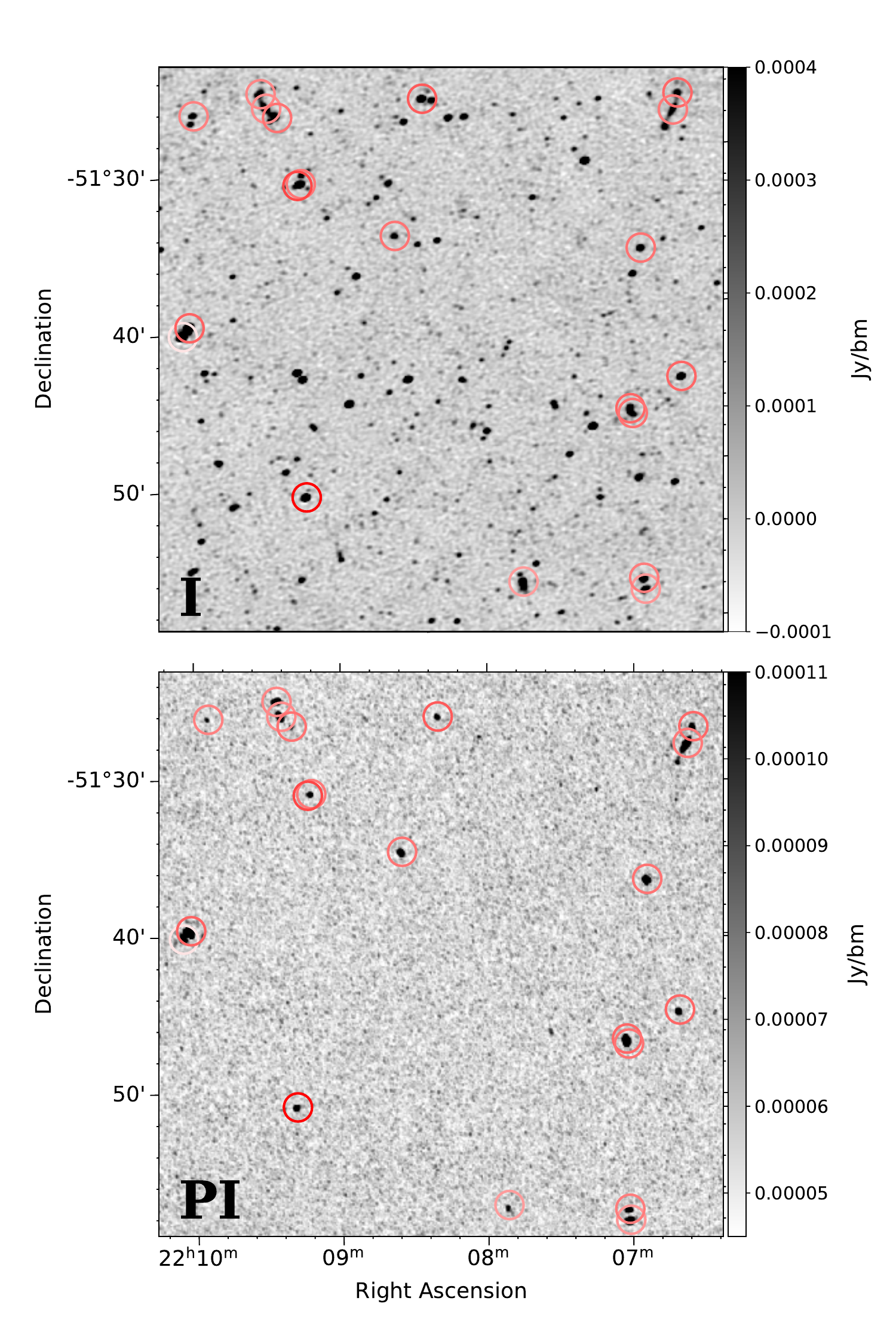}
    \caption{\textit{Top panel}: total intensity image of a region of the EL observation. Polarized component RMs are indicated with circles with color corresponding to the sign and magnitude of the RM as described in Figure \ref{fig: comb RMgrid}. \textit{Bottom panel:} same as for the top panel but in peak polarized intensity.}
\label{fig: comp types}
\end{figure}

In Figure \ref{fig: comp types} we show an example region of the EL observation in total intensity (top panel) and polarized intensity (bottom panel). We see examples of the typical types of sources that will be observed by POSSUM: compact, single component sources (e.g. far right center of the top panel), close double components (e.g. far left center of the top panel), and multi-component objects (e.g. the three-component object at the top left of the top panel). We note that the individual polarized components of double or multi-component objects are plotted separately in Figure \ref{fig: RMgrids}.

\begin{figure}
\centering
    \subfloat[Plot of distance to the next nearest neighbor (NNN) versus nearest neighbor (NN) for the polarized components in the EC RM catalog. The points are colored according to whether only the NN is polarized (green solid circles), only the NNN is polarized (green open circles), both the NN and NNN are polarized (black solid circles), or neither the NN or NNN are polarized (black open circles). Dashed lines are plotted at 70 and 100 arcsec to highlight different populations of sources, described in the text.]{
    \includegraphics[width=0.46\textwidth]{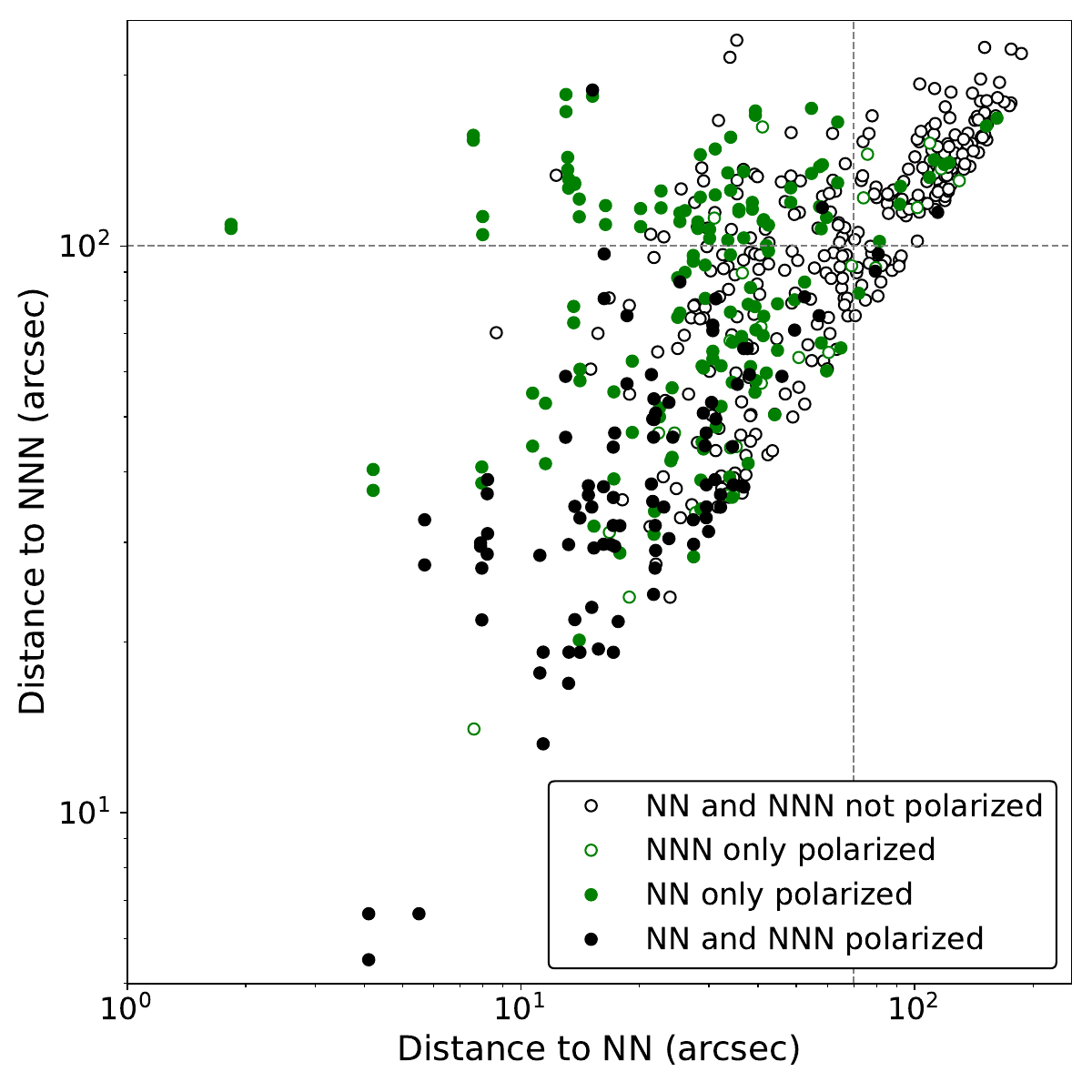}
    \label{fig: NNN vs NN}}\newline
    \subfloat[Absolute difference in RM of polarized components in the EC RM catalog and their nearest neighbor (NN) versus distance. The dashed line at 70 arcsec separates small and large separation pairs. The data are colored by polarized fraction, with low polarized fraction ($p < 0.03$) components in orange and high polarized fraction components ($p \geq 0.03$) in green, and the solid colored lines indicate the median value of the two populations of pairs.]{
    \includegraphics[width=0.46\textwidth]{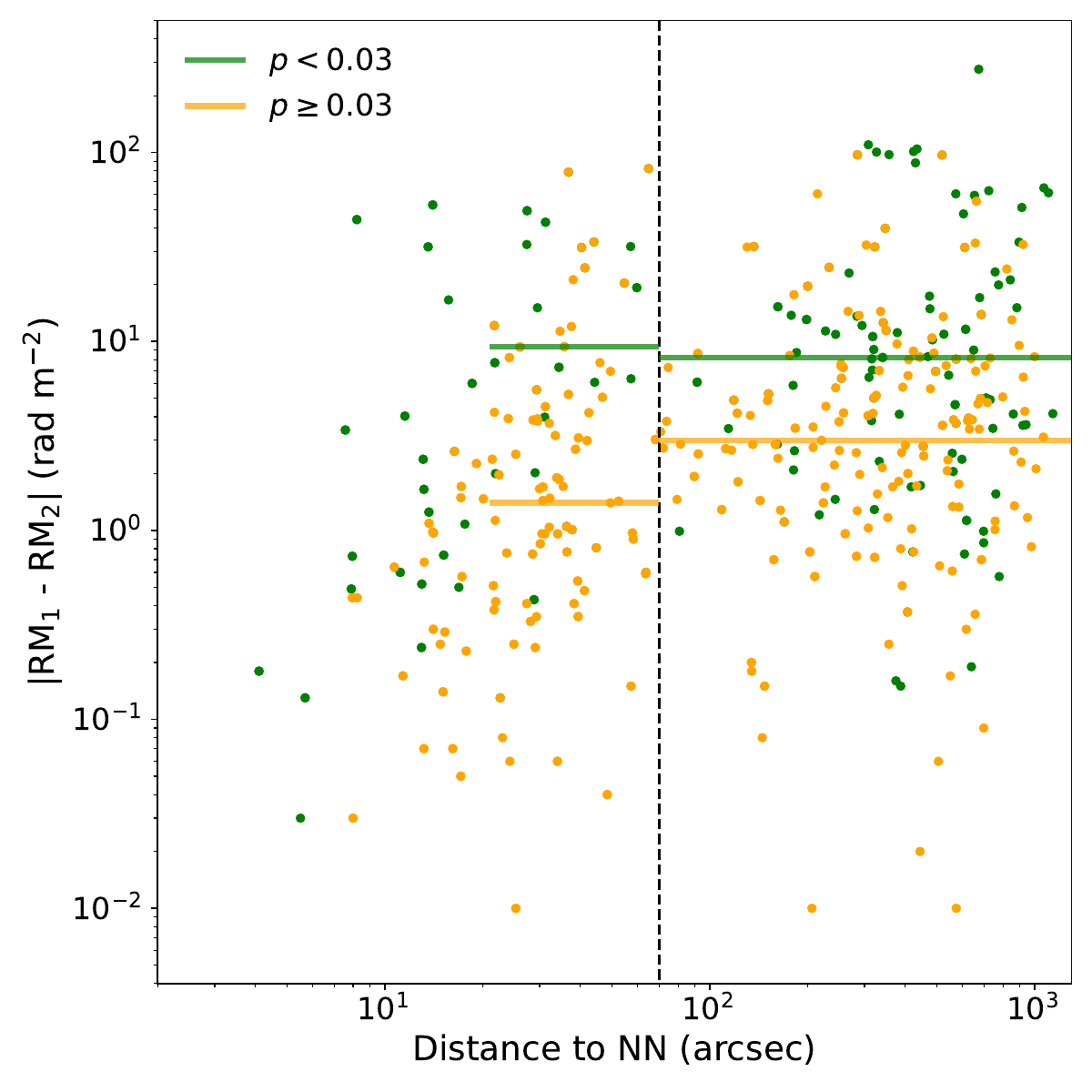}
    \label{fig: RMdiff vs sep}}
\caption{}
\label{fig: NN plots}
\end{figure}

In Figure \ref{fig: NNN vs NN} we plot the distance of each polarized component in the EC RM catalog to its next nearest neighbor (NNN) versus the distance to its nearest neighbor (NN) in the source finder catalog and determine the polarization level of thpse neighbors. We plot dashed lines at 70 arcsec and 100 arcsec to divide the plot into three approximate populations of sources. In the upper right quadrant of the plot we have 128 components where the NN and NNN are both fairly far from the polarized component. The majority of this population are isolated, compact sources, and only 16 (13\%) of the components have a NN or NNN that is polarized. The upper left quadrant has a population of 109 components where the NN is within $\sim$1 arcmin while the NNN is at a larger distance ($\sim$2 arcmin). These components are a population of isolated double sources, where the majority NNs are the other lobe of a double-lobed radio galaxy (we note that we cannot say conclusively that two close components are part of a physical source an not randomly coincident on the sky without crossmatching with optical or infrared data). Of this population, 67 (61\%) of these components have a NN and/or NNN that is polarized (43 components show only the NN is polarized). The lower left quadrant has the third population of 294 components where the NN and NNN are both quite close to the polarized component. This population is predominantly extended, multi-component sources and 186 (63\%) of these components have a NN or NNN that is polarized (in the case of 93 components, both the NN and the NNN is polarized).

Figure \ref{fig: NNN vs NN} indicates that the majority of close doubles and multi-component objects have two or more components that are polarized. 
When using RM grids to probe structures such as the Galactic magnetic field or the ICM, two or more components with very small angular separations and nearly identical RMs do not provide the user with additional, independent information. In this work, we have counted these components as individual RMs, which may be somewhat misleading when quoting RM sky densities.
If we count just one RM of the 61\% of close doubles, one RM of the 31\% of the multi-component objects with on polarized neighbor, and one RM of the 31\% of multi-component objects with two polarized neighbors, we get a total of 411 independent RMs in the EC observation, as opposed to the 553 RMs we currently report, which is a 25\% reduction in RM sky density.

In Figure \ref{fig: RMdiff vs sep} we plot the absolute difference in RM between each of the polarized components in the EC observation and their NN as a function of distance between the two components. The data are colored by low or high polarized fraction, and we separate the data into two regions: small separation and large separation, based on the close double population identified in Figure \ref{fig: NNN vs NN}. Component pairs with a separation less than the resolution of the observation (21 arcsec) are more likely to have the same RM, due to the signal from the two components blending. This will bias the median value of the RM difference for the small-separation pairs, so we exclude these pairs from our analysis here.

The median absolute difference in RM of the small separation population in the low (high) polarization regimes is 9.4 (1.4) rad m$^{-2}$, while the median absolute difference in RM of the large separation population in the low (high) polarization regimes is 8.2 (3.0) rad m$^{-2}$. The median value of the low polarized fraction population decreases for large-separation pairs, while the median of the high polarized fraction population increases. We expect the median difference in RM to be lower at small separations because at these distance the component pairs should predominantly be part of the same physical object (e.g. radio galaxies), which have been shown to have similar or smaller RM differences between components than physically unrelated component pairs \citep{Vernstrom+19,O'Sullivan+20}. The decrease in the low fractional polarization median at large separations may be due to the smaller number of pairs contributing to the low polarized fraction median at low separations. 
At both small and large separations, the median RM difference of the low polarized fraction population is substantially greater than the high polarized fraction population, indicating a dependence in RM difference on polarized fraction. We investigate this polarized fraction dependence in the data further in Section \ref{subsubsec: pf vs snr}.

\section{Discussion}
\label{sec: disc}

\subsection{Rotation measure grids}

\subsubsection{Comparison to previous observations}\label{subsubsec: compare Fsky}

\begin{figure*}
    \centering
    \subfloat[EC RM grid.]{
    \includegraphics[width=0.485\textwidth]{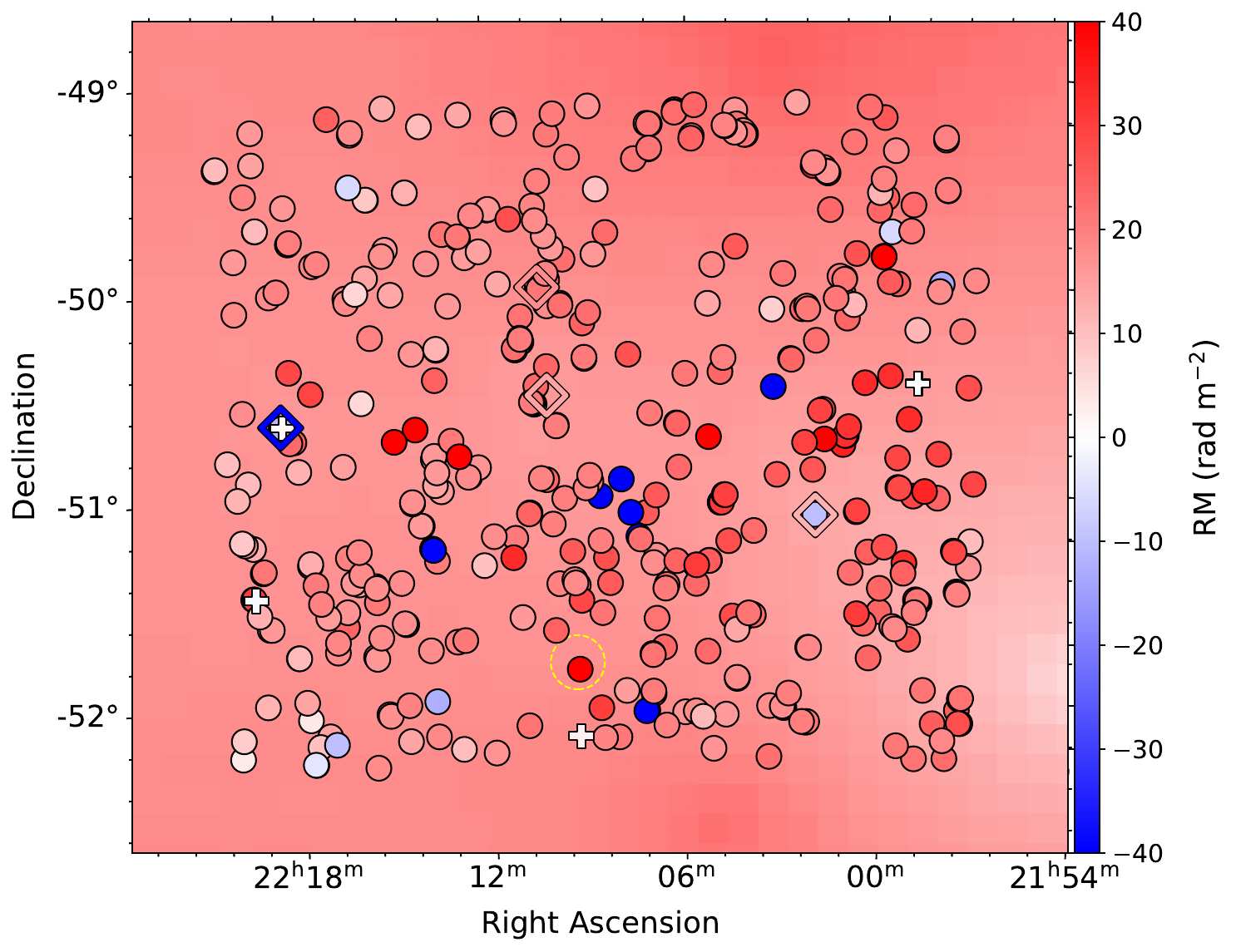}
        \label{fig: Huts comb RM grid}}
    \subfloat[GL observation RM grid.]{
    \includegraphics[width=0.482\textwidth]{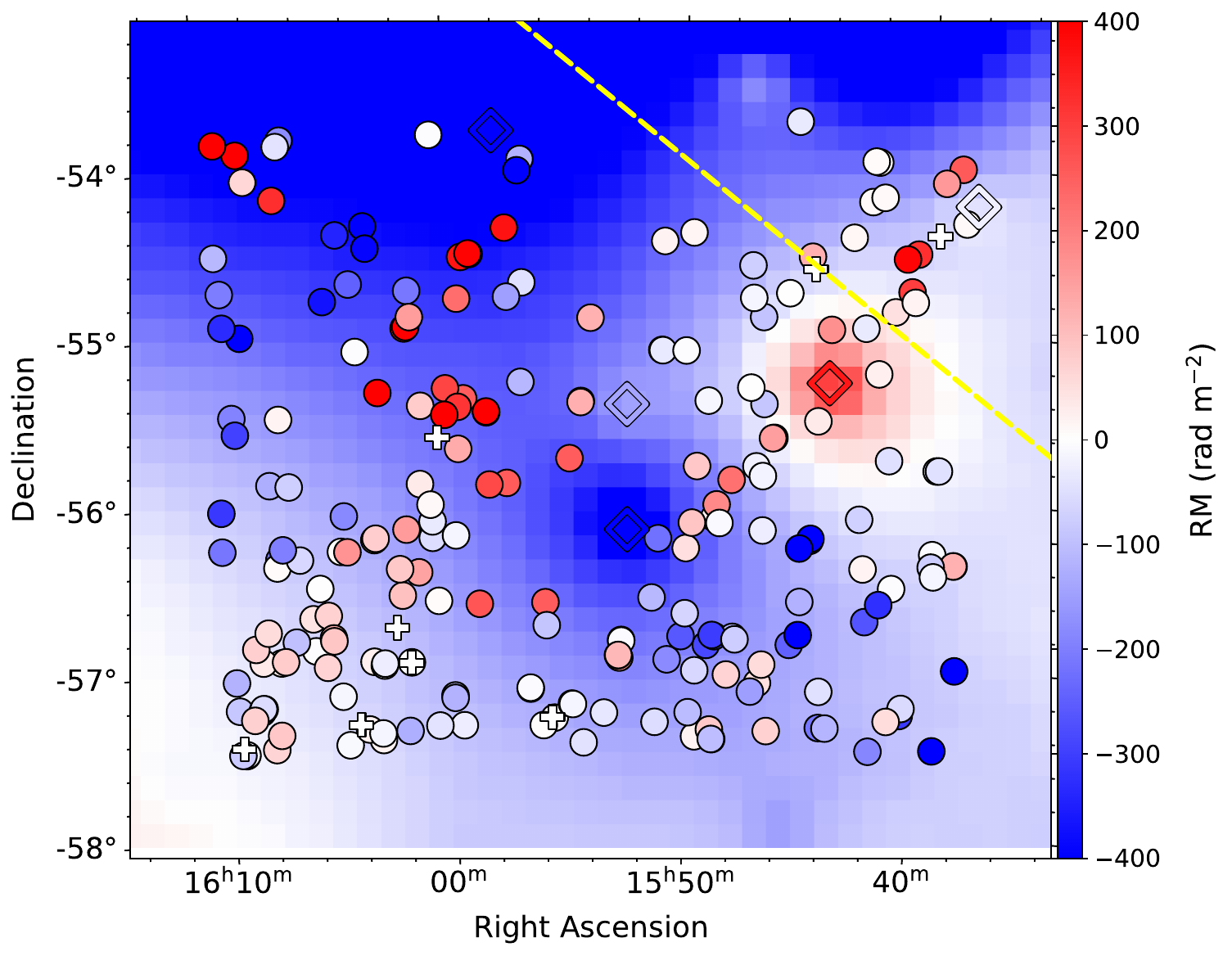}
        \label{fig: Huts gal RM grid}}
    \caption{RM grids of the median-filtered EC (\textit{left}) and GL (\textit{right}) observations plotted over the \citet{Hutschenreuter+22} all-sky Faraday depth map in each region. The diamond shaped markers are the components used to construct the \citet{Hutschenreuter+22} Faraday depth map in each region, and are included for reference only. The dashed yellow line in the GL observation lies along $b = 0^{\circ}$ in the Galactic plane.}
    \label{fig: Huts RM grids}
\end{figure*}

Here we compare our prototype RM grids to the current Faraday depth all-sky map from \citet{Hutschenreuter+22}. The Faraday depth map is constructed from a catalog of RMs compiled by \citet{VanEck+23} from various surveys, the largest of which is from the NVSS \citet{Taylor2009} catalog. The \citet{VanEck+23} catalog and has an average density of $\sim$1.35 RMs per square degree, however the southern sky is typically more poorly sampled (see Section \ref{sec:intro}), and the sky density of the catalog in the EC and GL fields are just 0.43 and 0.23 RMs per square degree, respectively.
In Figures \ref{fig: Huts comb RM grid} and \ref{fig: Huts gal RM grid} we plot the RM grids from Section \ref{subsec: RM grids results} and Figure \ref{fig: RMgrids} for the EC and GL observations, respectively, overlaid on the \citet{Hutschenreuter+22} Faraday depth sky\footnote{Cutouts of the \citet{Hutschenreuter+22} all-sky Faraday depth map can be accessed at \url{http://cutouts.cirada.ca/rmcutout/}.}. We plot only the Faraday simple components in our RM grids for comparison. The diamond shaped markers in both plots are the polarized background sources used by \citet{Hutschenreuter+22} to construct the Faraday depth map along the lines of sight of our two observations.
There are four components from the \citet{Hutschenreuter+22} data set in the EC observation and five components in the GL observation. The dashed yellow line in Figure \ref{fig: Huts gal RM grid} traces $b = 0^{\circ}$ through the Galactic plane. We only discuss the EC RM grid here and not the EL or EM RM grids because, as discussed in Section \ref{subsec: data quality}, the EL and EC data are very similar and the large EM RM uncertainties make that grid less informative on its own.

Our RM grid components are subject to variations in RM intrinsic to the source, which have been shown to be $\sim$8--11 rad m$^{-2}$ for Galactic latitudes $\gtrsim \lvert 30^{\circ} \rvert$ \citep{Mao+10,Schnitzeler2010,Stil+11,Rudnick&Owen2014}, while the Faraday depth sky map uses Bayesian inference to smooth out these variations. Thus, a direct comparison between our discrete RM grid and the continuous Faraday depth background is not straightforward. We can comment on some of the most obvious differences between the two maps, however, beginning with the difference in the polarized component sky density. Our RM grids are generally consistent with the Faraday depth background, however the Faraday depth background has limited angular resolution and RM grids are able to show smaller-scale fluctuations. The median-filtered EC observation and the GL observation have 428 and 231 RMs which to map the Faraday depth, respectively, compared to 5 and 4 RMs in the \citet{Hutschenreuter+22} data sets for these observations.
This significant increase in polarized component sky density that POSSUM will achieve will increase the angular resolution of RM structure in future Faraday depth sky maps from degrees to tens of arcminutes.

In the case of the EC observation, our RM grid shows more variation in RM than the Faraday depth sky background. This may be partly due to the intrinsic variation in background components mentioned above, which we have not smoothed out and which the \citet{Hutschenreuter+22} map has, however we also see overall larger positive RMs in the lower right region of Figure \ref{fig: Huts comb RM grid} and overall smaller positive RMs in the upper left region. This might suggest there is a fluctuation in the coherent magnetic field strength or structure on a scale larger than the field of view of the EC observation. The median RM value of the smoothed Faraday depth sky map in this field is +17.4 rad m$^{-2}$ with a median uncertainty of 5.7 rad m$^{-2}$, while our median RM and median uncertainty is +20.4 and 1.1 rad m$^{-2}$, respectively (see Tables \ref{tab: RMgrid stats} and \ref{tab: numbers}), showing that the overall RM of the two maps agree. We also see more components with large negative RMs present in our data than in the \citet{Hutschenreuter+22} data set. These components illustrate the improvement in our ability to probe Faraday structure on the sky with denser sampling. Structure in the Faraday depth sky on angular scales smaller than seen in \citet{Hutschenreuter+22} have been previously predicted in simulations by \citet{Sun&Reich09}. We discuss these large negative RM components in more detail in Section \ref{subsubsec: large neg RMs}.

The differences between our RM grid and the background Faraday depth sky in the GL observation in Figure \ref{fig: Huts gal RM grid} are more pronounced. In the bottom right region of the figure, the Faraday depth sky map of \citet{Hutschenreuter+22} has a typical value of $\sim -50$ rad m$^{-2}$, while our data include several components with RM $\lesssim - 300$ rad m$^{-2}$. This difference is most likely due to the difference in angular resolution of our two data set. The vastly improved RM sky density in our data allows us to probe variations in the Faraday depth sky on smaller angular scales than \citet{Hutschenreuter+22}. 

The most striking difference is the presence of many positive RMs in our RM grid that are not present in the Faraday depth map (with the exception of the single RM close to the Galactic plane line). There also appears to be a ripple-like pattern to the RM values moving diagonally from the top left of the RM grid to the bottom right. This could possibly be due to a large-scale solar ripple artefact like we see in the EL observation, however a ripple pattern is not visible in the Stokes I MFS image or in the peak polarized intensity image of the GL observation (as it is in the unfiltered EL observation). We also show in Section \ref{subsec: medfilt application 10635} that the median filter does a good job of removing the ripple artefact from the data, so we believe this ripple pattern in the RMs is likely real. An oscillatory pattern in RM as a function of Galactic longitude and position with respect to Galactic arms was observed by \citet{Brown+07} (see Figure 3 therein). We estimate an oscillation period of 3$^{\circ}$ in our RM grid, which is smaller than that seen in the binned data of \citet{Brown+07}, although they do see fluctuations in individual RMs on these scales. We also note that their data set had just 148 components spread over 100$^{\circ}$ in Galactic latitude, allowing for limited angular resolution.

\subsubsection{Structure functions}\label{subsubsec: SFs}

The Galactic foreground is believed to be the dominant contribution to the Faraday rotation of polarized extragalactic radio components \citep{Schnitzeler2010,Hutschenreuter+22}. Tangled magnetic fields due to turbulence are present throughout the Galactic ISM \citep{Armstrong+95,Gaensler+11,Xu&Zhang17} due to energy injection from processes such as supernova explosions. Tangled magnetic fields can cause depolarization of synchrotron emission from background components (see Section \ref{subsec: complexity descrip}).

We can probe the turbulent power of the magnetized ISM on different scales by plotting the variance in RM at different angular separations on the sky using the RM structure function (SF; e.g. \citealt{Haverkorn+06,Stil+11}):

\begin{equation}
	\mathrm{SF}(\delta\theta) = \frac{1}{N} \sum_i [\mathrm{RM}(\theta) - \mathrm{RM}(\theta + \delta\theta)]^2_i  \, ,
\end{equation}

\noindent where $N$ is the number of component pairs at separation $\delta\theta$ on the sky.

The SF is a sum of all contributions to the variance along the line of sight:

\begin{equation}
	\mathrm{SF}(\delta\theta) = 2 \sigma^2_{int} + 2 \sigma^2_{IGM}(\delta\theta) +  2 \sigma^2_{Gal}(\delta\theta) +  2 \sigma^2_{noise} \, ,
\end{equation}

\noindent where $\sigma^2_{int}$ is the variance that is intrinsic to the source, $\sigma^2_{IGM}$ is the variance due to the IGM, $\sigma^2_{Gal}$ is the variance due to the Galactic ISM, and $\sigma^2_{noise}$ is the variance due to measurement uncertainty. $\sigma^2_{noise}$ should be subtracted from the SF, leaving the sum of the astrophysical variances.

When dominated by $\sigma^2_{Gal}$, the SF is expected to follow a power-law distribution \citep{Kolmogorov1991} due to a cascade of power from the largest scales ($\sim$100 pc; \citealt{Armstrong+95,deAvillez&Breitschwerdt07}) to smaller scales. The SF can probe turbulence down to the smallest scales where there is a sufficient number of component pairs for a reliable measure of variance.  Previous work has shown that the slope of the SF often has a break at an angular scale near $\theta \approx 1^{\circ}$ \citep{Haverkorn+08, Roy+08,Stil+11,Anderson+15}, requiring separate power laws to describe the slope above and below this break. The exceptional RM sky density that POSSUM will achieve will allow for a more complete and accurate probe of turbulent power at scales below this break, for which previous studies have typically had few measurements \citep{Haverkorn+06,Mao+10,Stil+11,O'Sullivan+20,Livingston+21}.

We calculate the SF for our EC and GL Pilot observations. \citet{Sun&Han04} showed that individual RMs that have magnitudes very different from the local average will significantly affect the shape and interpretation of a SF. 
To avoid this issue in the EC observation, we follow the suggestion of \citet{Sun&Han04} and remove any component RM that is more than 3$\sigma$ from the local average, which we set as the the average of all RMs within a 10 arcmin radius of the component. This removes five of the seven large negative RM components in the observation. We do not perform this 3$\sigma$ cut on the RMs in the GL observation because when we do so, the large magnitudes and sign changes on smaller scales leaves very few RMs left to calculate the structure function with. We interpret the GL structure function with caution because of this.

\begin{figure}
\centering
    \includegraphics[width=0.47\textwidth]{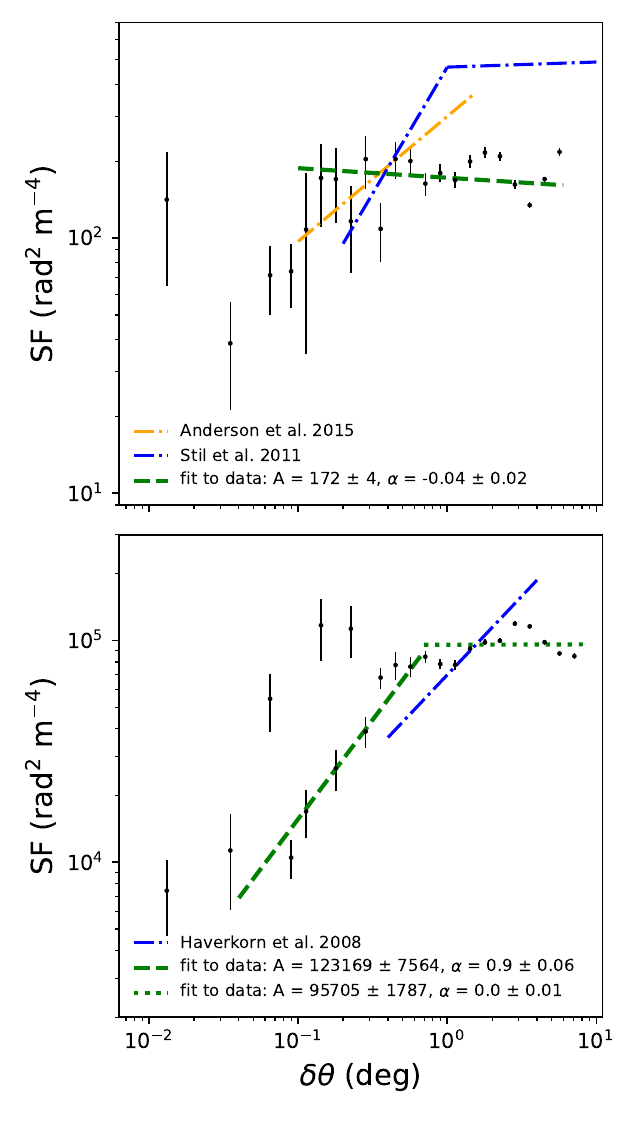}
    \caption{\textit{Top}: Structure function for the EC observation. \textit{Bottom}: Structure function for the GL observation. In both panels, the best fit power law model is shown by the green dashed line and previous SF measurements are included in dot-dashed lines for comparison. The \citet{Stil+11} SF in the top panel is for the South Galactic Pole.}
\label{fig: SFs}
\end{figure}

We subtract $\sigma_{noise}^2$ from SF($\theta$) using the method described in Appendix A of \citet{Haverkorn+04}, and we present the results in Figure \ref{fig: SFs}. The data are binned such that the first data point contains all components with separations 0.006--0.019$^{\circ}$, which will contain components in the isolated doubles population identified in Figure \ref{fig: NNN vs NN}.
The smallest number of component pairs in a given bin is 78 in the EC SF and 25 in the GL observation. The uncertainty in each bin is calculated as the standard error in the mean. POSSUM will provide the best sampling of structure in the RM sky at separations $\delta\theta < 1^{\circ}$ to date.

The SF in the EC field is shown in the top panel of Figure \ref{fig: SFs}. The data appears approximately constant above $\sim$0.1$^{\circ}$, and begins to drop off below this threshold. We fit the binned data above $\delta\theta = 0.1^{\circ}$ with a power law SF($\delta\theta$) = $A \delta\theta^{\alpha}$ using a least squares fit, and the uncertainties on the best fit parameters are calculated using a Monte Carlo simulation with 5000 iterations. The best fit model is the dashed green lines in Figure \ref{fig: SFs}. The SFs of selected previous comparable works are also included in the Figure as dot-dashed lines.

The best fit parameters to the power law model of the outer scale ($\delta\theta \geq 0.1^{\circ}$) of the EC SF are $A = 172 \pm 4$ rad$^2$ m$^{-4}$ and $\alpha = -0.04 \pm 0.02$. The nearly flat slope of the power law fit is expected for this observation, since we see little variation in the RMs in our RM grid, with the exception of the large negative RMs discussed in Section \ref{subsubsec: large neg RMs}, which have been mostly been excluded from the calculation of our SF by our $\lvert \mathrm{RM} \rvert \leq \mu \pm 3\sigma$ requirement. This tells us that on angular scales $\theta \sim 0.1^{\circ}$ -- 10$^{\circ}$, the RM fluctuations are approximately scale free. It also suggests that variation in RM on these scales is dominated by $2\sigma^2_{int}$, given that the typical estimates of $\sigma_{int}$ of 8--11 rad m$^{-2}$ match our amplitude $A = 172$ rad$^2$ m$^{-4}$ well.

The slope of our fit is similar to the outer scale slope of $\alpha = 0.02 \pm 0.04$ found by \citet{Stil+11} in the South Galactic Pole (SGP; see the dashed blue line in top panel of Figure \ref{fig: SFs}) using the \citet{Taylor2009} catalogue.  Their fit to $\delta\theta < 1^{\circ}$ is significantly steeper than ours, with greater uncertainty, at $\alpha = 0.99 \pm 0.58$. \citet{Anderson+15} (the dashed orange line in top panel of Figure \ref{fig: SFs}) found a slope of $\alpha = 0.49 \pm 0.1$ at $\delta\theta < 1.5^{\circ}$ in a 30 deg$^2$ region of sky at a similar Galactic latitude to the EC observation ($b \sim -55^{\circ}$). This is also steeper than the slope we find in our EC observation, and is in agreement with \citet{Stil+11}. The authors suggest that a coincidence on the sky between HI structures and enhanced complexity in the sources may indicate regions with turbulent structure or RM gradients in their field, which could account for the steeper slope and increased turbulent power on scales $\theta > 1^{\circ}$ compared to our findings.

The nearly-flat and low amplitude of the SF in the EC observation suggest that any Faraday complexity that we are seeing in the components in this observation is more likely due to extragalactic Faraday structures than Galactic ones.
The difference in both amplitude and slope between our SF and that of \citet{Anderson+15} illustrates the difference in SF for two seemingly similar fields.

At $\delta\theta \lesssim 0.1^{\circ}$ the SF in the EC observation decreases in amplitude rather sharply. At these small angular separations we begin to be dominated by close pairs, and we expect to see a higher correlation between RMs of close neighbors (see discussion of independent sightlines in Section \ref{subsec: src types}). We see this in bins 2--4, however the bin with pairs at the smallest angular separations has a much higher amplitude. From manual inspection of the pairs in this bin, we find that 9 of the 93 pairs have $\Delta$RM$^2$ $> 100$ rad$^2$ m$^{-4}$. One pair was coincident with an imaging artefact, three pairs appeared to be randomly coincident on the sky, two pairs were part of more complicated multi-component objects, and three appeared to be physically related components from the same source. We suggest that the higher amplitude of this bin compared to the subsequent three is due to the inclusion of random, physically unrelated pairs and complex multi-components objects that may or may not be physically related, both of which would be dominated by $\sigma_{int}$. If we remove the 9 pairs with $\Delta$RM$^2$ $> 100$ rad$^2$ m$^{-4}$ from the calculation, the amplitude of the SF in this bin is 33 rad$^2$ m$^{-4}$, suggesting that the majority of the pairs in this bin are indeed physical pairs.

The SF in the GL observation is shown in the bottom panel of Figure \ref{fig: SFs}. The SF shows a break at $\delta\theta \sim 0.7^{\circ}$. The line of sight of the GL observation passes along the outer edge of the Norma spiral arm (although it does not appear to pass through it) at a distance of $\sim$6–-11 kpc from the Sun, through the Scutum-Centaurus spiral arm at a distance of $\sim$1–2 kpc and again at $\sim$14–-15 kpc, and through a secondary spiral arm at a distance of $\sim$17.5--19 kpc \citep{Churchwell+09}\footnote{Distances were estimated from the illustration of the Milky Way by Robert Hurt of the Spitzer Science Center, based on radio, infrared, and visible wavelength data and in consultation with Robert Benjamin at the University of Wisconsin-Whitewater.}. Any one, or all, of these spiral arms may be contributing to the foreground diffuse emission seen in the GL observation. This suggests that the 0.7$^{\circ}$ break in the GL structure function corresponds to an angular scale somewhere in the range of 12--232 pc. The diffuse polarized emission of the ISM is not smooth, and therefore we expect the polarized intensity, and its contribution to a background component’s RM, to decrease with increasing distance. This argument favors the nearer Scutum-Centaurus spiral arm as the dominant contributor to the break in the SF at 0.7$^{\circ}$. A distance of 1--2 kpc corresponds to a physical scale of 12-24 pc, which is consistent with the typical turbulent scale in spiral arms of $\sim$17 pc found by \citet{Haverkorn+06}.

We fit the data above and below the break at $\delta\theta \sim 0.7^{\circ}$ with separate power laws. We exclude the first bin from the fit of the smaller angular scales because the SF appears to begin to flatten out below 0.04$^{\circ}$. 
The best fit parameters to the power law model for the inner inertial range ($0.04^{\circ} \leq \delta\theta \leq 0.7^{\circ}$) of the SF are $A =$ 123 169 $\pm$ 7564 rad$^2$ m$^{-4}$ and $\alpha = 0.90 \pm 0.06$. The fit to the outer scale ($\delta\theta > 0.7^{\circ}$) of the SF, where the data becomes nearly constant, gives $A =$ 95 705 $\pm$ 1787 rad$^2$ m$^{-4}$ and $\alpha = 0.00 \pm 0.01$. 
\citet{Haverkorn+08} plotted SFs using RMs from the Southern Galactic Plane Survey \citep{McClure-Griffiths+05} along two spiral arms and through three inter-arm regions. The slope of their fit to the inter-arm region nearest on the sky to our GL observation had $\alpha = 0.71 \pm 0.17$ (we estimate $A \approx$ 70 000) for $\delta\theta \approx 0.4$--$4^{\circ}$. Our slopes are nearly in agreement within uncertainties.

In contrast to the EC observation, we see significant turbulent power in the GL observation, and it is likely that much of the Faraday complexity that we see in the components in this observation is also due to Galactic Faraday structures and turbulence. The inner scale of the two fields is also in contrast. While the RM difference of close pairs in the EC observation are typically on the order of $\sigma_{int}$ or lower, close pairs in the GL observation show significantly more power (SF($\delta\theta = 0.01$) $\sim$ 7000 rad$^2$ m$^{-4}$). The GL SF is still dominated by the $\sigma^2_{Gal}$ contribution at small scales.

We show here the ability of POSSUM to probe turbulent structure on smaller angular scales than previously possible with sparse RM sky density. We are limited by small regions of sky in our Pilot observations, however with the full POSSUM survey it will be possible to calculate structure functions for many regions of sky and with significantly more component pairs at small separations. This will better inform our understanding of intrinsic variation in RM between radio galaxies, turbulence in the ISM, and how turbulent Faraday structure changes with position on the sky.

\subsection{Collection of amplified negative RMs in the EC observation}\label{subsubsec: large neg RMs}

We see one example of what denser sampling of the polarized sky can show us in the EC observation in Figure \ref{fig: Huts comb RM grid}, where there are seven large negative RMs ($\lesssim -40$ rad m$^{-2}$; one is mostly obscured by a nearby positive component) that are in contrast with the generally positive and moderate RM of the rest of the observation. In the full catalog for the EC observation, there are 11 components with RM $\lesssim -40$ rad m$^{-2}$ (see Figure \ref{fig: RMgrids}), with a mean uncertainty of 1.6 rad m$^{-2}$. A sign change indicates a reversal in the average line of sight coherent magnetic field, which suggests there is some Faraday structure affecting the geometry of the magnetic field in this direction of sky. A single negative RM ($-132$ rad m$^{-2}$) is detected in the \citet{Hutschenreuter+22} data for this field, however the sparse sampling would make understanding the size and effect of the Faraday structure producing these large negative RMs difficult. The large negative RM components in our data appear to cluster together on the sky, with a projected distance of $\sim$2.7$^{\circ}$ between the two most distant components. 
Looking at the Faraday spectra of higher-S/N$_{\mathrm{pol}}$ Faraday complex positive RM components adjacent to the central large negative RM components, we see clean peaks present between $\sim -60$ and $-80$ rad m$^{-2}$ in all but two of them. This points to a common Faraday structure along this line of sight shared by these components and not individual Faraday structures with different magnetic properties and structure.

There are three broad possible explanations for the difference in magnitude and sign of these RMs relative to the components around them: 1. they are not real and are simply noise or some unknown instrumental effects, 2. there is a common extragalactic Faraday structure local to these components, or 3. there is a common foreground, Galactic Faraday structure.

We rule out the first scenario by visual inspection of the components’ Stokes $QU$ and polarization angle spectra, where we see the expected behaviour of polarized components, in particular an observable linear relationship between polarization angle and wavelength squared that suggests that these are real detections. Moreover, the negative RM present in the \citet{Hutschenreuter+22} data set also supports the argument that there are real negative RM components in this observation.

The second scenario would require a Faraday structure local to the large negative RM components that is at a greater distance to us than the foreground positive RMs that surround them. A search of known objects within a radius of 1$^{\circ}$ from the central large negative RM components in our RM grid returns the galaxy cluster RXC J$2209.2-5149$ \citep{Piffaretti+11}, identified as part of supercluster 167 (as A3836B) by \citet{Chow-Martinez+14}.
The ICM magnetic field is known to be turbulent and amplified with respect to the surrounding IGM \citep{Donnert+18}. Previous studies have observed increased RM magnitude and sign changes due to the intervening ICM \citep{Bonafede+10,Bohringer+16,Anderson+21}. In particular, \citet{Bonafede+10} see similar behavior to our work when studying the Coma cluster, with predominantly small-positive RMs with a median of 32 rad m$^{-2}$ and two large negative RMs, the largest of which is $-256$ rad m$^{-2}$. However, this work measures just 7 sources passing through the cluster and has significant error bars on the RM measurements (4--50 rad m$^{-2}$).

\citet{Piffaretti+11} estimate RXC J$2209.2-5149$ is at $z = 0.11$ with a radius $R_{500} = 0.9239$ Mpc (the radius at which 500 times the critical density of the Universe at $z = 0.11$ is encompassed), which would give it an angular radius of $\sim 0.13^{\circ}$. We indicate the position and $R_{500} = 0.13^{\circ}$ extent of the cluster by a dashed yellow circle in Figure \ref{fig: Huts comb RM grid}, and we can see it does not lie near the central of the collection of large negative RMs and none of the negative components lie within the indicated region. This suggests that any structure lying between foreground positive RM components and background negative RM components would need to have a very large physical extent in order to span a projected diameter of $\sim2.7^{\circ}$, which makes this scenario less likely.

The third scenario requires a Faraday structure that produces an increase in magnitude and sign change in only select component RMs in this region of sky. 
Previous studies of H II regions and stellar bubbles have shown RM amplifications of up to $\lvert$RM$\rvert \sim 1000$ rad m$^{-2}$ and sign changes of background components with lines of sight through these objects \citep{Harvey-Smith+11,Savage+13,Purcell+15,Costa+16,Costa&Spangler2018}. The angular extent of our region of large negative RMs is within the range of 0.002--1.6 degrees of the sample of H II regions observed by the Wide-Field Infrared Survey Explorer (WISE; \citealt{Anderson+14}), however the mean value of their sample was just 0.028 degrees, much smaller than our defined radius. The largest magnitude negative RM in our sample is $-251$ rad m$^{-2}$, which is less than the typical RM amplification see in the other studies. It is unlikely that this Faraday structure is an H II region since most of the RMs near the large negative RMs show no amplification or sign changes, and the line of sight of the observation is looking south of the Galactic plane ($b \sim -51^{\circ}$), where we do no expect to see H II regions. Additionally, inspection of this region of sky in the Southern H$\alpha$ Sky Survey Atlas (SHASSA; \citealt{Gaustad+01}) and the Wisconsin H$\alpha$ Mapper (WHAM) Sky Survey \citep{Haffner+03} shows no discrete H$\alpha$ structures.

Another possibility for scenario 3 is that the sign reversal is due to a distant spiral arm. \citet{Shanahan+19} observed strong excess Faraday rotation in background sources passing through the Sagittarius spiral arm, with $-310$ rad m$^{-2}$ $<$ RM $<$ $+4219$ rad m$^{-2}$ at Galactic longitude 39$^{\circ}$--52$^{\circ}$ and $b \sim 0$. These RMs are much larger than the typical values we see in our data, however their observations are much closer to the Galactic center. At an estimated distance of 1.2 kpc to the Scutum-Centaurus arm, the line of sight of the EC observation passes $\sim$1.5 kpc below the Galactic plane, which is perhaps too low to cause the sign reversal we see in our RMs. However, it would require further analysis to rule this scenario out.

Scenario 3 appears to be the most likely, however we have not found a suitable physical setup to properly describe our observations. Decisively determining what the Faraday structure causing the amplification and sign change in our RMs is and whether it is Galactic or extragalactic will require a more detailed study of the complexity of the large negative RM components and the positive RM components near them. Broader-bandwidth observations and $QU$ fitting would shed light on the structure of the Faraday complexity of these components to determine whether they lie behind a common Faraday structure, and obtaining redshift for these components would help determine their relative distances.

\section{Summary and conclusion}
\label{sec: conclusion}

We have presented linearly polarized component catalogs and prototype rotation measure grids using Pilot observations for the POSSUM survey with ASKAP. We have used these data products to investigate polarized component and Faraday simple component sky densities and RM precision and their dependence on frequency, bandwidth, Galactic latitude, and the presence of foreground polarized diffuse emission in anticipation of the full POSSUM survey. We highlighted the need for separating foreground diffuse emission from background components before extracting polarization spectra and calculating polarization parameters such as RM, and we assess the effect of applying the median filter to each of our four observations. We summarize our key results below.

\begin{itemize}
    \setlength{\parskip}{0pt}
    \setlength{\itemsep}{0pt plus 1pt}
    
    \item We describe a median filter method for separating foreground polarized diffuse emission. We compare the extracted polarization properties of 5000 simulated compact components both with and without the application of the median filter in an observation with extensive diffuse emission. We find that with median-filtering, 99.5\% of measured RMs are within 3$\sigma$ of the true value, compared with 97.9\% without filtering, and  with a typical reduction in the measured polarized intensity of 5\% $\pm$ 5\%. In addition, we find an 80\% decrease in cases where the foreground diffuse emission RM is measured instead of the background component RM. Further testing is required for the impact of the filter on extended components. We show that the median filter method can also be used to remove large-scale imaging artefacts from observations, such as the ripple seen in the EL observation.

    \item We present RM catalogs for our EL, EM, EC, and GL observations. The polarized sky densities in the median-filtered (unfiltered) EL, EM, EC, and GL catalogs are 42.0 (45.0), 31.4 (31.6), 48.0 (51.5), and 20.2 (31.4) components per square degree, respectively. The GL observation sky density is less than half of the EL and EC observations due to source and sidelobe confusion associated with complex emission in the Galactic plane which reduces the number of detected total intensity components.     The median RM uncertainties in the median-filtered (unfiltered) EL, EM, EC, and GL catalogs are 1.55 (1.56), 12.82 (12.74), 1.06 (1.13), and 1.89 (2.71) rad m$^{-2}$, respectively. The relatively large uncertainty in the EM catalog show that extracting meaningful RM values in the mid-band is difficult.

    \item We estimate that the full POSSUM survey will return an RM catalog with upwards of 877 000 polarized radio components with combined-band observations, or upwards of 775 000 polarized radio components with low-band observations. With the Faraday complexity thresholds that we apply in this work to select Faraday simple components, we estimate that the POSSUM catalog will contain $\sim$675 000 combined-band RMs, or $\sim$637 000 low-band RMs, that would be suitable for use in the construction of an RM grid following the method we use in this work. Over 20 000 square degrees of sky, this is an average sky density of 33.8 RMs per square degree with combined-band observations or 31.9 RMs per square degree with low-band observations. However, we estimate that $\sim$25\% of these components may not contribute independent RMs to an RM grid, which would reduce the sky densities to 25.4 and 23.9 RMs per square degree for the combined-band and low-band observing methods, respectively.

    \item From our analysis of expected RM uncertainties and polarized component sky densities, we determine that the combined-band observing strategy returns the highest polarized component sky density and the smallest RM uncertainties (we discuss the disadvantages of this strategy in Section \ref{subsec: survey densities}). While we estimate that the total number of measured RMs would be higher if low- and mid-band observations are analyzed separately (647 total RMs in the two bands versus 553 RMs in the combined-band) due to resolution and beam depolarization effects, the large uncertainties on mid-band RMs make this scenario inferior. If combined-band is not an option, low-band observation should be used. The mid-band data should not be used alone for the full POSSUM survey, but should be combined with the low-band data.

    \item The median-filtered (unfiltered) EL, EM, EC, and GL observations have Faraday simple component sky densities of 35.1 (37.8), 30.6 (30.7), 37.2 (40.5), and 13.5 (23.6) components per square degree, respectively. While the EL, EM, and EC observations see a 3--23\% reduction compared with their respective total median-filtered polarized component sky densities, the GL observation sees a 34\% reduction due to the enhanced Faraday complexity in the Galactic plane as compared to the extragalactic line of sight of the Pilot I observations.

    \item We present RM grids for our median-filtered EL, EM, EC, and GL observations and compare them to the \citet{Hutschenreuter+22} Faraday depth sky map. The EL, EM, and EC RM grids show reasonable agreement, although we see greater RM variation across our RM grid than is present in the Faraday depth sky map due to the vast increase in RM sky density. The GL observation RM grid shows significantly more variation in RM than the Faraday depth sky map, including an overall sign change from negative to positive to negative RMs diagonally across the observation, indicating a large-scale magnetic field reversal on the scale of $\sim3^{\circ}$. The RM sky density of our RM grids is 1--2 orders of magnitude greater than the data set used to construct the currently best available Faraday depth sky map in these regions of sky.

\end{itemize}

The results of the analysis done here suggest many avenues for future work, both with this data and with full POSSUM survey data. One major result in this work is the significantly lower polarized component sky density near the Galactic plane. One goal of RM grids is to understand the geometry and properties of the Galactic magnetic field, and observations near the Galactic plane will be key in doing this because this is where the field is most complicated. Investigating better methods of source finding in the presence of bright, complicated diffuse emission is needed. Additional testing of the median filter method for extended sources is also needed to better understand the effects on recovered RMs and polarized intensity loss in these cases.

We show in Sections \ref{subsec: src types} and \ref{subsubsec: pf vs snr} that the data have a dependence on polarized fraction which needs to be explored in more detail. This has implications for certain science goals such as RM grids, and suggests that components with low polarized fractions need to be used with caution. Further work on measuring and understanding Faraday complexity in POSSUM data is also needed. \citet{Thomson+23} highlight issues with effects of noise and Ricean bias in the current method of calculating $\sigma_{\mathrm{add}}$, and understanding how complexity varies on and of the Galactic plane is also important for a variety of POSSUM science. Determining whether excess complexity in components in the median-filtered GL observation is due to residual contamination from diffuse emission or the whether it is due to variations in Galactic magnetic field on component or source scales would provide further testing for the median filter and would help us deepen our understanding of complexity in Galactic plane regions.

In this work, we approach Faraday complexity as a factor that will influence the construction of  RM grids, and we interpret our results in the context of the current capabilities of POSSUM observations. We note, however, that the measured complexity of a component is influenced by several factors, including spatial resolution of the observation, Faraday resolution ($\delta\phi$), signal-to-noise, and $W_{max}$ (see Equation \ref{eq: W max}; \citealt{Rudnick&Cotton2023}). Extrapolating toward a true fraction of intrinsically Faraday-simple components could be done with a comparison of a subset of our POSSUM data with higher angular resolution and broader-band observations (resulting in higher Faraday resolution) from telescopes such as MeerKAT and ATCA, and with lower-frequency ($<$800 MHz) observations with the future Square Kilometre Array. We suggest this as a valuable future work.

Additionally, the two Faraday complexity metrics used in this work aim to quantify the level of complexity present in the polarization spectra of components, but they are not designed to identify the origins of the physical complexity. While investigating the origins of the Faraday complexity present in our data is beyond the scope of this paper, a tool designed for this purpose would be ideal for the interpretation of future POSSUM and other radio survey data. QU fitting is a common tool used to attempt to determine the physical origins of Faraday complexity, although it has been shown to have difficulties in the presence of ISM turbulence and Faraday thick structures (e.g. Sun et al. 2015, Basu et al. 2019). We highlight the need for tools that can reliably extract meaningful information on the origins of the Faraday complexity of a component for future POSSUM science.

\section*{Acknowledgements}

We thank the anonymous reviewer for their careful review of the manuscript and their thoughtful feedback.

The Dunlap Institute is funded through an endowment established by the David Dunlap family and the University of Toronto. B.M.G. acknowledges the support of the Natural Sciences and Engineering Research Council of Canada (NSERC) through grant RGPIN-2022-03163, and of the Canada Research Chairs program. SPO acknowledges support from the Comunidad de Madrid Atracción de Talento program via grant 2022-T1/TIC-23797.

This scientific work uses data obtained from Inyarrimanha Ilgari Bundara / the Murchison Radio-astronomy Observatory. We acknowledge the Wajarri Yamaji People as the Traditional Owners and native title holders of the Observatory site. CSIRO’s ASKAP radio telescope is part of the Australia Telescope National Facility (\url{https://ror.org/05qajvd42}). Operation of ASKAP is funded by the Australian Government with support from the National Collaborative Research Infrastructure Strategy. ASKAP uses the resources of the Pawsey Supercomputing Research Centre. Establishment of ASKAP, Inyarrimanha Ilgari Bundara, the CSIRO Murchison Radio-astronomy Observatory and the Pawsey Supercomputing Research Centre are initiatives of the Australian Government, with support from the Government of Western Australia and the Science and Industry Endowment Fund. The POSSUM project (\url{https://possum-survey.org}) has been made possible through funding from the Australian Research Council, the Natural Sciences and Engineering Research Council of Canada, the Canada Research Chairs Program, and the Canada Foundation for Innovation.

This research has made use of the CIRADA cutout service at URL cutouts.cirada.ca, operated by the Canadian Initiative for Radio Astronomy Data Analysis (CIRADA). CIRADA is funded by a grant from the Canada Foundation for Innovation 2017 Innovation Fund (Project 35999), as well as by the Provinces of Ontario, British Columbia, Alberta, Manitoba and Quebec, in collaboration with the National Research Council of Canada, the US National Radio Astronomy Observatory and Australia’s Commonwealth Scientific and Industrial Research Organisation. 

This research used the facilities of the Canadian Astronomy Data Centre operated by the National Research Council of Canada with the support of the Canadian Space Agency.

This research made use of: SciPy \citep{SciPy2020}; Astropy\footnote{http://www.astropy.org}, a community-developed core Python package and an ecosystem of tools and resources for astronomy \citep{astropy:2022}; Montage, funded by the National Science Foundation under Grant Number ACI-1440620, and was previously funded by the National Aeronautics and Space Administration's Earth Science Technology Office, Computation Technologies Project, under Cooperative Agreement Number NCC5-626 between NASA and the California Institute of Technology.

\clearpage   
\appendix

\vspace{-5mm}
\section{Example $QU$ spectra and Faraday depth functions}\label{app: example spectra}

We provide some example total intensity and polarized spectra and Faraday spectra for a selection of components from our four observations. We include examples of components with low and high polarized intensity, that are both Faraday simple and Faraday complex, and both with and without the median filter applied.

\begin{figure*}[h!]
    \centering
    \subfloat[Median-filtered EL Faraday simple component J$221024-505540$. \textit{Top left:} Stokes $I$ spectrum (black) and model fit from 1D RM synthesis with \texttt{RM-Tools} (red dashed). \textit{Top center:} Stokes $q$ (blue), $u$ (red), and polarized fraction (black) spectra. \textit{Top right:} Polarization angle $\psi$ as a function of $\lambda^2$. \textit{Bottom:} Clean Faraday spectrum (black) with clean peaks (green) and 8$\sigma$ GES clean threshold (red dashed).]{
    \includegraphics[width=0.84\textwidth]{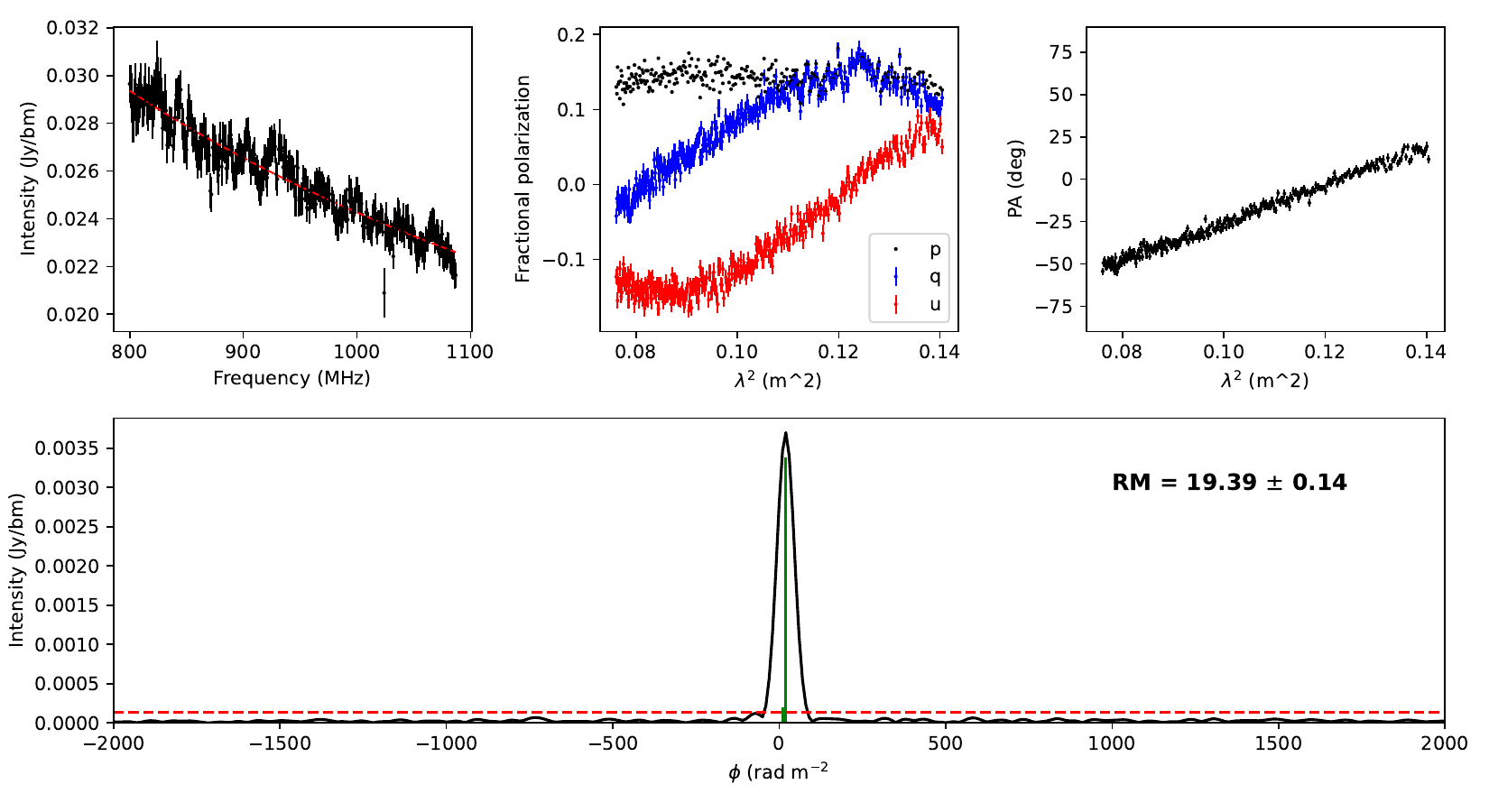}
        \label{fig: low medfilt bright simple}} \\
    \subfloat[As for \ref{fig: low medfilt bright simple}, but in the EM, convolved to the resolution of EL and EC (21 arcsec).]{
    \includegraphics[width=0.84\textwidth]{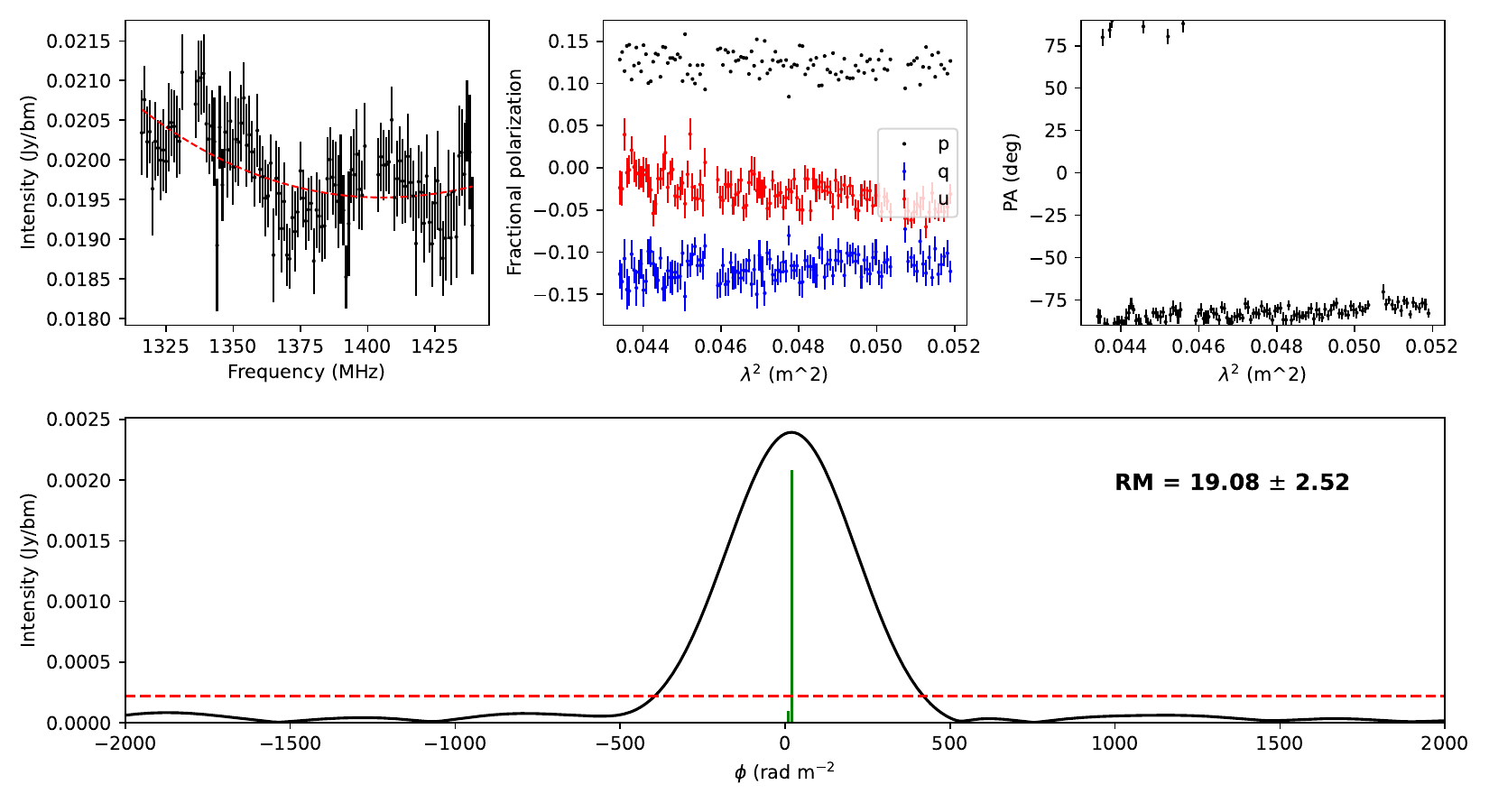}
        \label{fig: mid medfilt bright simple}}
\caption{}
\end{figure*}
\begin{figure*}\ContinuedFloat
    \centering
    \subfloat[As for \ref{fig: low medfilt bright simple}, but for the EC data.]{
    \includegraphics[width=0.97\textwidth]{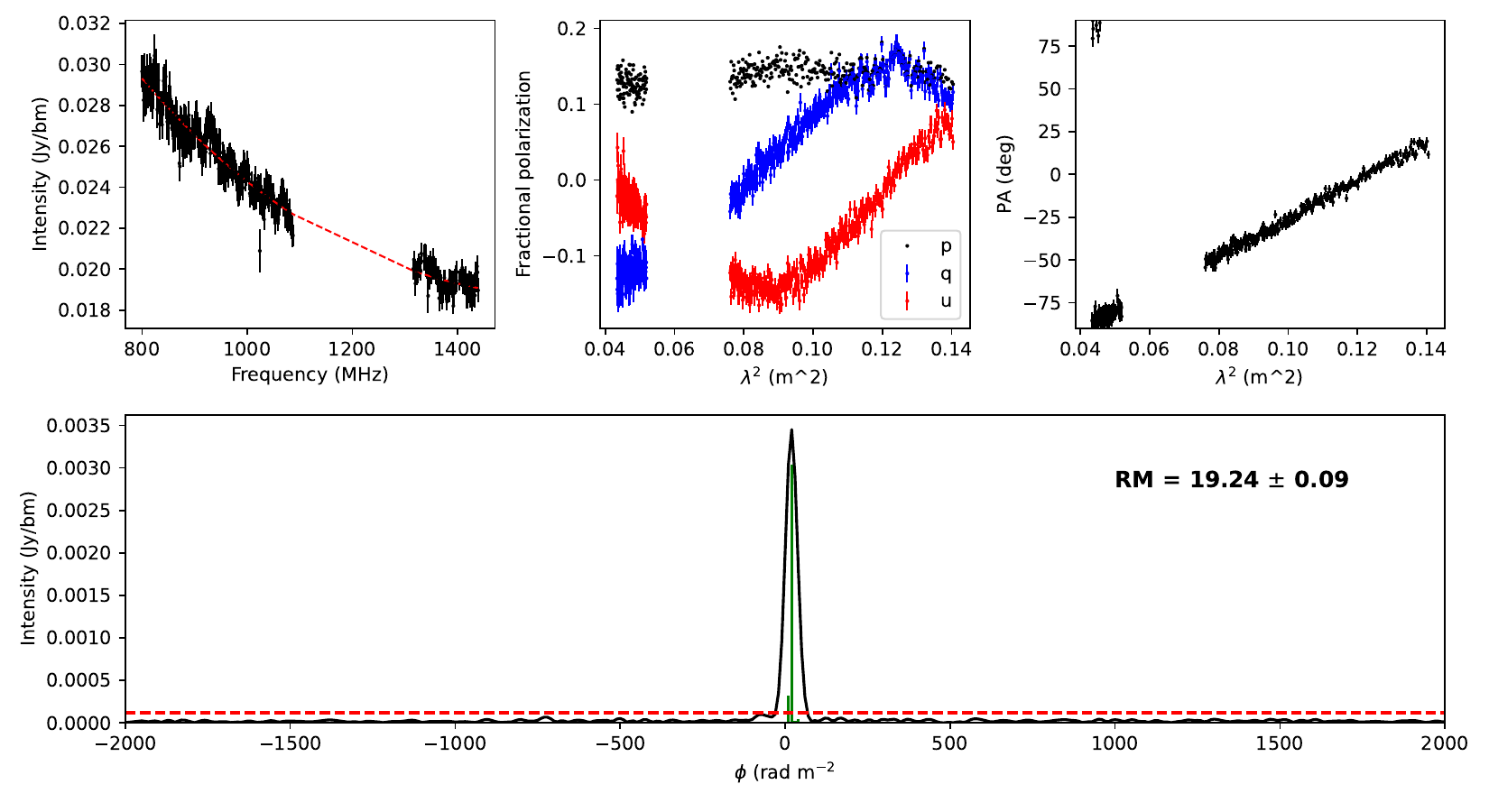}
        \label{fig: comb medfilt bright simple}}\\
    \subfloat[Unfiltered GL observation component J$154230-545826$. Plots are as described in Figure \ref{fig: low medfilt bright simple}. This is an example of where foreground diffuse emission is brighter in polarized intensity than a faint background component. We confirm that this is in fact a situation of diffuse emission dominating the background component by visual inspection of the diffuse and component maps from the median filter process. The clean peak at $\phi$ = +31.7 rad m$^{-2}$ is due to foreground diffuse emission, while the clean peak at clean peak at $\phi$ = $-32.3$ rad m$^{-2}$ is the background component RM. The clean peak associated with the diffuse emission is brighter in polarized intensity than the peak associated with the background component, so 1D RM synthesis returns the diffuse emission $\phi$ as the component RM.]{
    \includegraphics[width=0.97\textwidth]{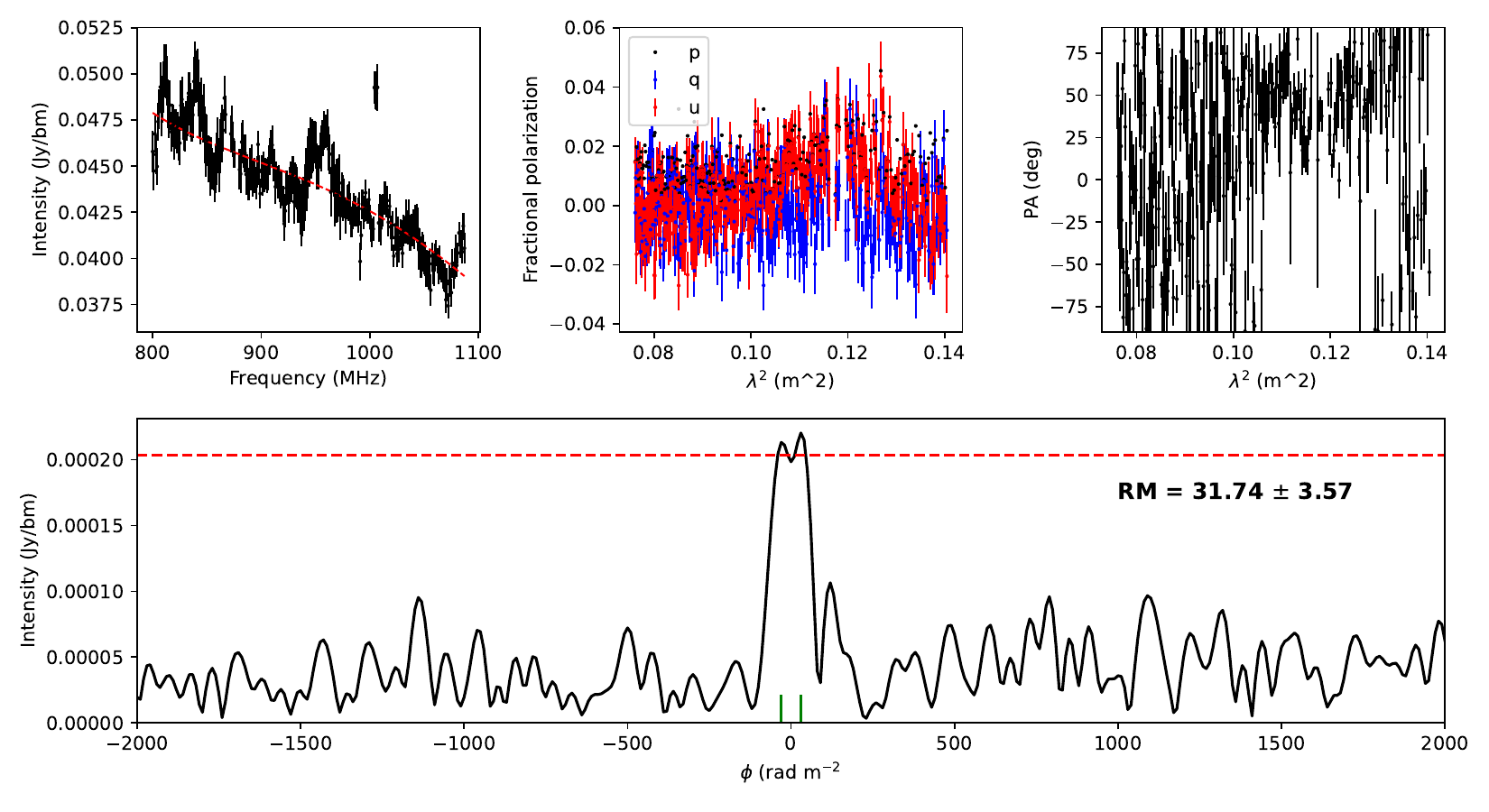}
        \label{fig: gal nofilt DE faint}}
\caption{}
\end{figure*}
\begin{figure*}\ContinuedFloat
    \centering
    \subfloat[As for \ref{fig: gal nofilt DE faint}, but after median filtering. The application of the median filter reduces the intensity of the diffuse emission peak at $\phi$ = +31.7 rad m$^{-2}$ to below the 8$\sigma$ GES clean threshold, leaving the component clean peak at $\phi$ = $-32.3$ rad m$^{-2}$ as the brightest detectable peak, as desired.]{
    \includegraphics[width=0.97\textwidth]{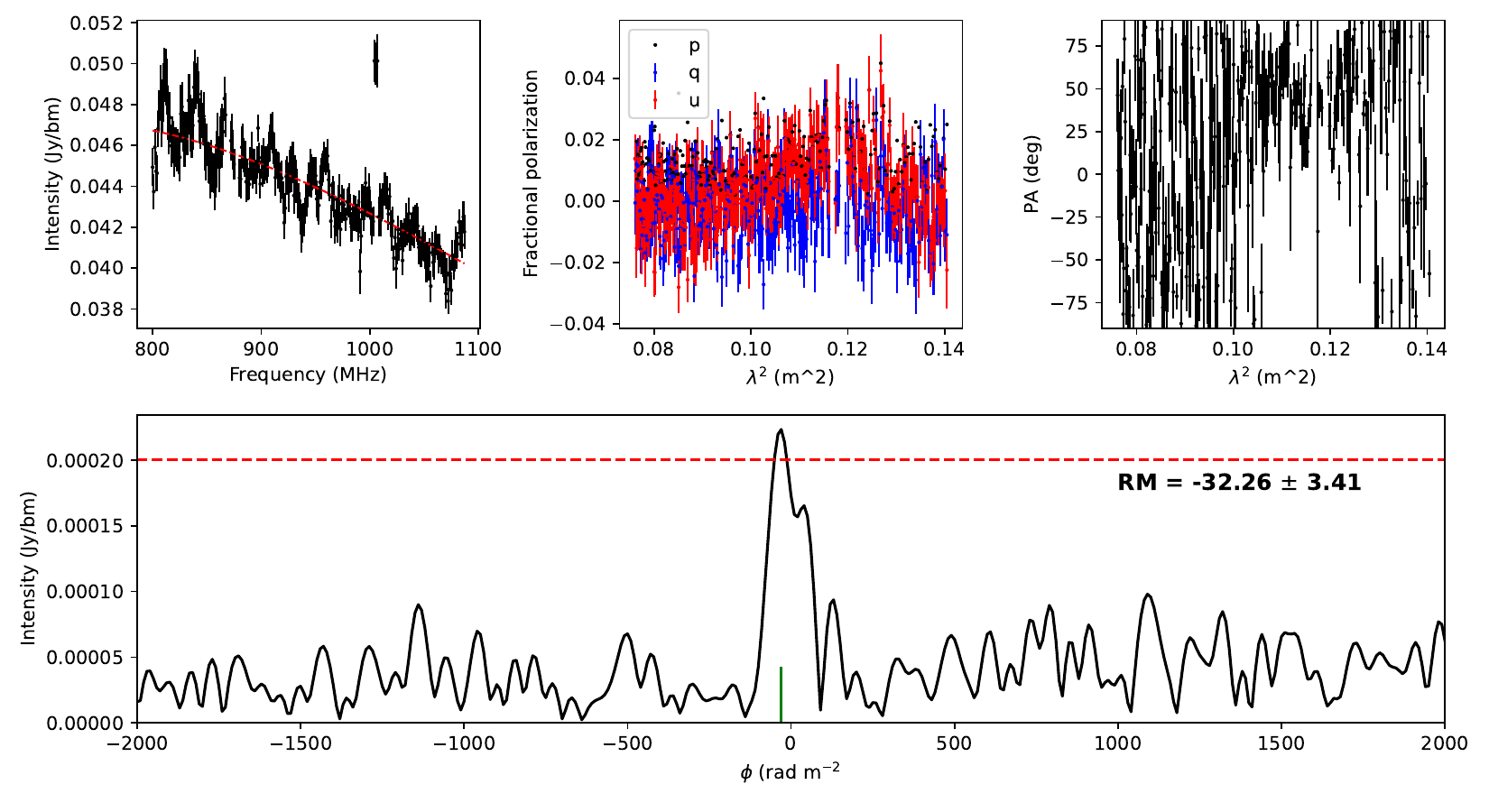}
        \label{fig: gal medfilt DE faint}}\\
    \subfloat[Unfiltered EL Faraday complex component J220307-494052. Plots are as described in Figure \ref{fig: low medfilt bright simple}. The dominant clean peak is at $\phi$ = 9.7 $\pm$ 1.0 rad m$^{-2}$.]{
    \includegraphics[width=0.97\textwidth]{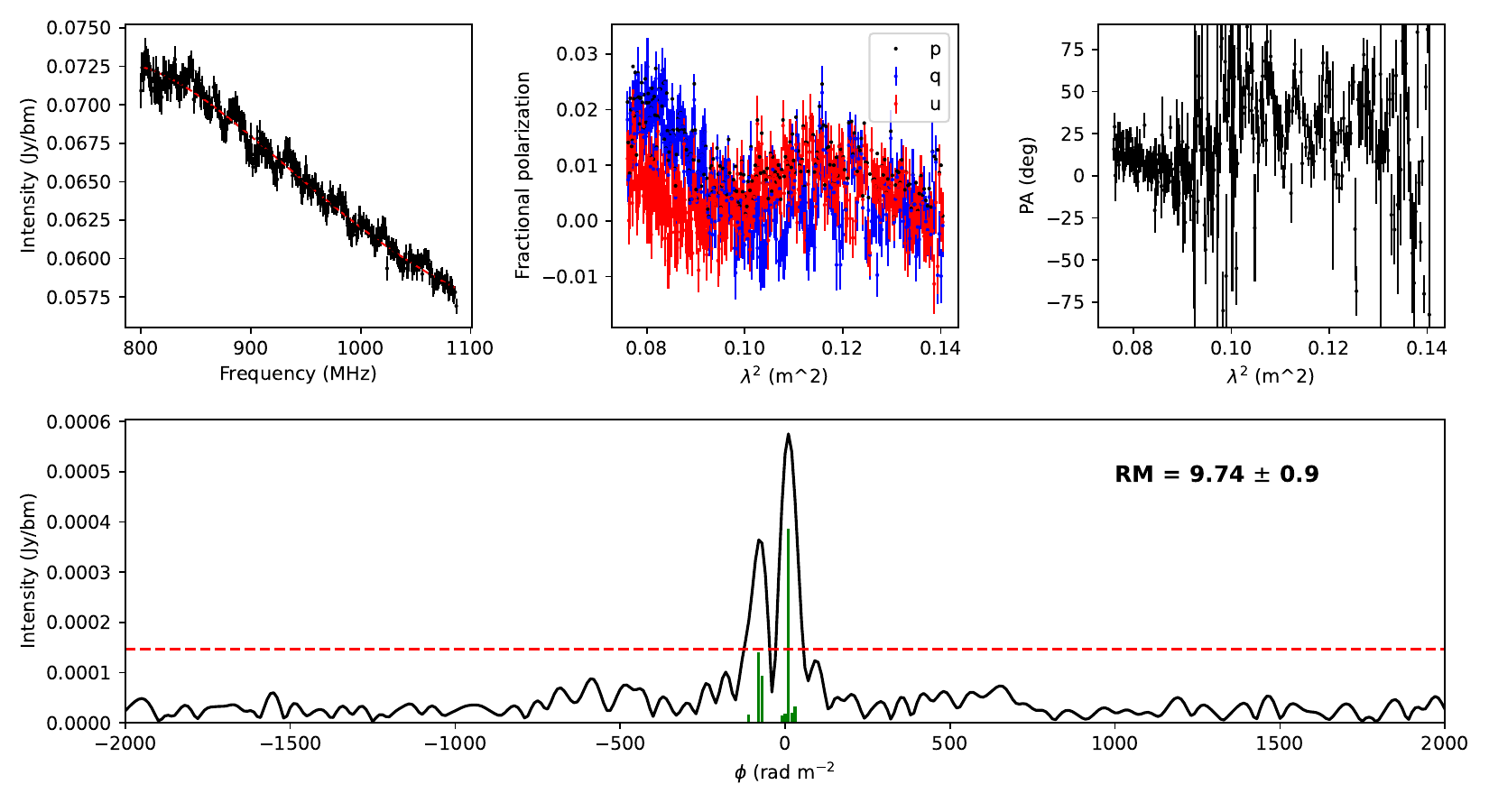}
        \label{fig: low no filt complex}}
\caption{}
\end{figure*}
\begin{figure*}\ContinuedFloat
    \subfloat[As for \ref{fig: low no filt complex}, but after median filtering. We see the expected reduction in polarized intensity due to the application of the median filter in the Stokes $qu$ spectra and Faraday spectrum as compared with \ref{fig: low no filt complex}. The dominant clean peak is at  $\phi$ = 9.4 $\pm$ 0.9 rad m$^{-2}$, which is consistent with \ref{fig: low no filt complex}.]{
    \includegraphics[width=0.97\textwidth]{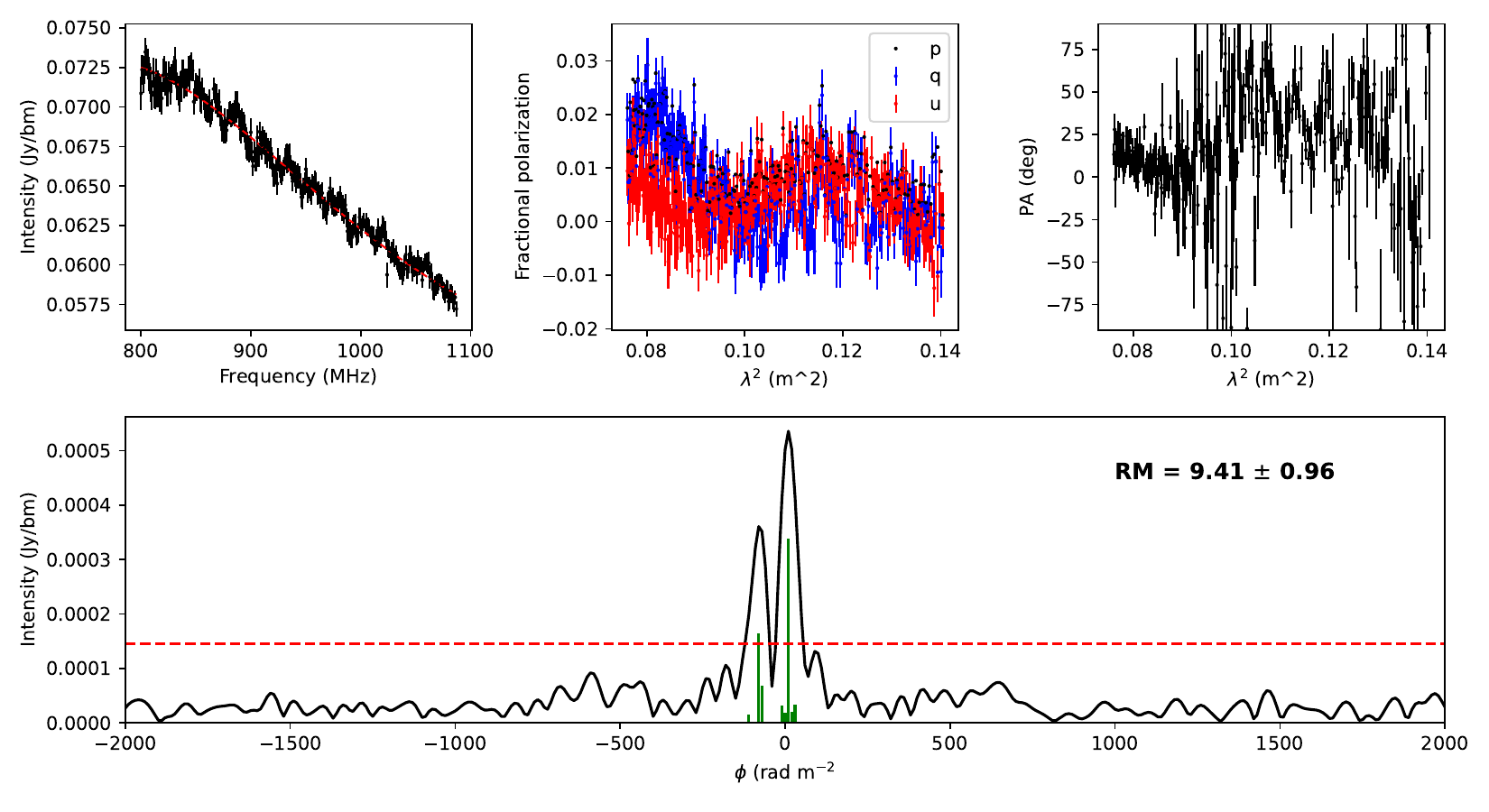}
        \label{fig: low medfilt complex}}
\caption{Example Stokes $Iqu$ spectra, polarization angles, and Faraday spectra for a variety of components from our four observations.}
\label{fig: example spectra}
\end{figure*}
\FloatBarrier

\clearpage

\newpage
\section{Column name descriptions and catalogs}
\label{app: cols and cats}

Here we define the table columns for our polarized component catalogs. We follow the \texttt{RM-\textsc{Table}} standard convention for RM catalogs \citep{VanEck+23} plus provide some additional columns. In Table \ref{tab: gal cat} we provide the units for each column and list the first two rows from the median-filtered GL polarized component catalog.

\begin{itemize}
    \setlength{\itemsep}{0pt}
    \setlength{\parskip}{0pt}
    \item \texttt{component\_name}: component name from the source finder catalog
    \item \texttt{cat\_id}: component identifier from the source finder catalog
    \item \texttt{ra}: right ascension of the component
    \item \texttt{ra\_err}: error in the right ascension
    \item \texttt{dec}: declination of the component
    \item \texttt{dec\_err}: error in declination
    \item \texttt{l}: Galactic longitude of the component
    \item \texttt{b}: Galactic latitude of the component
    \item \texttt{pos\_err}: positional uncertainty of the component; this is the larger value of \texttt{ra\_err} and \texttt{dec\_err}
    \item \texttt{int\_flux\_I}: integrated Stokes $I$ flux density of component from the source finder catalog
    \item \texttt{int\_flux\_I\_err}: error in integrated Stokes $I$ flux density
    \item \texttt{peak\_int\_I}: peak Stokes $I$ intensity of component from the source finder catalog
    \item \texttt{peak\_int\_I\_err}: error in peak Stokes $I$ intensity
    \item \texttt{spectral\_index}: spectral index of Stokes $I$; this is taken to be the $C_1$ from the \texttt{RM-Tools} model fit to the Stokes $I$ spectrum (see Equation \ref{eq: stokes I model})
    \item \texttt{spectral\_index\_err}: error in spectral index of Stokes $I$
    \item \texttt{maj\_axis\_source finder}: semi-major axis of the component from the source finder catalog
    \item \texttt{maj\_axis\_source finder\_err}: error in the semi-major axis of the component from the source finder catalog
    \item \texttt{min\_axis\_source finder}: semi-minor axis of the component from the source finder catalog
    \item \texttt{min\_axis\_source finder\_err}: error in the semi-mmior axis of the component from the source finder catalog
    \item \texttt{pos\_angle\_source finder}: position angle of the component from the source finder, increasing east from north
    \item \texttt{pos\_angle\_source finder\_err}: error in the position angle of the component from the source finder catalog
    \item \texttt{decon\_maj\_axis\_source finder}: deconvolved semi-major axis of the component from the source finder catalog
    \item \texttt{decon\_maj\_axis\_source finder\_err}: error in the deconvolved semi-major axis of the component from the source finder catalog
    \item \texttt{decon\_min\_axis\_source finder}: deconvolved semi-minor axis of the component from the source finder catalog
    \item \texttt{decon\_min\_axis\_source finder\_err}: error in the deconvolved semi-minor axis of the component from the source finder catalog
    \item \texttt{decon\_pol\_angle\_source finder}: deconvolved position angle of the component from the source finder catalog
    \item \texttt{decon\_pol\_angle\_source finder\_err}: error in the deconvolved position angle of the component from the source finder catalog
    \item \texttt{rm}: rotation measure
    \item \texttt{rm\_err}: error in the RM, $\delta$RM (see Equation \ref{eq: deltaRM})
    \item \texttt{polint}: peak polarized intensity, $P$ (see Equation \ref{eq: P, psi, p}); this value is corrected for polarization bias (see \texttt{pol\_bias} column description)
    \item \texttt{polint\_err}: error in the peak polarized intensity
    \item \texttt{fracpol}: fractional polarization, $p$ (see Equation \ref{eq: P, psi, p})
    \item \texttt{fracpol\_err}: error in fractional polarization; this is currently not an output of \texttt{RM-Tools}
    \item \texttt{polangle}: polarization angle at \texttt{reffreq\_pol}, $\psi$ (see Equation \ref{eq: P, psi, p}), increasing east from north
    \item \texttt{polangle\_err}: error in polarization angle 
    \item \texttt{derot\_polangle}: derotated polarization angle, increasing east from north
    \item \texttt{derot\_polangle\_err}: error in derotated polarization angle
    \item \texttt{reffreq\_I}: reference frequency for Stokes $I$, calculated as the mean value of the unflagged frequency channels
    \item \texttt{stokesI}: Stokes $I$ intensity at the reference frequency (\texttt{reffreq\_I}), derived from the \texttt{RM-Tools} Stokes I model fit (see Equation \ref{eq: stokes I model})
    \item \texttt{stokesI\_err}: error in Stokes $I$ flux at the reference frequency (\texttt{reffreq\_I})
    \item \texttt{stokesQ}: Stokes $Q$ intensity at peak rotation measure (\texttt{rm})
    \item \texttt{stokesQ\_err}: error in Stokes $Q$
    \item \texttt{stokesU}: Stokes $U$ intensity at peak rotation measure (\texttt{rm})
    \item \texttt{stokesU\_err}: error in Stokes $U$
    \item \texttt{reffreq\_pol}: reference frequency for polarization, chosen to correspond to $\lambda^2_0$ (see Equation \ref{eq: ref lamsq})
    \item \texttt{rmsf\_fwhm}: full width half maximum of the rotation measure spread function, $\delta\phi$ (see Equation \ref{eq: rmsf fwhm})
    \item \texttt{noise\_chan}: the median noise in the combined Stokes $QU$ frequency channels
    \item \texttt{dFSth}: the theoretical noise in the Faraday spectrum, calculated from the per-channel errors
    \item \texttt{snr\_pol}: signal-to-noise in polarization, S/N$_{\mathrm{pol}}$ (see Equation \ref{eq: s/n_pol}), measured as the peak polarized intensity divided by \texttt{dFSth}
    \item \texttt{stokesI\_fit\_coeffs}: coefficients of the \texttt{RM-Tools} model fit to the Stokes $I$ spectrum, listed from highest to lowest order ($C_3$, ..., $C_0$; see Equation \ref{eq: stokes I model})
    \item \texttt{stokesI\_fit\_chisqred}: reduced chi-squared statistic of the model fit to the Stokes $I$ spectrum
    \item \texttt{sigmaAdd}: $\sigma_{\mathrm{add}}$ complexity metric (see Equation \ref{eq: sigadd total} and Section \ref{subsec: sigadd metric})
    \item \texttt{sigmaAdd\_plus\_err}: plus error in $\sigma_{\mathrm{add}}$ complexity metric, $\delta\sigma_{\mathrm{add},+}$
    \item \texttt{sigmaAdd\_minus\_err}: minus error in $\sigma_{\mathrm{add}}$ complexity metric, $\delta\sigma_{\mathrm{add},-}$
    \item \texttt{rm\_width}: width in Faraday depth, which is reported as the second moment of the clean peaks, M$_2$
    \item \texttt{rm\_width\_err}: error in the width in Faraday depth; this is currently not an output of \texttt{RM-Tools}
    \item \texttt{Nchan}: number of frequency channels used to determine the RM of the component
    \item \texttt{beam\_maj}: semi-major axis of the synthesized beam
    \item \texttt{beam\_min}: semi-minor axis of the synthesized beam
    \item \texttt{beam\_bpa}: position angle of the synthesized beam, increasing east from north
    \item \texttt{reffreq\_beam}: reference frequency for the synthesized beam
    \item \texttt{minfreq}: lowest frequency used to calculate the RM
    \item \texttt{maxfreq}: highest frequency used to calculate the RM
    \item \texttt{channelwidth}: channel width; this is 1 MHz for all components
    \item \texttt{rm\_method}: method used to determine the RM; this is `RM Synthesis - Fractional polarization' for all components
    \item \texttt{complex\_flag}: indication of whether component is considered complex or not, `Y' for Yes or `N' for No
    \item \texttt{complex\_test}: Faraday complexity test used; this is `\texttt{sigmaAdd or M2}' for all components
    \item \texttt{ionosphere}: method used to correct for ionospheric effects; this is `None' for all components since no correction was applied to any of the Pilot observations
    \item \texttt{Ncomp}: integer number of measured RM values for the component; this is `1' for all components since we do not identify `sources' in this work, and we report an RM for each component identified by the source finder
    \item \texttt{pol\_bias}: method used to correct for polarization bias; this is the method from \citet{George+12} used by \texttt{RM-Tools} for all components
    \item \texttt{flux\_type}: method used to extract the Stokes spectra for the component; this is `box' for all components (see Section \ref{subsec: spec extract})
    \item \texttt{aperture}: size of the integration aperture over which the Stokes spectra have been integrated
    \item \texttt{telescope}: name of the telescope by which the data were observed; this is `ASKAP' for all components
    \item \texttt{int\_time}: integration time of observation in seconds; this is 36 000 seconds for all components
    \item \texttt{epoch}: median epoch of observation used to determine the RM; this is the time that the observation was half-complete
    \item \texttt{interval}: interval of the observation used to determine the RM; this is the same as the integration time for all components
    \item \texttt{leakage}: estimate of the residual leakage local to the component in the observation after on- and off-axis leakage correction has been applied
    \item \texttt{beamdist}: distance of the component from the (nearest) primary beam center 
    \item \texttt{catalog}: name of the catalog; that is this paper DOI for all components
    \item \texttt{dataref}: references to the source of the data used to determine the RM; that is this paper for all components
    \item \texttt{type}: component classification; this is `Unknown' for all components
    \item \texttt{stokesV}: Stokes $V$ intensity; this is `NaN' for all components
    \item \texttt{stokesV\_err}: error in Stokes $V$ intensity; this is `NaN' for all components
    \item \texttt{notes}: any notes pertaining to the component
\end{itemize}

\newpage

\begin{longtable}{| p{.26\textwidth} | p{.08\textwidth} | p{0.27\textwidth} p{0.27\textwidth} |}
\caption{First two rows of the median-filtered GL polarized component catalog, ordered by descending S/N$_{\mathrm{pol}}$.}\label{tab: gal cat} \\
\hline
Column name & Units & Row 1 & Row 2 \\ \hline
component\_name & --- & `J$160309-550549$' & `J$155227-552650$'  \\ 
cat\_id & --- & `SB43773\_component\_22a' & `SB43773\_component\_51a' \\ 
ra & deg & 240.7877 & 238.1146 \\ 
ra\_err & arcsec & 0.00 & 0.01 \\ 
dec & deg & $-55.0970$ & $-55.4473$ \\ 
dec\_err & arcsec & 0.00 & 0.01 \\ 
l & deg & 328.0805 & 326.6921 \\ 
b & deg & $-1.8762$ & $-1.1585$ \\ 
pos\_err & arcsec & 0.00 & 0.01 \\ 
int\_flux\_I & mJy & 357.242 & 167.999 \\ 
int\_flux\_I\_err & mJy & 0.135 & 0.289 \\ 
peak\_flux\_I & mJy/beam &  342.944 & 162.728 \\ 
peak\_flux\_I\_err & mJy/beam & 0.076 & 0.163 \\ 
spectral\_index & --- & $-0.741$ & $-0.686$ \\ 
spectral\_index\_err & --- & 0.002 & 0.004  \\ 
maj\_axis\_sourcefinder & arcsec & 18.60 & 18.38 \\ 
maj\_axis\_sourcefinder\_err & arcsec & 0.00 & 0.02 \\ 
min\_axis\_sourcefinder & arcsec & 18.15 & 18.20 \\ 
min\_axis\_sourcefinder\_err & arcsec & 0.00 & 0.02 \\ 
pos\_angle\_sourcefinder & deg & 71.53 & 10.08 \\ 
pos\_angle\_sourcefinder\_err & deg & 0.01 & 0.07 \\ 
decon\_maj\_axis\_sourcefinder & arcsec & 4.68 & 3.73 \\ 
decon\_maj\_axis\_sourcefinder\_err & arcsec & 0.0 & 0.0 \\ 
decon\_min\_axis\_sourcefinder & arcsec & 2.31 & 2.66 \\ 
decon\_min\_axis\_sourcefinder\_err & arcsec & 0.04 & 0.13 \\ 
decon\_pos\_angle\_sourcefinder & deg & 71.53 & 10.08 \\ 
decon\_pos\_angle\_sourcefinder\_err & deg & 0.01 & 0.07 \\ 
rm & rad m$^{-2}$ & 247.79 & $-113.08$ \\ 
rm\_err & rad m$^{-2}$ & 0.05 & 0.06 \\ 
polint & Jy/beam & 0.01300 & 0.01325 \\ 
polint\_err & Jy/beam & 2e-5 & 3e-5 \\ 
fracpol & --- & 0.038 & 0.083 \\ 
fracpol\_err & --- & --- & --- \\ 
polangle & deg & 119.70 & 116.60 \\ 
polangle\_err & deg & 0.05 & 0.05 \\ 
derot\_polangle & deg & 115.72 & 61.72 \\ 
derot\_polangle\_err & deg & 0.27 & 0.31 \\ 
reffreq\_I & Hz & 940029423 & 935681011 \\ 
stokesI & Jy/beam & 0.34533 & 0.16006 \\ 
stokesI\_err & Jy/beam & 7e-5 & 8e-5 \\ 
stokesQ & Jy/beam & $-0.00655$ & $-0.00784$ \\ 
stokesQ\_err & Jy/beam & 2e-5 & 3e-5 \\ 
stokesU & Jy/beam & $-0.01107$ & $-0.01048$ \\ 
stokesU\_err & Jy/beam & 2e-5 & 3e-5 \\ 
reffreq\_pol & Hz & 940029436 & 935681024 \\ 
rmsf\_fwhm & rad m$^{-2}$ & 62.9 & 60.9 \\ 
noise\_chan & Jy/beam & 3.6e-4 & 4.2e-4 \\ 
dFSth & Jy/beam & 2e-5 & 3e-5 \\ 
snr\_pol & --- & 588.81 & 527.14 \\ 
stokesI\_fit\_coeffs & --- & 0.000, $-1.427$, $-0.741$, 0.344 & 
0.000, $-1.972$, $-0.686$, 0.159 \\ 
stokesI\_fit\_chisqred & --- & 62.00 & 10.49 \\ 
sigmaAdd & --- & 4.58 & 9.04 \\ 
sigmaAdd\_plus\_err & --- & 0.16 & 0.27 \\ 
sigmaAdd\_minus\_err & --- & 0.17 & 0.29 \\ 
rm\_width & rad m$^{-2}$ & 101.5 & 40.7 \\ 
rm\_width\_err & rad m$^{-2}$ & --- & --- \\ 
Nchan & --- & 278 & 278 \\ 
beam\_maj & arcsec & 16.5 & 16.5 \\ 
beam\_min & arcsec & 16.5 & 16.5 \\ 
beam\_pa & deg & 0.0 & 0.0 \\ 
reffreq\_beam & Hz & 0 & 0 \\ 
minfreq & Hz & 799990740 & 799990740 \\ 
maxfreq & Hz & 1086990740 & 1086990740 \\ 
channelwidth & Hz & 1000000 & 1000000 \\ 
rm\_method & --- & `RM Synthesis - \newline Fractional polarization' & `RM Synthesis - \newline Fractional polarization' \\ 
complex\_flag & --- & `Y' & `Y' \\ 
complex\_test & --- & `Sigma\_add or Second\_moment' & `Sigma\_add or Second\_moment' \\ 
ionosphere & ---  & `None' & `None' \\ 
Ncomp & --- & 1 & 1 \\ 
pol\_bias & --- & `2012PASA...29..214G' & `2012PASA...29..214G' \\ 
flux\_type & --- & `box' & `box' \\ 
aperture & deg  & 0.0028 & 0.0028 \\ 
telescope & ---  & `ASKAP' & `ASKAP' \\ 
int\_time & s & 36001 & 36001 \\ 
epoch & MJD & 59829.4 & 59829.4 \\ 
interval & days & 0.417 & 0.417 \\ 
leakage & ---  & 0.006 & 0.008 \\ 
beamdist & deg & 0.343 & 0.113 \\ 
catalog & --- & `10.3847/1538-3881/ad2fc8' & `10.3847/1538-3881/ad2fc8' \\ 
dataref & --- & --- & --- \\ 
type & --- & `Unknown' & `Unknown' \\ 
stokesV & Jy/beam & NaN & NaN \\ 
stokesV\_err & Jy/beam & NaN & NaN \\ 
notes & --- & `' & `' \\ \hline
\end{longtable}

\newpage
\section{Rotation measure grids of observations with no median filter applied}\label{app: nofilt RMgrids}

Here we prove the RM grids of our four observations with out the application of the median filter. In particular, we note the increased number of components in the lower left and right regions of Figure \ref{fig: gal nofilt RMgrid}. Many of these polarized component detections are due to foreground diffuse emission and not true polarized background components.

\begin{figure*}[!h]
    \centering
    \subfloat[EL observation.\label{fig: SB10635 nofilt RMgrid}]{\includegraphics[width=0.71\textwidth]{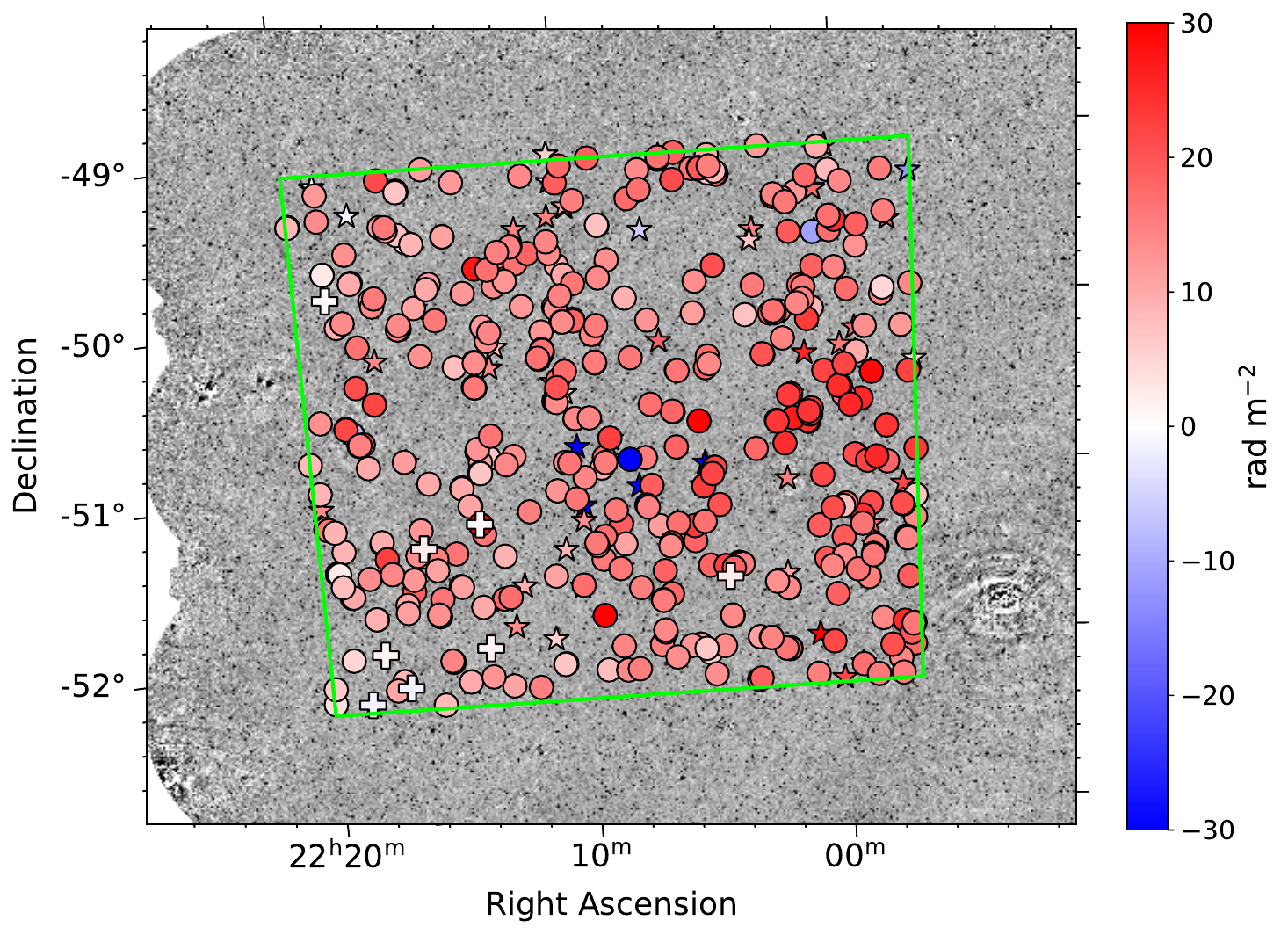}}\\%
    \subfloat[EM observation.]{
    \includegraphics[width=0.71\textwidth]{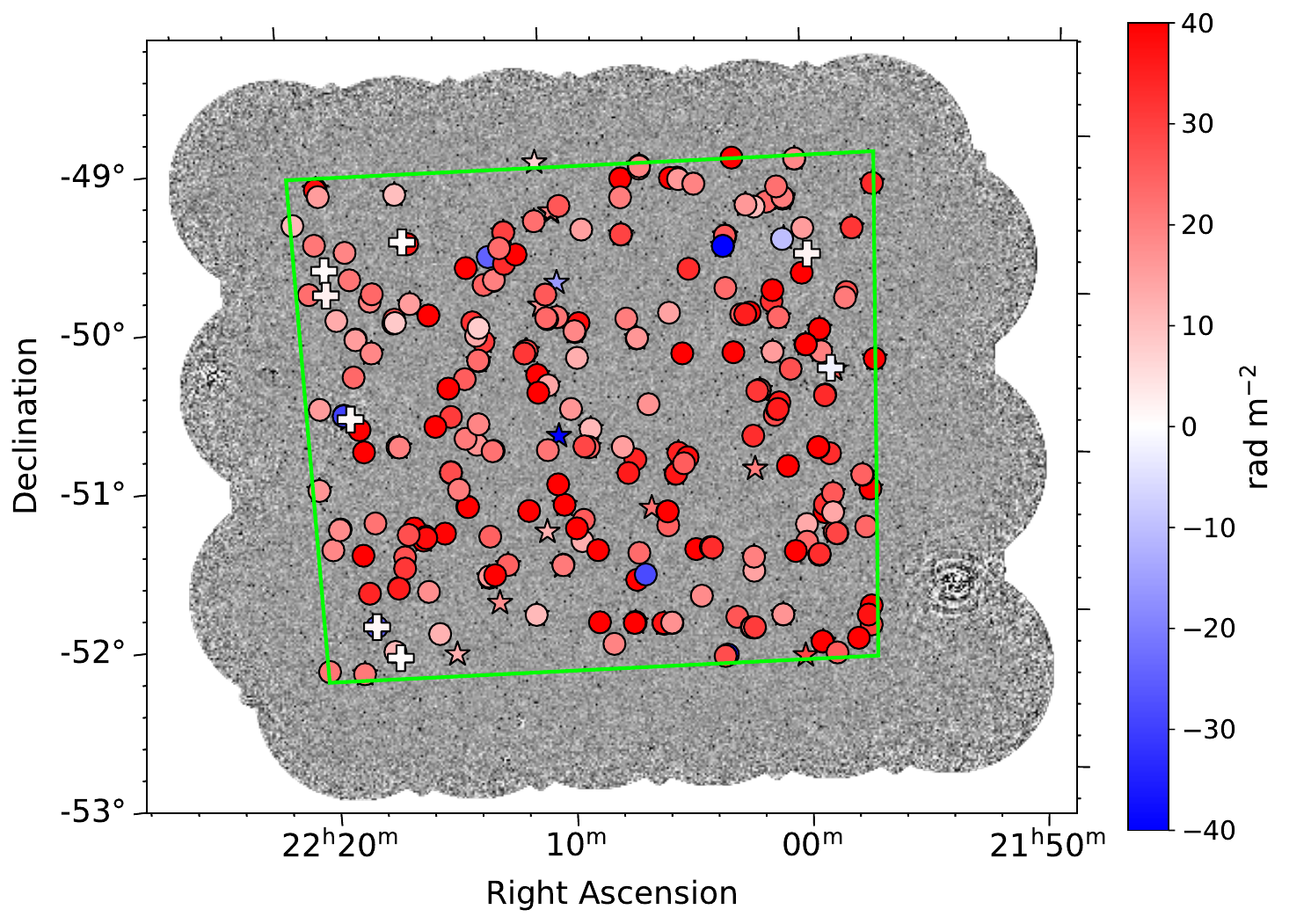}
        \label{fig: SB10043 nofilt RMgrid}}
\caption{}
\end{figure*}
\begin{figure*}\ContinuedFloat
    \centering
    \subfloat[EC observations.]{
    \includegraphics[width=0.8\textwidth]{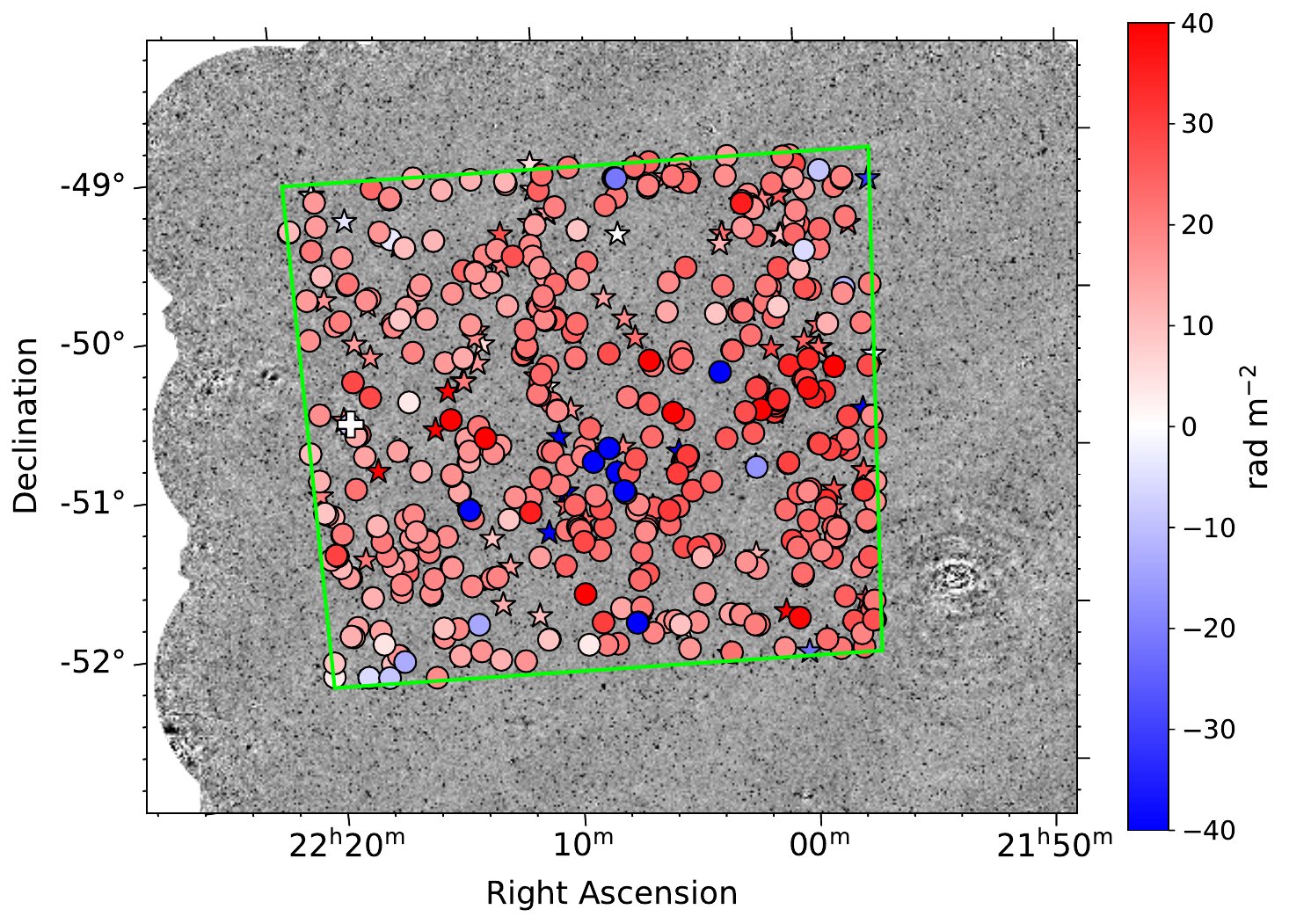}
        \label{fig: comb nofilt RMgrid}}\\
    \subfloat[GL observation.]{
    \includegraphics[width=0.8\textwidth]{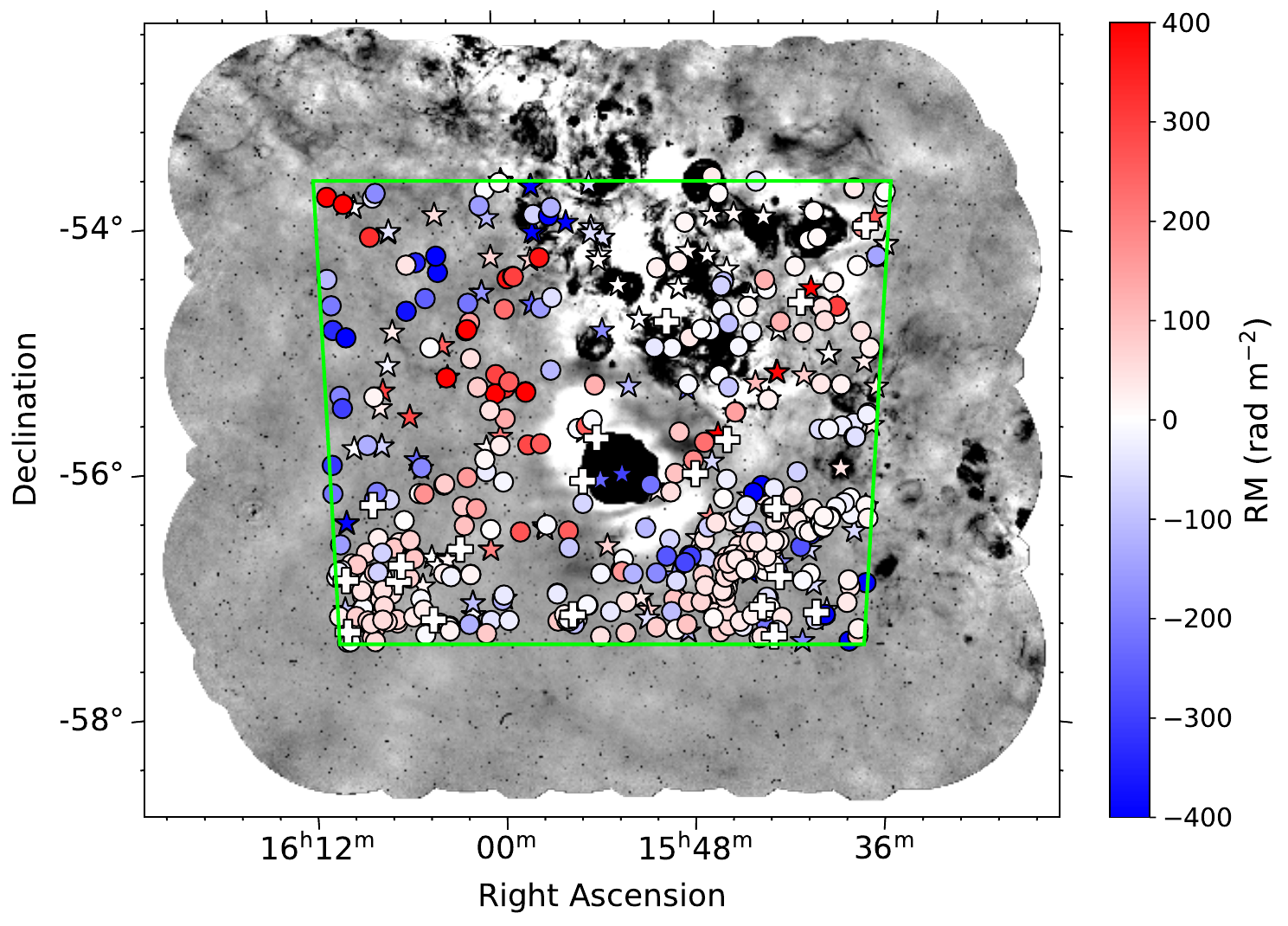}
        \label{fig: gal nofilt RMgrid}}
\caption{RM grids of the EL, EM, EC, and GL observations with no median filter applied. The markers indicate the presence of a polarized component at that location in the observation. Positive RMs are red and negative RMs are blue, while the depth of the color indicates the magnitude of the RM. Circle and star markers denote Faraday simple and complex components, respectively. Crosses indicate Faraday simple components where the RM is consistent with zero within uncertainty.}
\label{fig: nofilt RMgrids}
\end{figure*}
\FloatBarrier

\bibliography{RMgrids_main}{}
\bibliographystyle{aasjournal}

\end{document}